\documentclass[epj]{svjour}

\usepackage[english]{babel}
\usepackage[utf8x]{inputenc}
\usepackage[T1]{fontenc}
\usepackage{tgtermes}
\usepackage{amsmath}
\usepackage{graphicx}
\usepackage[colorinlistoftodos]{todonotes}
\usepackage[colorlinks=true, allcolors=blue]{hyperref}
\usepackage{color, colortbl}
\usepackage{braket}
\usepackage{booktabs}
\usepackage{amsfonts}
\usepackage{verbatim}
\usepackage{caption}
\usepackage{subcaption}

\def\beq {\begin{equation}}
\def\eeq {\end{equation}}
\def\w {\omega}

\def\bfr {\mathbf{r}}

\begin{document}
\title{Spectroscopy of the Hubbard dimer: the spectral potential}
\author{Marco Vanzini\inst{1,2}  \and Lucia Reining\inst{1,2} \and Matteo Gatti\inst{1,2,3}
}                     
%
%
\newcommand{\lsi}{Laboratoire des Solides Irradi\'es, \'Ecole Polytechnique, CNRS, CEA,  Universit\'e Paris-Saclay, F-91128 Palaiseau, France}
\newcommand{\etsf}{European Theoretical Spectroscopy Facility (ETSF)}
\newcommand{\soleil}{Synchrotron SOLEIL, L'Orme des Merisiers, Saint-Aubin, BP 48, F-91192 Gif-sur-Yvette, France}

\institute{ \lsi  \and \etsf \and \soleil}
\date{Received: date / Revised version: date}
%
\abstract{The spectral potential is the dynamical generalization of the Kohn-Sham potential. It targets, in principle exactly, the spectral function in addition to the electronic density. 
Here we examine the spectral potential in one of the simplest solvable models exhibiting a non-trivial interplay between electron-electron interaction and inhomogeneity, namely the asymmetric Hubbard dimer. 
We discuss a general strategy to introduce approximations, which consists in calculating the spectral potential in the homogeneous limit (here represented by the symmetric Hubbard dimer) and importing it in the real inhomogeneous system through a suitable ``connector''. The comparison of  different levels of approximation to the spectral potential with the exact solution of the asymmetric Hubbard dimer gives insights about the advantages and the difficulties of this connector strategy for applications in real materials.
}


\dedication{ }

\titlerunning{The spectral potential for the Hubbard dimer}
\authorrunning{Marco Vanzini et al.}

\maketitle

\section{Observables from reduced quantities}

From the solution of the time-independent 
many-electron Schr\"odinger equation it is in principle possible to obtain
the many-electron wavefunction $\Psi(\bfr_1,\ldots,\bfr_N)$
and hence the expectation value of   
any observable 
calculated with $\Psi$. 
Even though the equation to solve is precisely known, 
its exact solution in practice is  feasible only for a very limited number $N$ of electrons \cite{Dirac1929}.
However, besides being hardly possible, the full knowledge of $\Psi$
is often not even desiderable \cite{Kohn1999}.
Whereas $\Psi$ is not itself a measurable quantity,
the evaluation of expectation values  generally amounts to
integrating over most of the degrees of freedom of $\Psi$, 
thus losing most of its detailed information.

Alternatively, observables can be obtained in principle with much less effort by working with reduced quantities. 
These are  functions of a smaller number of degrees of freedom (independent of $N$), 
themselves obtainable from expectation values of $\Psi$.
Three prominent examples of reduced quantities are the electronic density $n(\bfr_1)$, the one-particle reduced density matrix $\gamma(\bfr_1,\bfr_2)$ and the one-particle Green's function $G(\bfr_1,\bfr_2,\w)$.
In each of these cases, one generally aims to express the searched observables as functionals of the corresponding basic variable,
giving rise, respectively, to density-functional theory (DFT) \cite{Hohenberg1964,Dreizler1990}, reduced-density-matrix functional theory (RDMFT) \cite{Gilbert1975}, 
and\linebreak Green's-function functional theory, often approximated within many-body perturbation theory (MBPT) \cite{Fetter1971,Gross1991,Martin2016}.

In those frameworks  the problem to solve becomes twofold: 
the explicit functional form is often not known for the observable of interest 
and the simplified equations needed to determine the key reduced quantity have to be approximated in practice.
The great advantage rests on the fact that the computational gain can be huge 
and the power of analysis of the results can be greatly enhanced 
thanks to the introduction of synthetic concepts, like effective particles, effective interactions, etc.
The balance between simplicity and accuracy is largely won, thus
explaining the great success of strategies based on the use of reduced quantities.

In many applications the target observable is the ground-state total energy $E_0(N)$, 
for which variational principles based on the corresponding basic variables exist in the three frameworks.
The three theories make use  of the different amount of information explicitly gained by calculating the corresponding key quantity.
In MBPT, knowing an approximated $G$ automatically determines $E_0$ (e.g. through the Galitskii-Migdal formula \cite{Galitskii1958}), 
whereas in RDMFT 
a piece of the energy functional $E[\gamma]$, namely the correlation contribution, 
is unknown and has to be approximated. 
In DFT the situation is apparently worse: the only contributions to $E[n]$ explicitly known in terms of $n$ are the external-potential and Hartree energies. 
Nevertheless, DFT is by far the most popular method to calculate the ground-state energy $E_0$ \cite{Martin2004}. 
One of the main reasons of its extraordinary success is the idea of Kohn and Sham \cite{Kohn1965} to introduce 
an auxiliary non-interacting system that is built in order to yield the exact density 
and, concomitantly, to reduce the amount of $E[n]$ that has to be approximated.
In such a way, simple approximations like the local-density approximation (LDA) \cite{Kohn1965}  are already quite accurate.
The main issue within DFT remains the possibility to systematically improve the existing approximations \cite{Pribram-Jones2015}.
 Designing systematically better approximations is instead easier within MBPT thanks to the larger  amount of information explicitly carried by $G$. 
However, being computationally much heavier, MBPT is much less competitive to calculate total energies, so in this context it is often used in an explorative way \cite{Dahlen2006,Hellgren2007,Caruso2013,Hellgren2015}.
On the other side, RDMFT could be a promising approach to deal with strong correlation \cite{hardy}, which is a notoriously difficult problem for DFT \cite{Cohen2012}.
The density matrix $\gamma$ can be diagonalized to give natural orbitals (the eigenvectors) and occupation numbers (the eigenvalues).
Since, at zero temperature, $\gamma$ is an idempotent function if and only if the corresponing $\Psi$ is a Slater determinant  
\cite{Gross1991}, fractional occupation numbers (i.e. not being strictly 0 or 1) are an explicit measure of electronic correlation.
Within RDMFT one  could hence benefit from this explicit information to build accurate approximations to deal with strong correlation,
with the hope to save computational time with respect to the Green's function framework.

Already from this short summary we can understand that there is always a trade-off between the computational cost that one is willing to afford and 
the amount of information that one needs to calculate explicitly in order to get accurate results.
Clearly, different problems lead to different choices. 
In the following we will address the particular question of the use of reduced quantities in the theoretical description of spectroscopy.

\paragraph{Spectroscopy}
In any spectroscopy experiment, an external perturbation drives the sample into an excited state \cite{Onida2002}.
Therefore, in order to analyse, understand and predict the measured spectra, besides the  ground state $E_0(N)$
one needs to know also the excitation energies of the system.
When the perturbation is simulated through  a time-dependent external potential, 
more efficient alternatives to the solution of  the time-dependent many-electron Schr\"odinger equation still exist. 
They are the extensions to the time-dependent situation of the theories
based on reduced quantities that we have just discussed.
In time-dependent density-functional theory (TDDFT) \cite{Runge1984,Marques2012},
time-dependent re\-duced-density-matrix functional theory 
\cite{Pernal2007,Pernal2015}
and in the Kel\-dysh-Green's function formalism \cite{Stefanucci2013}
all the corresponding basic variables   become explicitly dependent on time.
In these frameworks one  can then obtain the reaction to the perturbation 
also beyond the linear-response regime and deal with non-equilibrium situations.

In the rest of the article we will instead focus our discussion on a different class of excitation spectra \cite{Onida2002}: 
the  removal and addition energies and intensities that are measured in direct and inverse photoemission (PES) experiments, respectively \cite{Huefner2013}. 
The excitation energies $\epsilon_{\lambda}$ are formally defined as the differences between the ground-state energy  $E_0(N)$ and the energy of an excited state $\lambda$ where one electron has been removed/ added from/to the $N$-electron system in the ground state:
\beq
\epsilon_{\lambda} =
\begin{cases}
 E_0(N) - E_{\lambda}(N-1) & \text{if } \epsilon_{\lambda} < \mu   \\
E_{\lambda}(N+1) - E_0(N) & \text{if } \epsilon_{\lambda} > \mu  
 \end{cases}
 \label{eq:add-remove}
\eeq
with $\mu$ the Fermi energy.
Addition/removal energies define the (quasiparticle) band structure of a solid, including its band gap,
and genuine correlation features beyond the independent-particle picture such as satellites \cite{Martin2016}. 
They characterise the electronic structure of a material and are hence of fundamental interest.

In extended systems the direct evaluation of the total-energy differences in Eq. \eqref{eq:add-remove}
is impractical. The eigenvalues of the Kohn-Sham single-particle hamiltonian are largely 
employed as addition/removal energies $\epsilon_{\lambda}$. 
However, this commonly adopted procedure is not rigorous \cite{Sham1966,Perdew1982,Sham1983}
(besides $\mu$ and the highest-occupied level in finite systems \cite{vonBarth1984,Levy1984}).
In practice, this misuse of Kohn-Sham eigenvalues is at the origin of the so-called ``Kohn-Sham band gap  underestimation'' in semiconductors and insulators \cite{vanSchilfgaarde2006,Martin2016}. 
Despite recent attemps \cite{Pernal2005,Sharma2013,Lathiotakis2014}, also within RDMFT the calculation of addition/removal spectra remains problematic \cite{DiSabatino2015,Kamil2016}, since, as in DFT, they are difficult to express as functionals of the density matrix (or its natural orbitals and occupation numbers).
Instead, the one-particle Green's function is explicitly designed  to give those excitations energies $\epsilon_{\lambda}$,
which are formally the poles of $G$ in the frequency domain. 
The spectral function defined as 
\beq
A(\bfr_1,\bfr_2,\w) = -\frac{1}{\pi}{\rm sign}\left(\omega-\mu\right)\text{Im} G(\bfr_1,\bfr_2,\w)
\label{eq:defA1}
\eeq
displays peaks at the energies $\epsilon_{\lambda}$:
\beq
A(\bfr_1,\bfr_2,\w) = \sum_{\lambda} f_{\lambda}^*(\bfr_1)f_{\lambda}(\bfr_2) \delta(\w-\epsilon_{\lambda}),
\label{eq:defA}
\eeq
where $f_{\lambda}$ are the Lehmann amplitudes\footnote{Note that expression \eqref{eq:defA1} is valid if the products $f_{\lambda}^*(\bfr_1)f_{\lambda}(\bfr_2)$ are real, e.g. when they are symmetric under the exchange $\bfr_1\leftrightarrow \bfr_2$ \cite{Farid1999}. It will be the case here.}.
The spectral function is thus the primary quantity to consider in order to analyse addition/removal spectra.
It is not surprising that popular approximations within MBPT, such as Hedin's GW approximation \cite{Hedin1965}, 
are today the state-of-the-art method 
for the calculation of excitation spectra \cite{Martin2016}.

Even though the MBPT approach to calculate the excitation energies $\epsilon_{\lambda}$ is generally successful \cite{Martin2016},  at the same time it is
intrinsically inefficient.
In particular, in the case of angle-integrated photoemission, the relevant frequency-dependent 
spectrum is given by just the trace of $A$ in \eqref{eq:defA}. 
Instead, in the standard approach one has first to calculate the full $G(\bfr_1,\bfr_2,\w)$,
although only a limited part of its information is  finally needed.
It is therefore highly desiderable to devise an alternative method that targets directly
the variable of interest \cite{Gatti2007}, which in this case we identify with the diagonal of the spectral function,
$A(\bfr_1,\\ \bfr_1,\w)$. 
In the context of band-structure theories this quantity is usually known as the local density of states (DOS).
Our choice is motivated by the fact that from $A(\bfr_1,\bfr_1,\w)$ 
one can obtain both the density $n(\bfr_1$) (from an integration over $\w$) 
and the integrated DOS (from an integration over $\bfr_1$), which can be related to PES experiments.
We can understand $A(\bfr_1,\bfr_1,\w)$ as a further reduced quantity, whose complexity is intermediate between 
$n(\bfr_1)$ and $G(\bfr_1,\bfr_2,\w)$.
Here the motivation is clearly driven by spectroscopy applications 
and the goal is to obtain addition/removal spectra in an efficient way.
We also note that there is a general interest in the possibility 
of defining also an energy functional in terms of $A(\bfr_1,\bfr_1,\w)$
\cite{Ferretti2014}. However, we will not touch upon this problem here.

In the rest of the article, we will discuss in details 
a possible strategy to obtain in practice useful approximations to the spectral potential, 
which is the generalization of the Kohn-Sham potential designed to directly yield the diagonal of the spectral function $A(\bfr_1,\bfr_1,\w)$ without passing through the Green's function $G(\bfr_1,\bfr_2,\w)$ as in \eqref{eq:defA1}. 
In Sec. \ref{sec:SSE}, following Ref. \cite{Gatti2007}, we will introduce the spectral potential 
on the basis of a generalization of the Sham-Schl\"uter equation \cite{Sham1983,Sham1985}.
In Sec. \ref{sec:asym} we  will then consider one of the simplest hamiltonians (defined on a lattice, not in real space) 
that is representative of a real material: the asymmetric Hubbard dimer with one electron.
This very simple hamiltonian already illustrates the general problem of the interplay between the electron-electron interaction (i.e. the source of correlation effects) 
and the crystal potential (i.e. the source of inhomogeneities in real materials). 
Its exact solution will provide the benchmark to examine a possible strategy of approximation for the spectral potential in realistic applications.
Sec. \ref{sec:connector} will illustrate the general strategy, 
which will be then followed in Secs. \ref{sec:sym} and \ref{sec:approx}, 
where approximations to the spectral potential will be constructed and tested at different levels.
Finally, Sec. \ref{sec:conclusion} contains a concise summary and an outlook.

\section{Effective potentials: the generalized Sham-Schl\"uter equation}
\label{sec:SSE}

The Kohn-Sham scheme \cite{Kohn1965}
is the paradigm of an auxiliary system: 
a non-interacting system with an effective potential, i.e.
the local and real Kohn-Sham potential $V_{{\rm KS}}(\bfr_1)$,
that is designed  to yield the quantity of interest, namely the density, and in principle nothing else.
We can also understand the one-particle Green's function formalism within MBPT in the same spirit.
In this case the effective ``potential'' is the self-energy $\Sigma(\bfr_1,\bfr_2,\w)$, which is a non-local, non-hermitian 
and frequency-dependent operator.
Similarly to $V_{{\rm KS}}(\bfr_1)$ for the density, the self-energy is supposed to give $G$ exactly.

In both DFT and MBPT, the exchange-correlation (xc) parts of the effective potentials, $V_{{\rm xc}}(\bfr_1)$ and $\Sigma_{{\rm xc}}(\bfr_1,\bfr_2,\w)$, respectively, 
have to be approximated.
Sham and Schl\"uter \cite{Sham1983,Sham1985} established a formal connection between the two.
First of all, one can formally define the Kohn-Sham Green's function $G_{{\rm KS}}$
as the resolvent of the Kohn-Sham hamiltonian,
$G_{{\rm KS}} = (\w - H_{{\rm KS}})^{-1} = (\w - H_{0}-V_{{\rm xc}})^{-1}$, where $H_0$ is the Hartree Hamiltonian.
The connection between the DFT and MBPT effective potentials is then given by the fact that the density can be obtained both from  $G$ and $G_{{\rm KS}}$ as:
\beq
\begin{aligned}
n(\bfr_1) =&\int \frac{d\w}{2\pi i} e^{i{\w}\eta} G(\bfr_1,\bfr_1,\w) \\ 
= &\int \frac{d\w}{2\pi i} e^{i{\w}\eta} G_{{\rm KS}}(\bfr_1,\bfr_1,\w).
\end{aligned}
\eeq
Plugging this condition into the Dyson equation relating $G$ to $G_{{\rm KS}}$ one finds \cite{Sham1983}
\beq
\begin{gathered}
\int d\w d\bfr_3d\bfr_4 \,  e^{i{\w}\eta} G_{{\rm KS}}(\bfr_1,\bfr_3,\w)[\Sigma_{{\rm xc}}(\bfr_3,\bfr_4,\w)   \\ 
 - V_{{\rm xc}}(\bfr_3)\delta(\bfr_3 - \bfr_4)]G(\bfr_4,\bfr_1,\w)  = 0, 
  \end{gathered}
\label{eq:SSE}
\eeq
which can be solved for $V_{{\rm xc}}(\bfr_1)$.
The Sham-Schl\"uter equation \eqref{eq:SSE} in its linearized form where $G$ is replaced by $G_{{\rm KS}}$ everywhere
has  often been used to derive approximations to $V_{{\rm xc}}$ (notably in the context of the optimized effective potential method \cite{Sharp1953})
or to study properties of $V_{{\rm xc}}$  (see e.g. \cite{Godby1986,Godby1987,Eguiluz1992,Niquet2003}) for given approximations to  $\Sigma_{{\rm xc}}$.
It has been employed also in other contexts: for example, in the framework of superconducting  DFT \cite{Lueders2005,Marques2005} or within TDDFT where it has been  extended to the time-dependent case  for $V_{{\rm xc}}(\bfr_1,t_1)$
 \cite{Leeuwen1996} .

One could wonder whether the same approach could be followed also in the case of RDMFT
to introduce an {\it ansatz} of a non-local and static effective potential $V_{{\rm DM}}(\bfr_1,\bfr_2)$\footnote{Note that for simplicity of notation $V_{{\rm DM}}$ here stands only for the xc part of the total effective potential.  
The same notation will be used also for $V_{{\rm SF}}$ in the following.}
 for the density matrix:
 \beq
 \begin{aligned}
\gamma(\bfr_1,\bfr_2) & =\int \frac{d\w}{2\pi i} e^{i{\w}\eta}  G(\bfr_1,\bfr_2,\w) \\ 
& =\int \frac{d\w}{2\pi i} e^{i{\w}\eta}  G_{{\rm DM}}(\bfr_1,\bfr_2,\w)
\end{aligned}
\eeq
with $G_{{\rm DM}} = (\w - H_{0}-V_{{\rm DM}})^{-1}$.
The answer must be negative\footnote{Note that the generalisation to a degenerate non-interacting system, where the ground state is an ensemble, is in principle possible \cite{Gilbert1975,Requist2008}. However, this choice also leads to pathologies \cite{Gilbert1975,Requist2008} and will be avoided here. Another possibility is finite-temperature RDMFT \cite{Baldsiefen2015}.}. 
The effective potential $V_{{\rm DM}}$ would define a non-interacting 
system. However, any non-interacting wavefunction  $\Psi$ 
can give rise only to idempotent density matrices, 
without the possibility to cover the general correlated case of fractional occupation numbers.
As a matter of fact, this implies that the generalized Sham-Schl\"uter equation for the density matrix:
\beq
\begin{gathered}
\int d\w d\bfr_3d\bfr_4 \,  e^{i{\w}\eta} G_{{\rm DM}}(\bfr_1,\bfr_3,\w)[\Sigma_{{\rm xc}}(\bfr_3,\bfr_4,\w)   \\ 
 - V_{{\rm DM}}(\bfr_3,\bfr_4)]G(\bfr_4,\bfr_2,\w)  = 0 
  \end{gathered}
\label{eq:SSE_rdm}
\eeq
should have no solution for $V_{{\rm DM}}$ (except for the  case of a static and real $\Sigma_{{\rm xc}}$, for which $V_{{\rm DM}}$
is trivially equal to $\Sigma_{{\rm xc}}$).
This is proved analytically in the Appendix \ref{sec:sse_dm} for the simple Hubbard dimer at half filling.

Here, instead, we will show that it is possible to use a generalized Sham-Schl\"uter equation \cite{Gatti2007} 
to define the local, real and frequency-dependent
spectral potential
$V_{{\rm SF}}(\bfr_1,\w)$ 
that directly yields another part of the full $G$ that is of interest for us, namely the diagonal of the spectral function:
\beq
\begin{aligned}
A(\bfr_1,\bfr_1,\w) & =-\tfrac{1}{\pi}{\rm sign}(\omega-\mu)\text{Im} G(\bfr_1,\bfr_1,\w) \\ 
& =-\tfrac{1}{\pi}{\rm sign}(\omega-\mu)\text{Im} G_{{\rm SF}}(\bfr_1,\bfr_1,\w)
\end{aligned}
\eeq
with $G_{{\rm SF}} = (\w - H_{0}-V_{{\rm SF}})^{-1}$.
For each frequency, the following equation
\beq
\begin{gathered}
{\rm Im}\int d\bfr_3d\bfr_4 \, G_{{\rm SF}}(\bfr_1,\bfr_3,\w)\bigl[\Sigma_{{\rm xc}}(\bfr_3,\bfr_4,\w)   \\ 
 - V_{{\rm SF}}(\bfr_3,\w)\delta(\bfr_3 - \bfr_4)\bigr]G(\bfr_4,\bfr_1,\w)  = 0 
  \end{gathered}
  \label{eq:SSE_SF}
\eeq 
defines $V_{{\rm SF}}(\bfr_1,\w)$  
in the same manner as the original Sham-Schl\"uter equation \eqref{eq:SSE} does for $V_{{\rm xc}}(\bfr_1)$.
The new potential defines another auxiliary system that represents
the dynamical generalization of the Kohn-Sham scheme of DFT.
By construction, it gives exactly both the ground-state density and the $A(\w)\equiv\int d\bfr_1A(\bfr_1,\bfr_1,\w)$ needed for angle-integrated photoemission.
So its use would allow one to bypass expensive MBPT calculations: the computational cost would be much reduced since
$V_{{\rm SF}}$ is real and local like $V_{{\rm KS}}$.
Eq. \eqref{eq:SSE_SF} has been solved for simple illustrative cases in \cite{Gatti2007,Ferretti2014,matteo_phd}.
In the following, we will use the spectral potential $V_{{\rm SF}}$ to determine the excitation energies of the Hubbard dimer
and we will make a comparison with the exact solution and various approximations to the self-energy.

\section{The asymmetric Hubbard dimer}
\label{sec:asym}

\begin{figure}[t]
	\centering
	\begin{minipage}{0.90\columnwidth}
		\centering
			\includegraphics[width=0.89\textwidth]{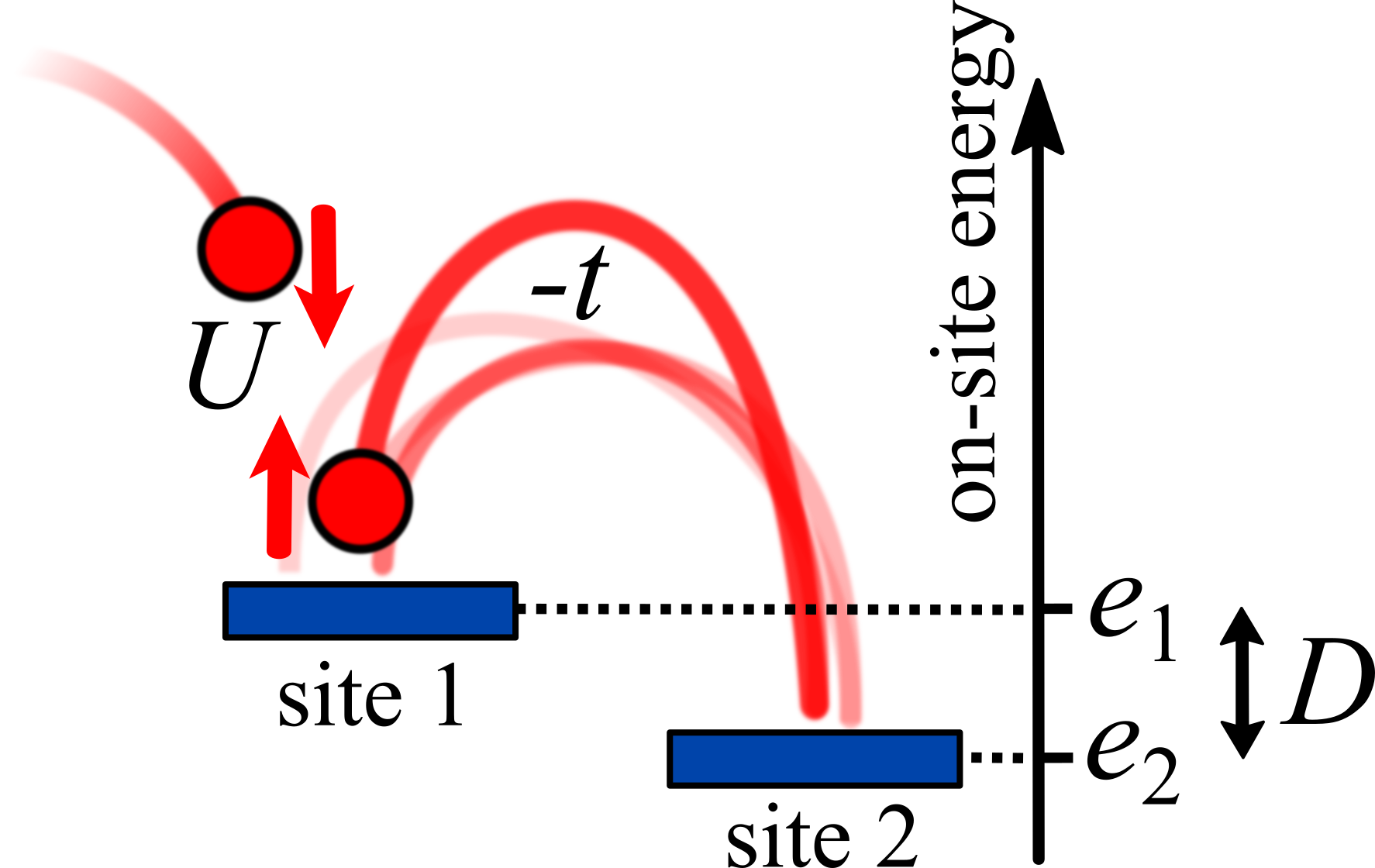}
		\caption{\emph{A schematic picture of the asymmetric Hubbard dimer, Eq. \eqref{eq:weibw}. The ground state is a single spin--up electron which jumps between site 1 and site 2, at energy $e_1$ and $e_2$, gaining an hopping energy $-t$. If an additional spin--down electron enters the system on the same site as the one in which the former electron is sitting, they interact paying the energy price $U$. }}
		\label{fig:model}
	\end{minipage}
\end{figure}

One of the simplest models 
that still exhibits a non--trivial interplay between the electron--electron interaction
and the interaction with an external potential 
is the asymmetric Hubbard dimer, occupied by a single spin--up electron  $N=1$.
Its Hamiltonian  
reads:
\begin{equation}
\hat{H}=-t\sum_{\sigma}\left(\hat{c}_{1\sigma}^{\dag}\hat{c}_{2\sigma}+\hat{c}_{2\sigma}^{\dag}\hat{c}_{1\sigma}\right)+
\sum_ie_i\hat{n}_{i}+
U\sum_{i}\hat{n}_{i\uparrow}\hat{n}_{i\downarrow}.
\label{eq:weibw}
\end{equation}
In the Hubbard model \eqref{eq:weibw} the electron--electron interaction is assumed to be only on--site: the electronic repulsion is $U$ for two electrons on the same site and 0 otherwise.
Varying the ratio between the Hubbard $U$ and the nearest-neighbour hopping parameter $t$ is the simplest way to capture the competition between the tendency of electrons to delocalise to reduce their kinetic energy 
and the opposite tendency to localise to reduce the cost of the electronic repulsion. This competition is the key to describe the Mott metal-insulator 
transition \cite{Georges2004}.
In the large $U/t$ limit, the Hubbard dimer also corresponds to a minimal-basis-set representation of the bond dissociation of simple diatomic molecules. For example, the symmetric dimer (i.e., for same on-site energies $e_1=e_2$ in \eqref{eq:weibw}) describes the dissociation of the $H_2$ molecule (or $H_2^+$ for the one-electron case), which is the paradigmatic case of static correlation in quantum chemistry (still a challenging problem within DFT \cite{Baerends2001,Cohen2012}).
Moreover, in the asymmetric dimer the external potential makes that the two sites 1 and 2 are at different energy $e_1$ and $e_2$ (we choose $e_1>e_2$).
The external potential thus introduces an inhomogeneity, mimicking the role of the crystal potential in a solid (which makes it different from the homogeneous electron gas). 
As in a real material, one cannot easily disentangle the effect of the inhomogeneity from the electron-electron interaction. 

Since the model is also exactly solvable, it has been recently often used 
(both in its asymmetric and symmetric versions and for $N=1$ or $N=2$) 
to study the general properties and benchmark different approximations in 
the various reduced-quantity frameworks that we have considered so far:
DFT \cite{Carrascal2015,Carrascal2012} and several of its extensions (thermal DFT \cite{Smith2016}, 
ensemble DFT \cite{Deur2017,Deur2018}, site occupation embedding theory \cite{Senjean2017}, 
and TDDFT \cite{Aryasetiawan2002,Baer2008,Li2008,Farzanehpour2012,Fuks2013,Fuks2014,Carrascal2018}),
RDMFT \cite{LopezSandoval2002,LopezSandoval2003,DiSabatino2015,Kamil2016,Mitxelena2017} (including its time-dependent version \cite{Requist2010}) and MBPT \cite{Romaniello2009,Romaniello2012,Olsen2014}.

In Eq. \eqref{eq:weibw} the on--site energy term $e_1\hat{n}_1+e_2\hat{n}_2$ can be recast in the form:
$\bar{E}\left(\hat{n}_1+\hat{n}_2\right)+({D}/{2})\left(\hat{n}_1-\hat{n}_2\right)$, 
with $\bar{E}=(e_1+e_2)/2$ the average energy and $D=e_1-e_2$ the difference. Through a redefinition of energy (a shift of $\hat{H}$) we can always choose  
$\bar{E}=0$, setting the zero of the energy axis; in this way, the parameters that define the system are $t$, $U$ and $D$, all positive. To simplify the notation, we measure the energy in units of $t$, defining the reduced quantities $\hat{H}/t\to\hat{H}$, $U/t\to U$ and $D/t\to D$. Thus, the Hamiltonian reads:
\begin{equation}
\hat{H}=-\sum_{\sigma}\left(\hat{c}_{1\sigma}^{\dag}\hat{c}_{2\sigma}+\hat{c}_{2\sigma}^{\dag}\hat{c}_{1\sigma}\right)+
\tfrac{D}{2}\left(\hat{n}_1-\hat{n}_2\right)+
U\sum_i\hat{n}_{i\uparrow}\hat{n}_{i\downarrow}
\label{eq:weihojbw},
\end{equation}
where $D$ can be regarded as an external potential that alters the symmetry \textit{site} $1\longleftrightarrow\text{ \textit{site} 2}$ and creates inhomogeneities in the dimer. 

In order to solve the Hamiltonian, we make a convenient change of basis from the site basis $\bigl\{\ket{i,\sigma}\bigr\}$, defined and ordered as  
$\bigl\{\ket{1,\uparrow},\ket{2,\uparrow},\ket{1,\downarrow},\ket{2,\downarrow}\bigr\}$, 
where $\ket{i,\sigma}$ represents an electron with spin $\sigma$ sitting on the site $i$,  to the bonding--antibonding basis $\bigl\{\ket{\pm,\sigma}\bigr\}\equiv\bigl\{\ket{-,\uparrow},\ket{+,\uparrow},\ket{-,\downarrow},\ket{+,\downarrow}\bigr\}$, where $-$ and $+$ stand for the bonding and antibonding states respectively. The latter can be obtained by diagonalizing the Hamiltonian \eqref{eq:weihojbw}:
\begin{equation}
\begin{aligned}
\ket{-,\sigma}&=\cos\rho\ket{1,\sigma}+\sin\rho\ket{2,\sigma}
\\
\ket{+,\sigma}&=\sin\rho\ket{1,\sigma}-\cos\rho\ket{2,\sigma}
\label{eq:twrhirn},
\end{aligned}
\end{equation}
with $\tan\rho=\frac{D}{2}+\sqrt{1+\frac{D^2}{4}}$.
The corresponding eigenvalues are $e_{\pm}=\pm\sqrt{1+\frac{D^2}{4}}$.
Only the lowest energy level $e_{-}$ is occupied, defining the ground state $\ket{\rm GS}\equiv\ket{-,\uparrow}$ and the Fermi energy $\mu=e_-=-\sqrt{1+\frac{D^2}{4}}$. 
The difference in the occupation of the two sites is given by
\beq
n_1-n_2=\frac{1-\tan^2\rho}{1+\tan^2\rho},
\eeq
which is $0$ for $D=0$ and tends to $-1$ for large $D$, as the assumption $e_1>e_2$ favours the occupation of the second site, lower in energy. 
This means that in the large $D$ limit the bonding state rather than a covalent bond (where the electron is equally shared by the two sites) represents a ionic situation (where the electron is localised on a specific site).

\begin{figure*}[t]
	\centering
	\begin{minipage}{0.99\textwidth}
		\centering
			\begin{subfigure}[t]{0.99\textwidth}
		\centering
		\includegraphics[width=0.19\textwidth]{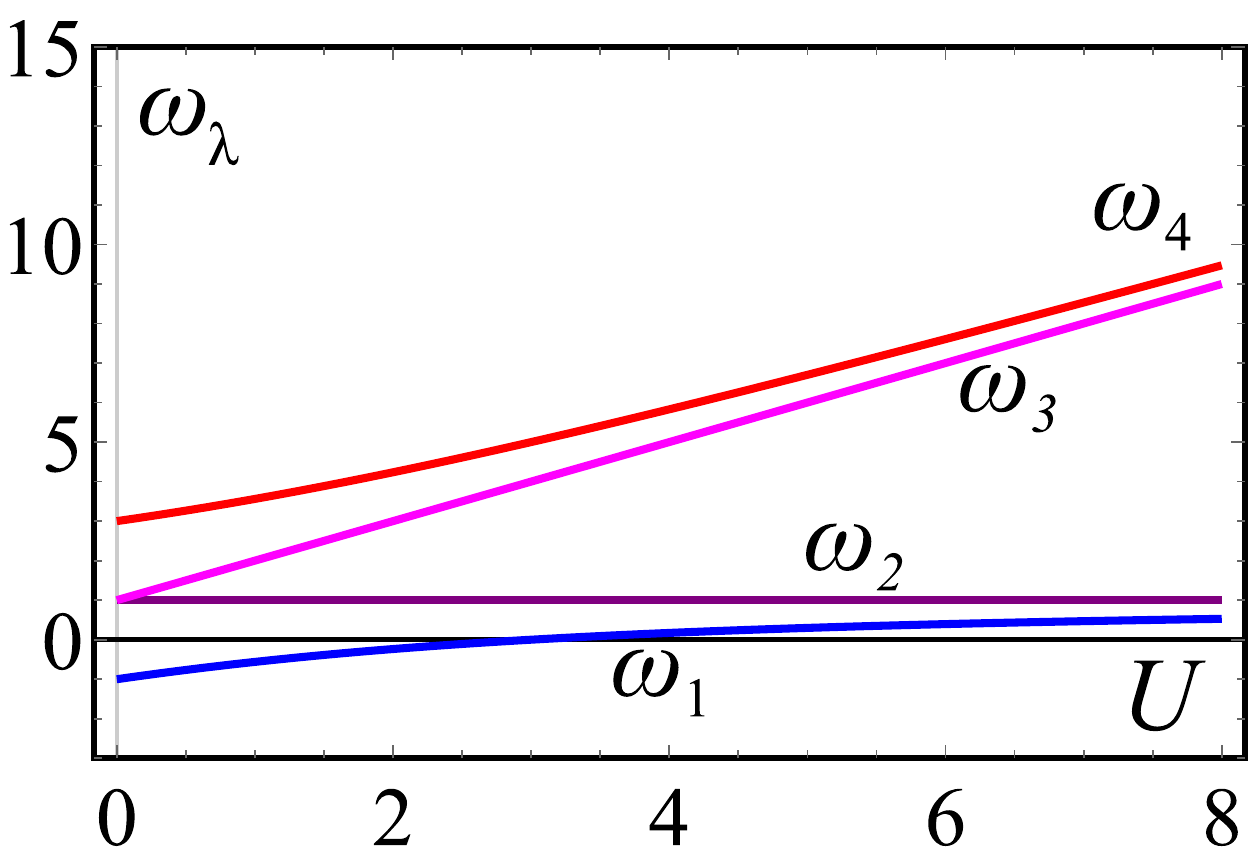}
		~
		\includegraphics[width=0.19\textwidth]{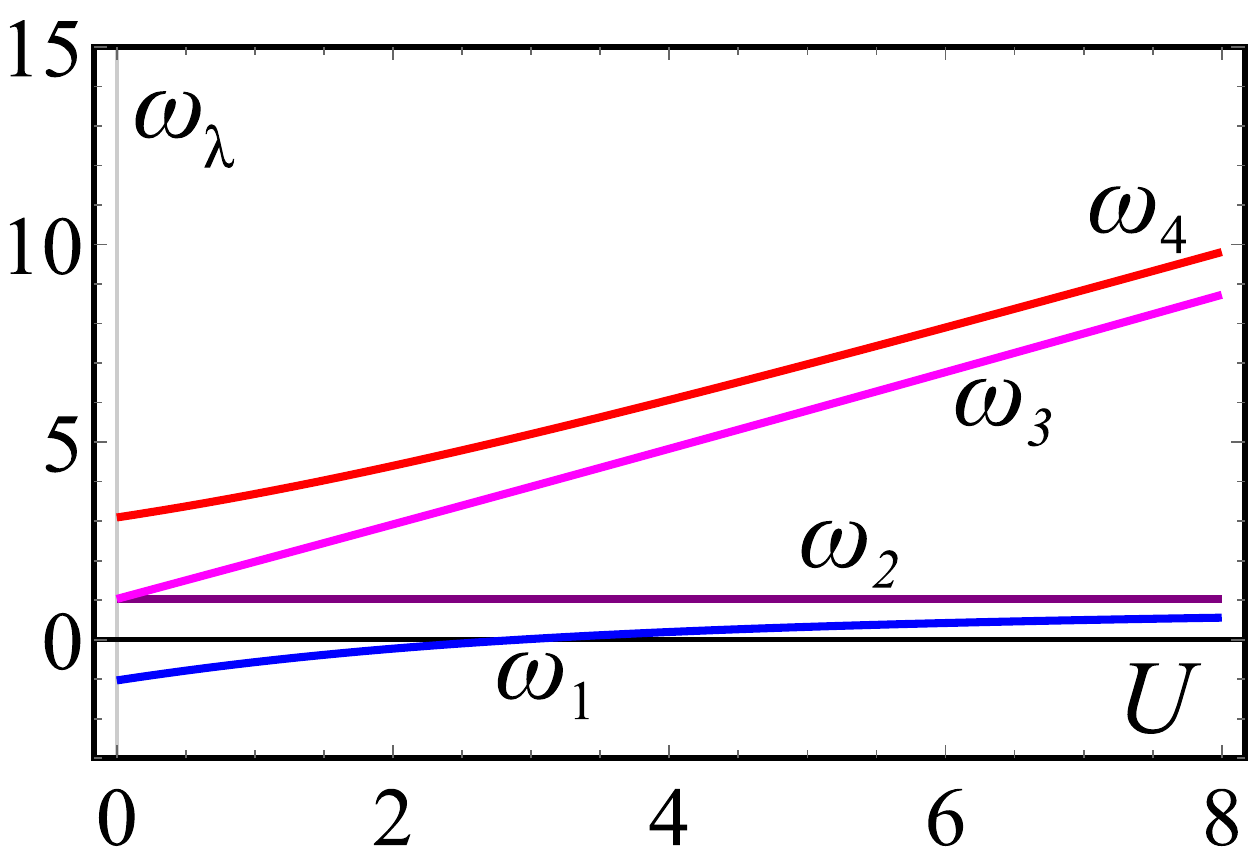}
		~
		\includegraphics[width=0.19\textwidth]{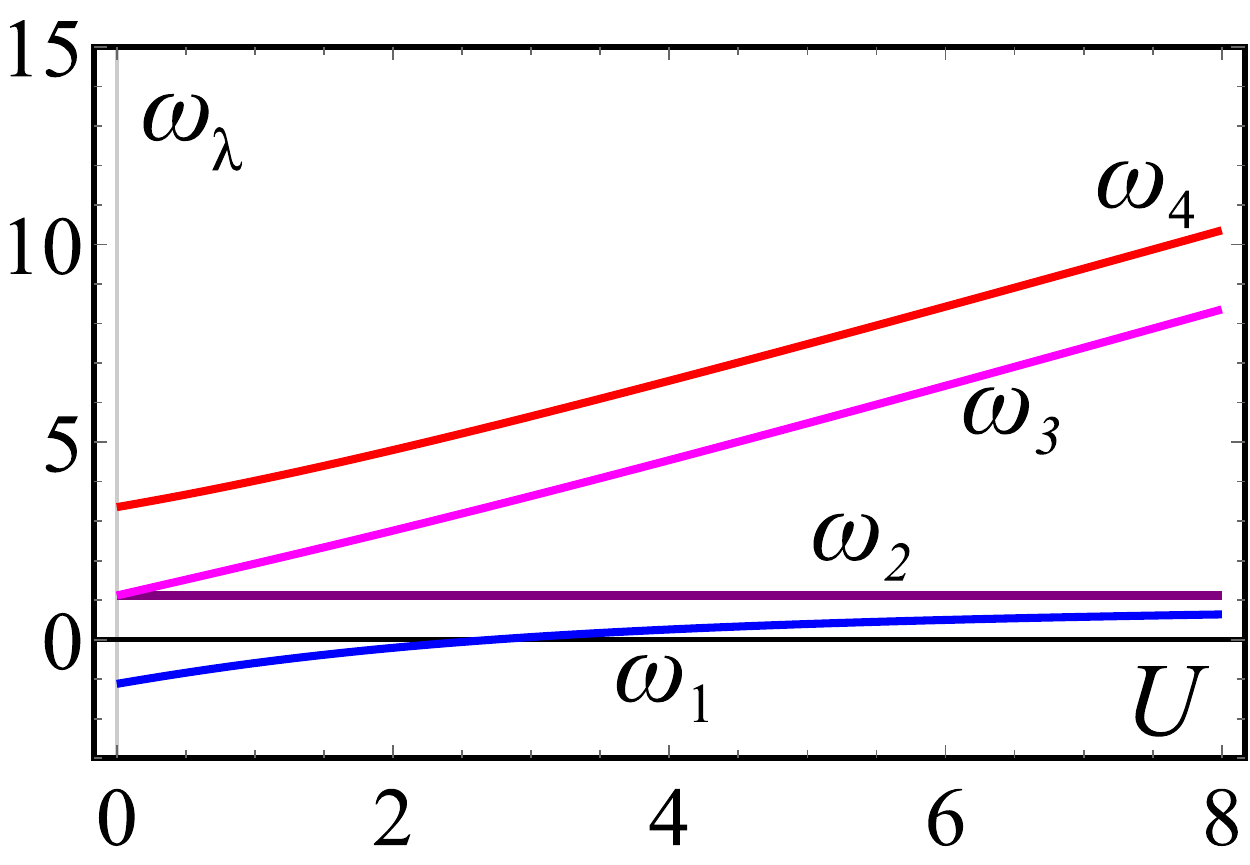}
		~
		\includegraphics[width=0.19\textwidth]{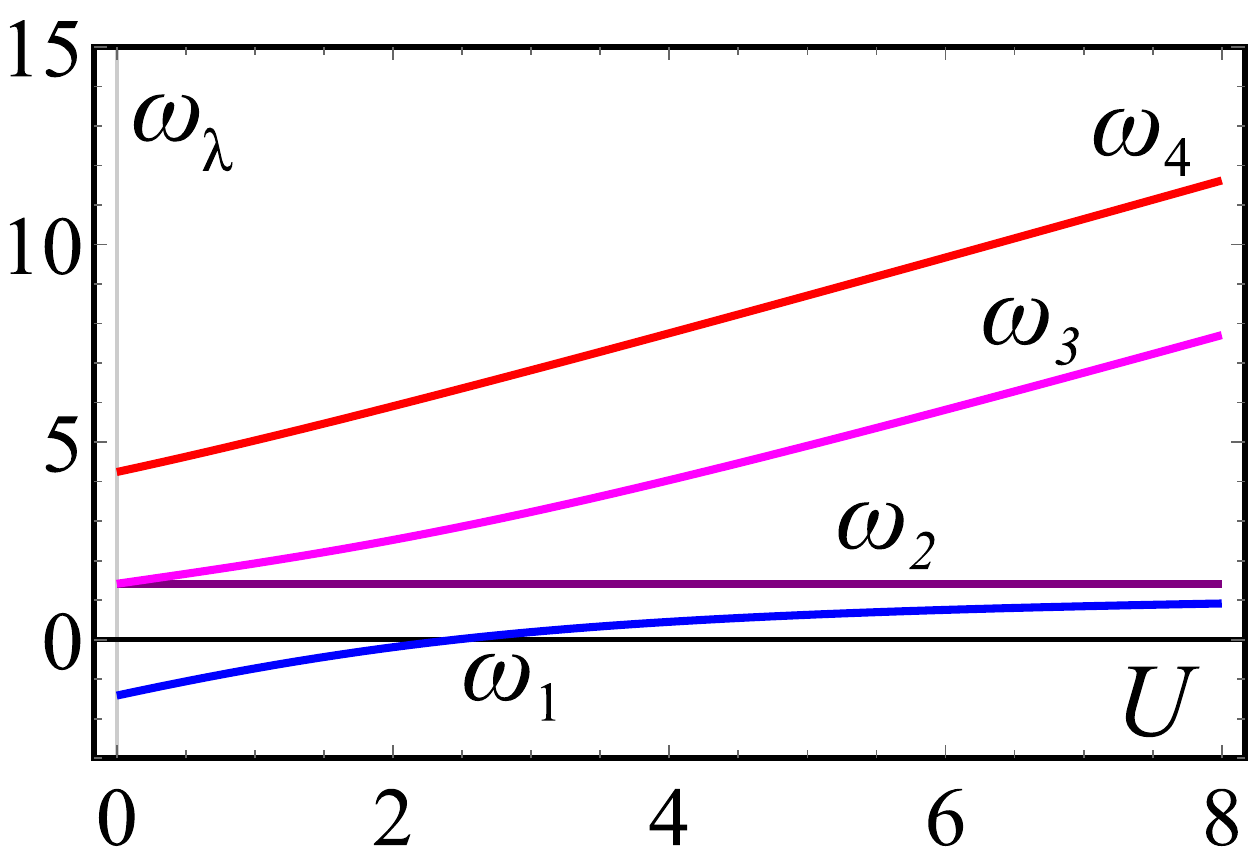}
		~
		\includegraphics[width=0.19\textwidth]{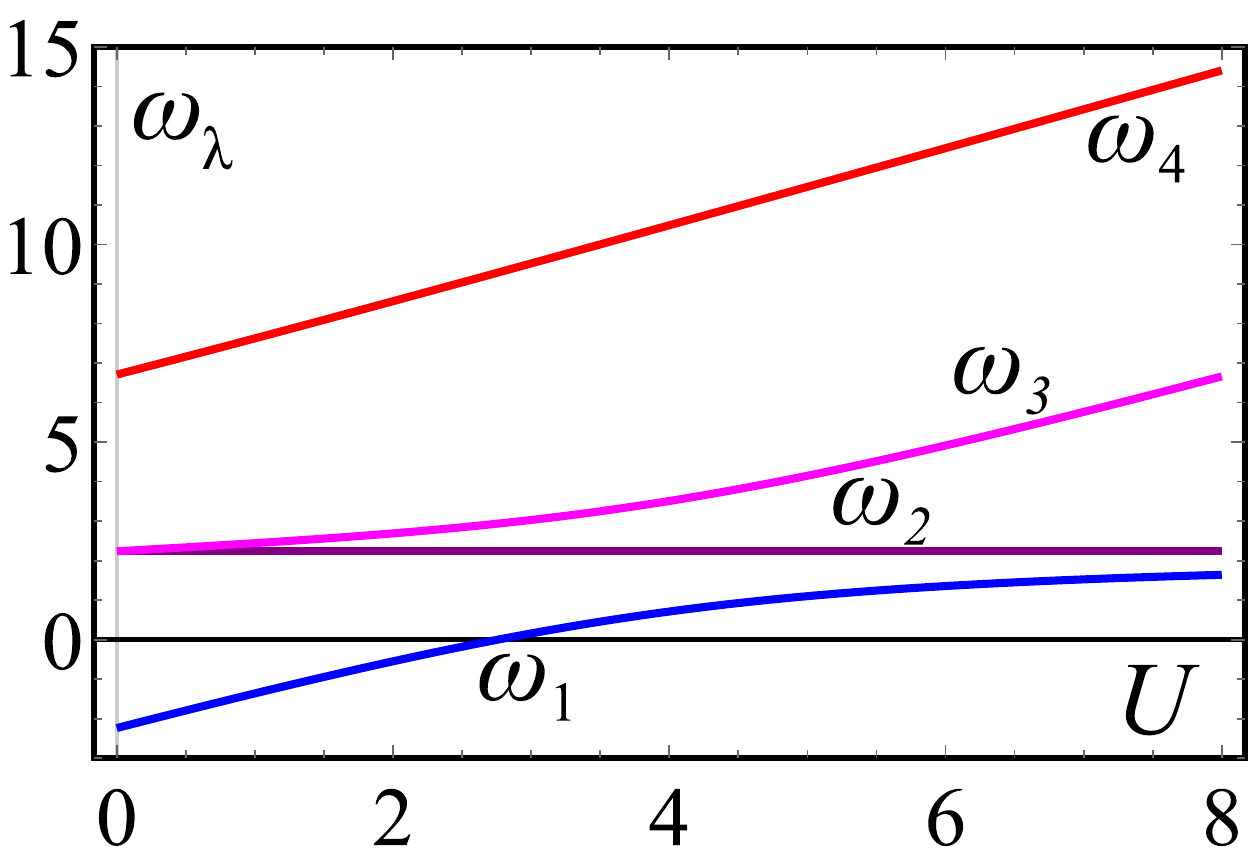}
		\\			$D=0.0$\qquad\qquad\qquad\quad\;\,$D=0.5$\quad\;\,\qquad\qquad\qquad$D=1.0$\qquad\qquad\qquad\quad\;\,$D=2.0$\qquad\quad\;\,\qquad\qquad$D=4.0$\\
		\caption{Position of the poles $\omega^{\lambda}$.}
		\label{fig:asymSF_poles}
	\end{subfigure}

\vspace{1.5em}
		\begin{subfigure}[t]{0.99\textwidth}
			\centering
			\includegraphics[width=0.19\textwidth]{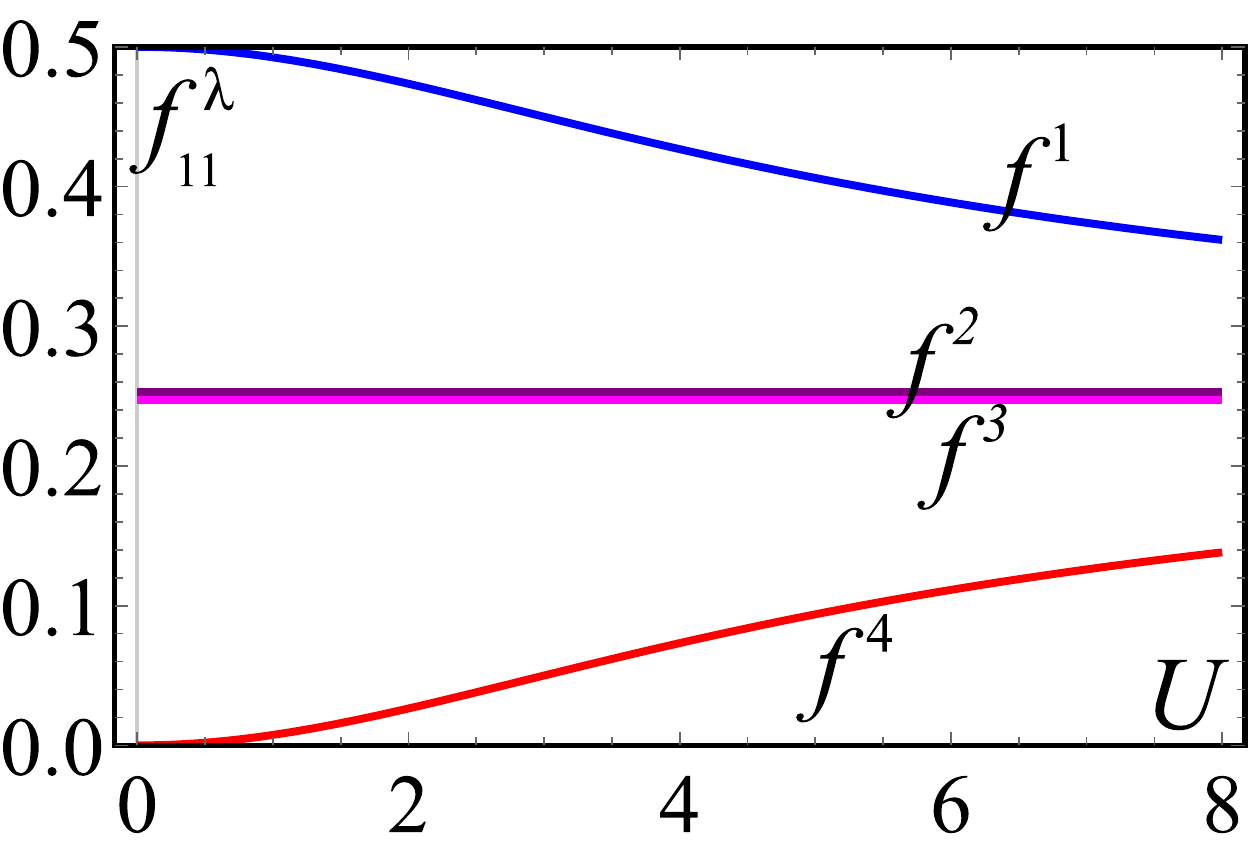}
			~
			\includegraphics[width=0.19\textwidth]{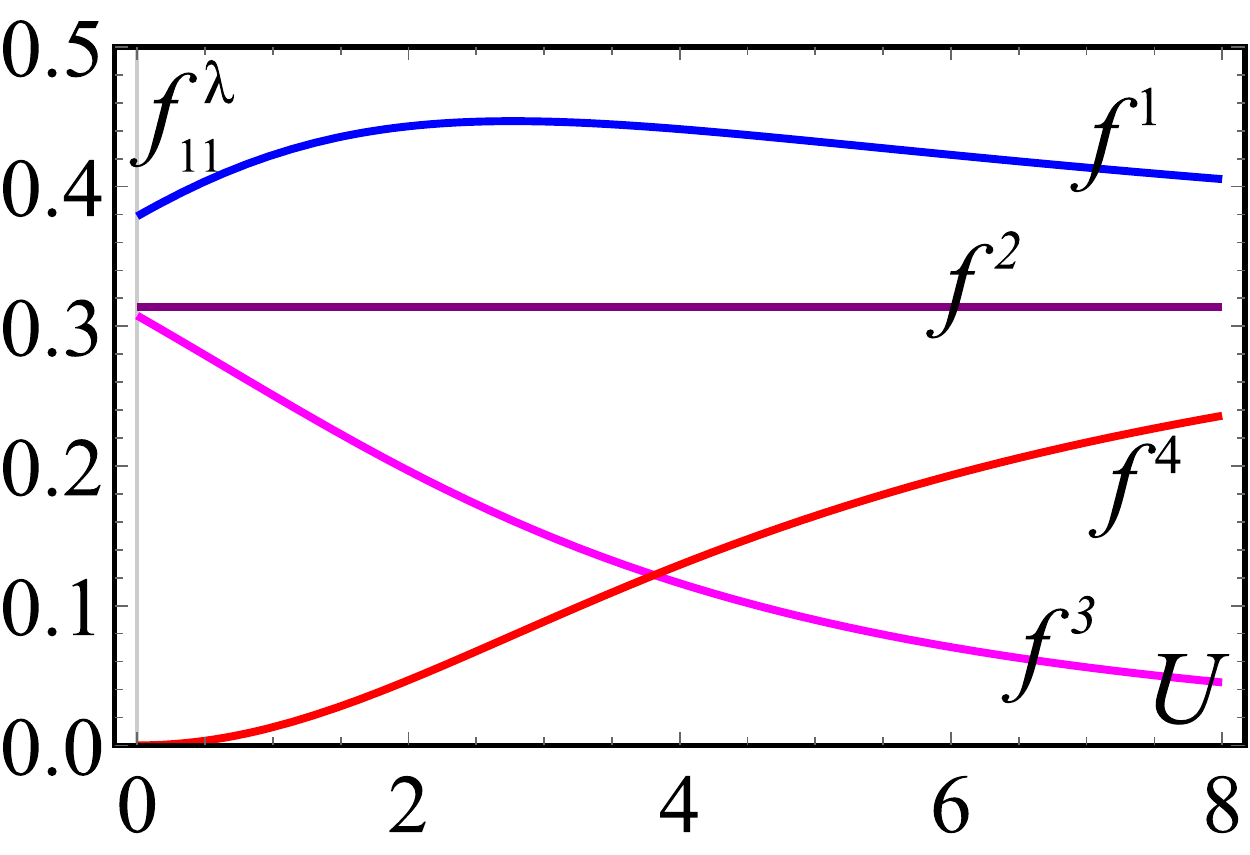}
			~
			\includegraphics[width=0.19\textwidth]{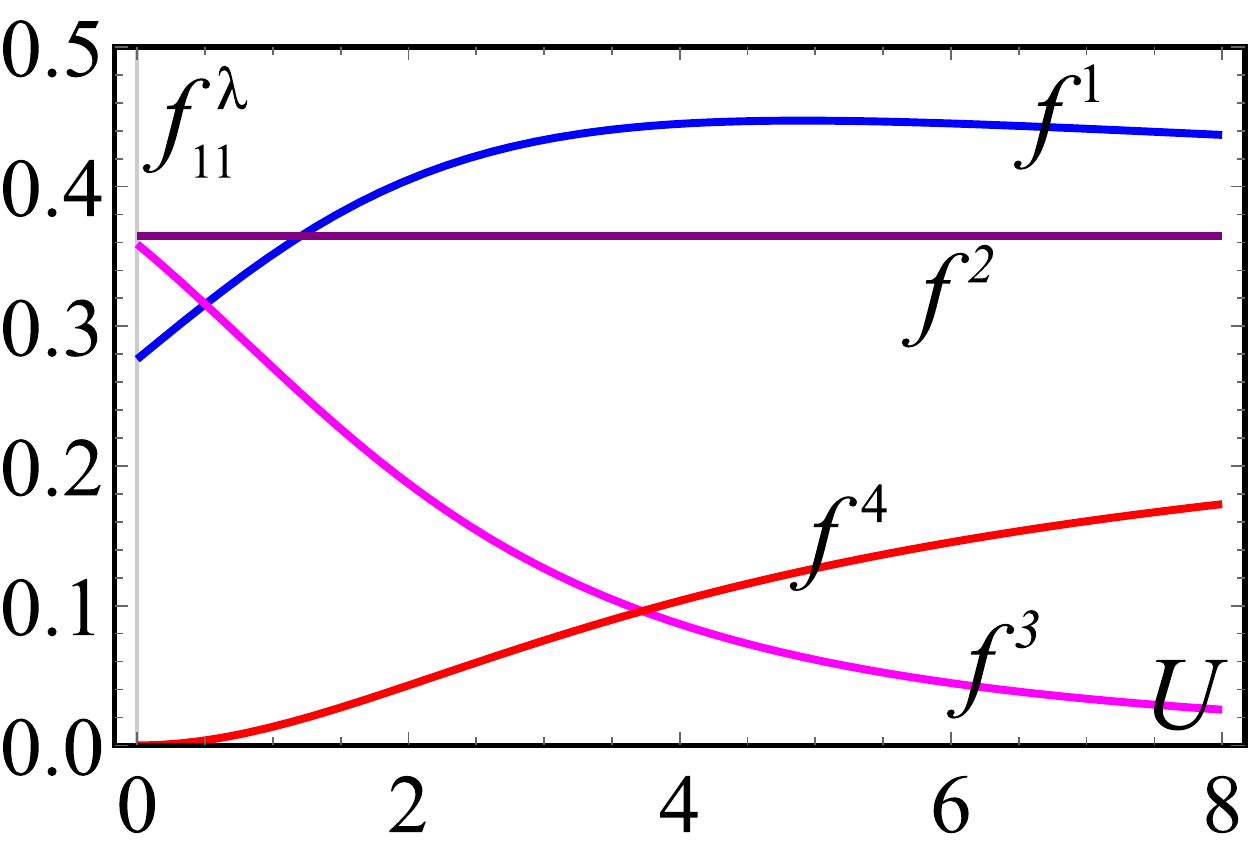}
			~
			\includegraphics[width=0.19\textwidth]{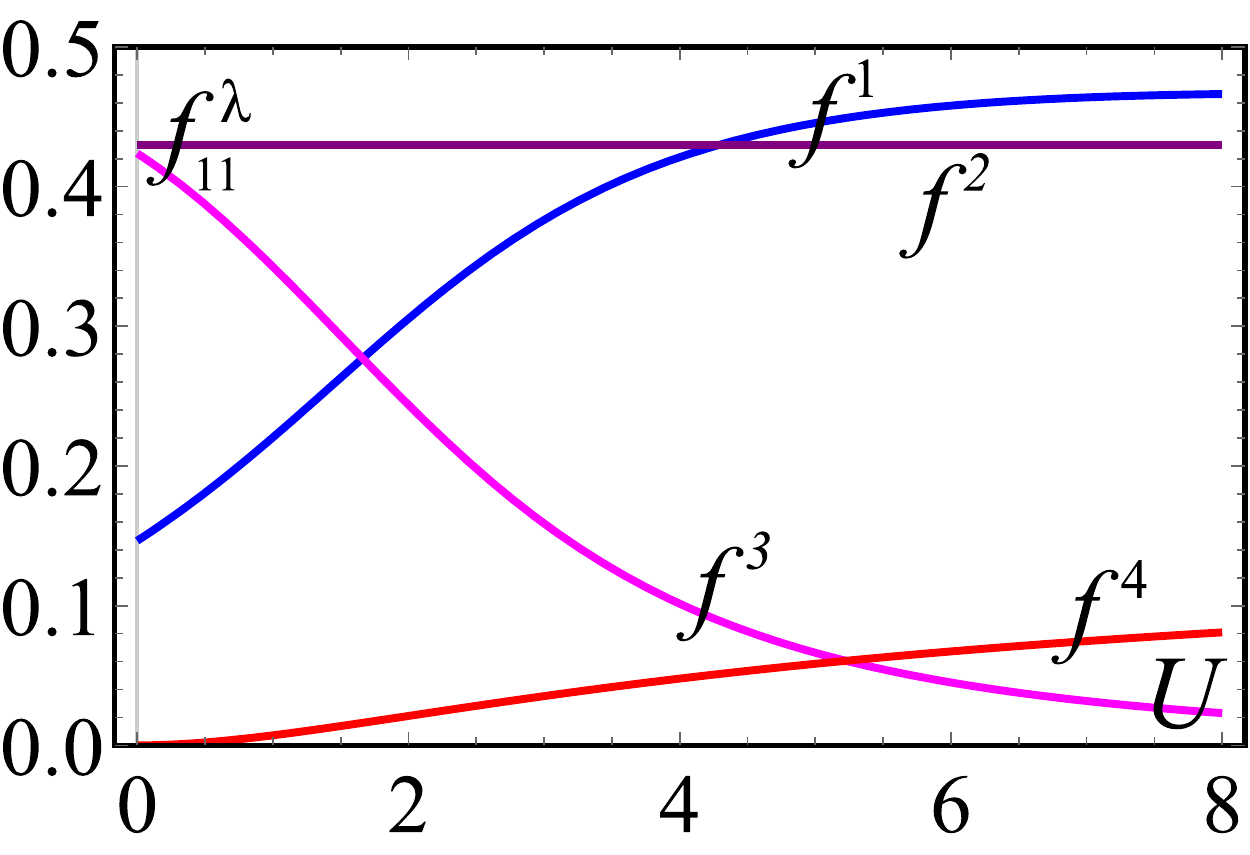}
			~
			\includegraphics[width=0.19\textwidth]{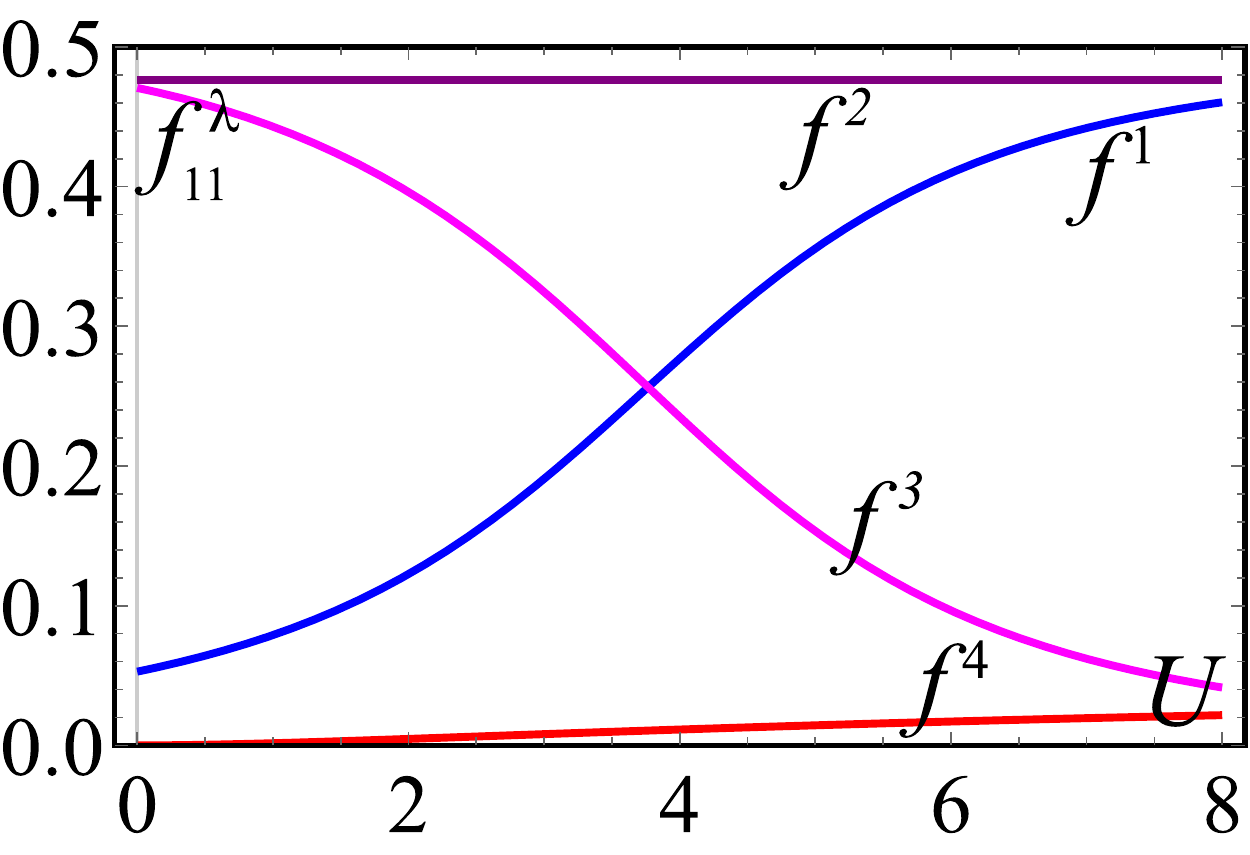}
			\\			$D=0.0$\qquad\qquad\qquad\quad\;\,$D=0.5$\quad\;\,\qquad\qquad\qquad$D=1.0$\qquad\qquad\qquad\quad\;\,$D=2.0$\qquad\quad\;\,\qquad\qquad$D=4.0$\\
			\caption{Weights $f_{11}^{\lambda}$.}
	        \label{fig:asymSF_amplitudes}
		\end{subfigure}
	
\vspace{1.5em}
\begin{subfigure}[t]{0.99\textwidth}
	\centering
	\includegraphics[width=0.19\textwidth]{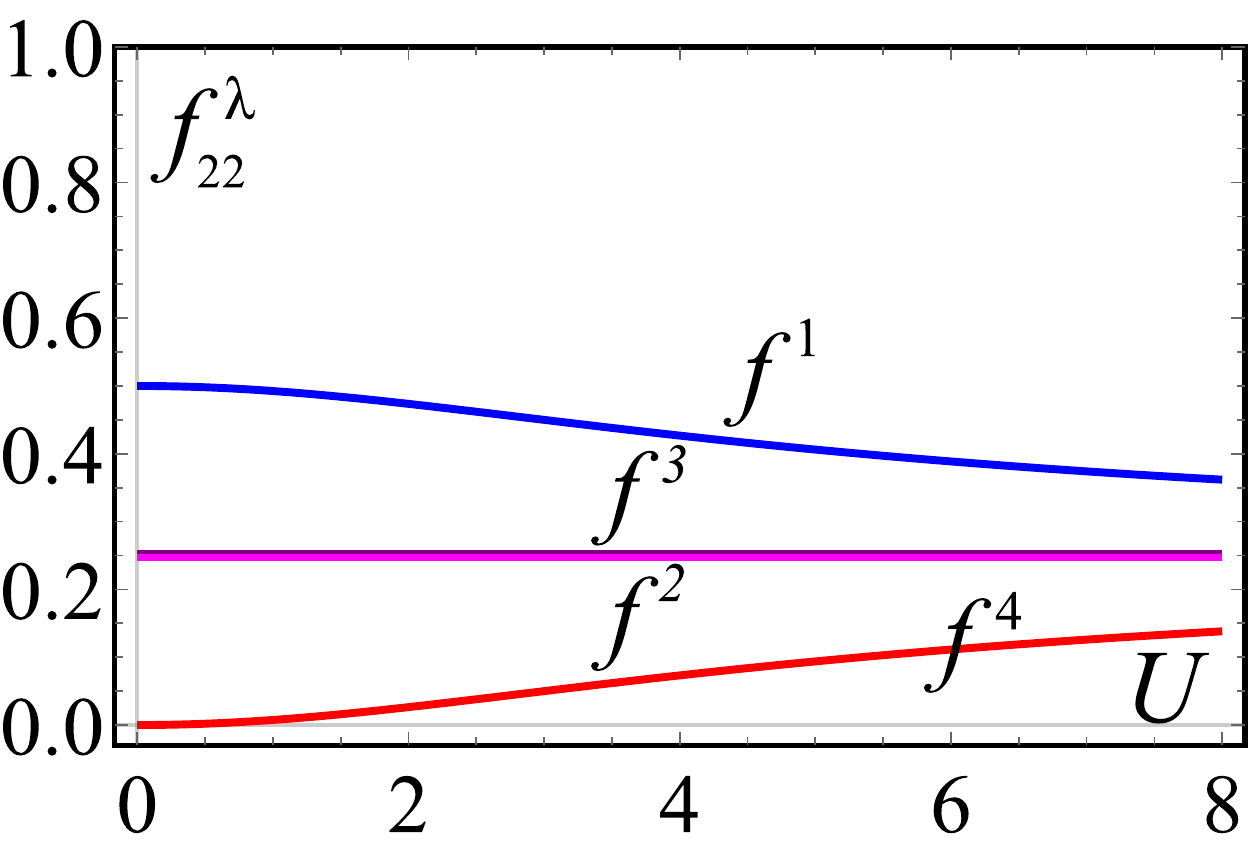}
	~
	\includegraphics[width=0.19\textwidth]{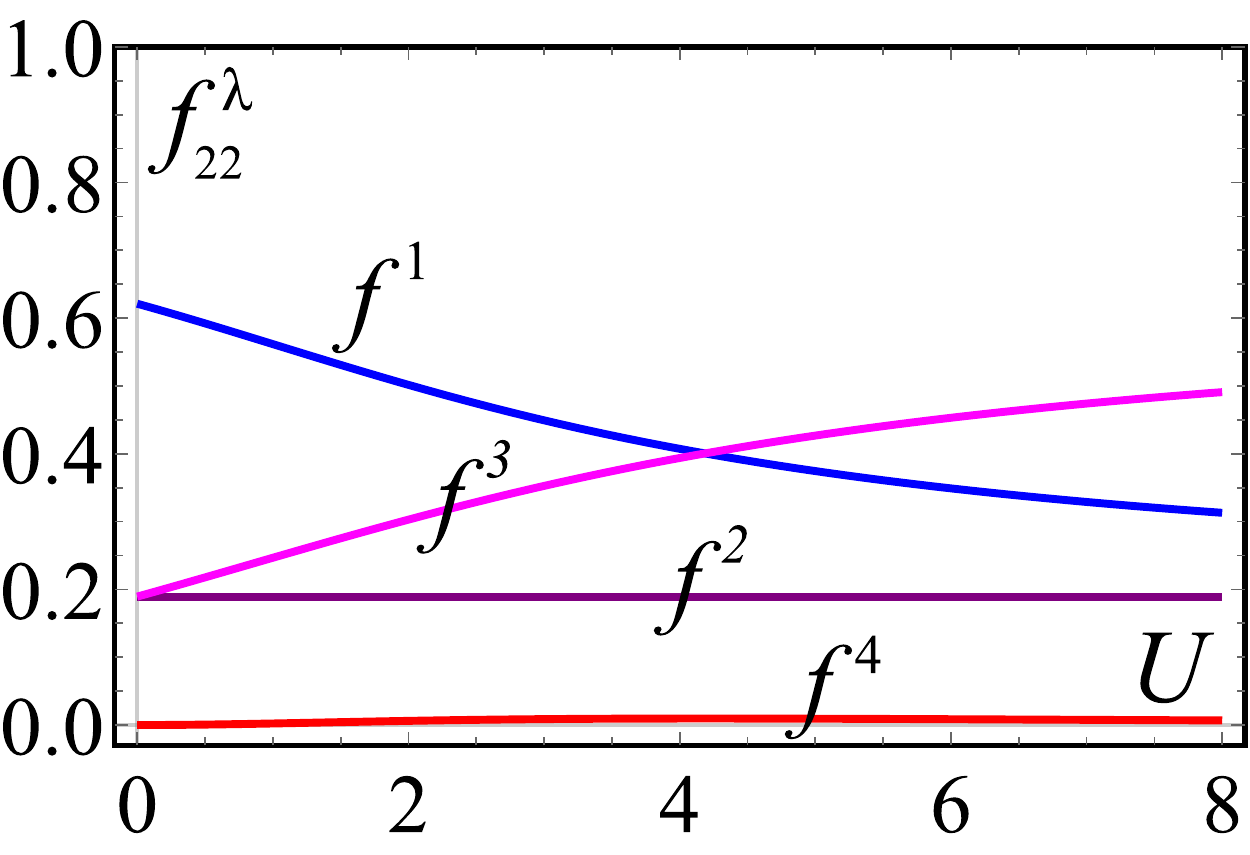}
	~
	\includegraphics[width=0.19\textwidth]{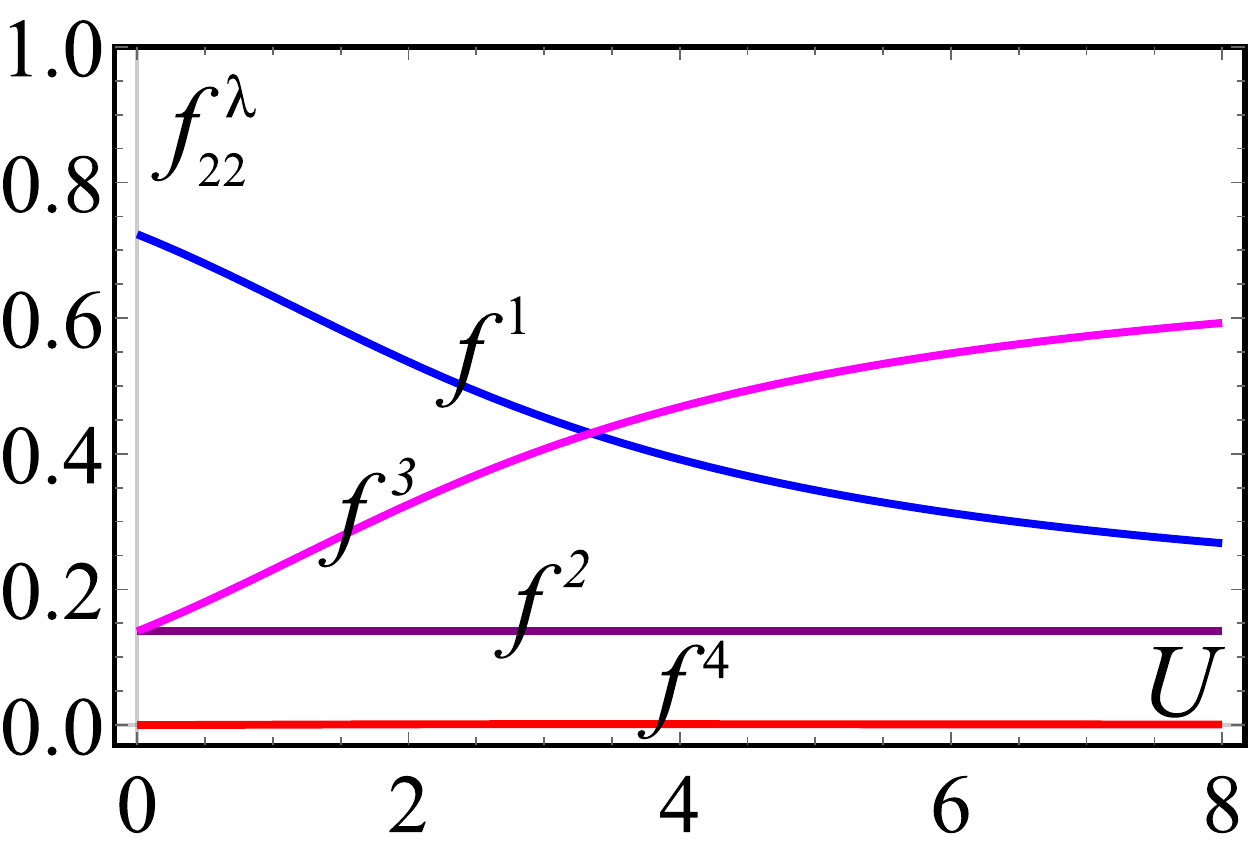}
	~
	\includegraphics[width=0.19\textwidth]{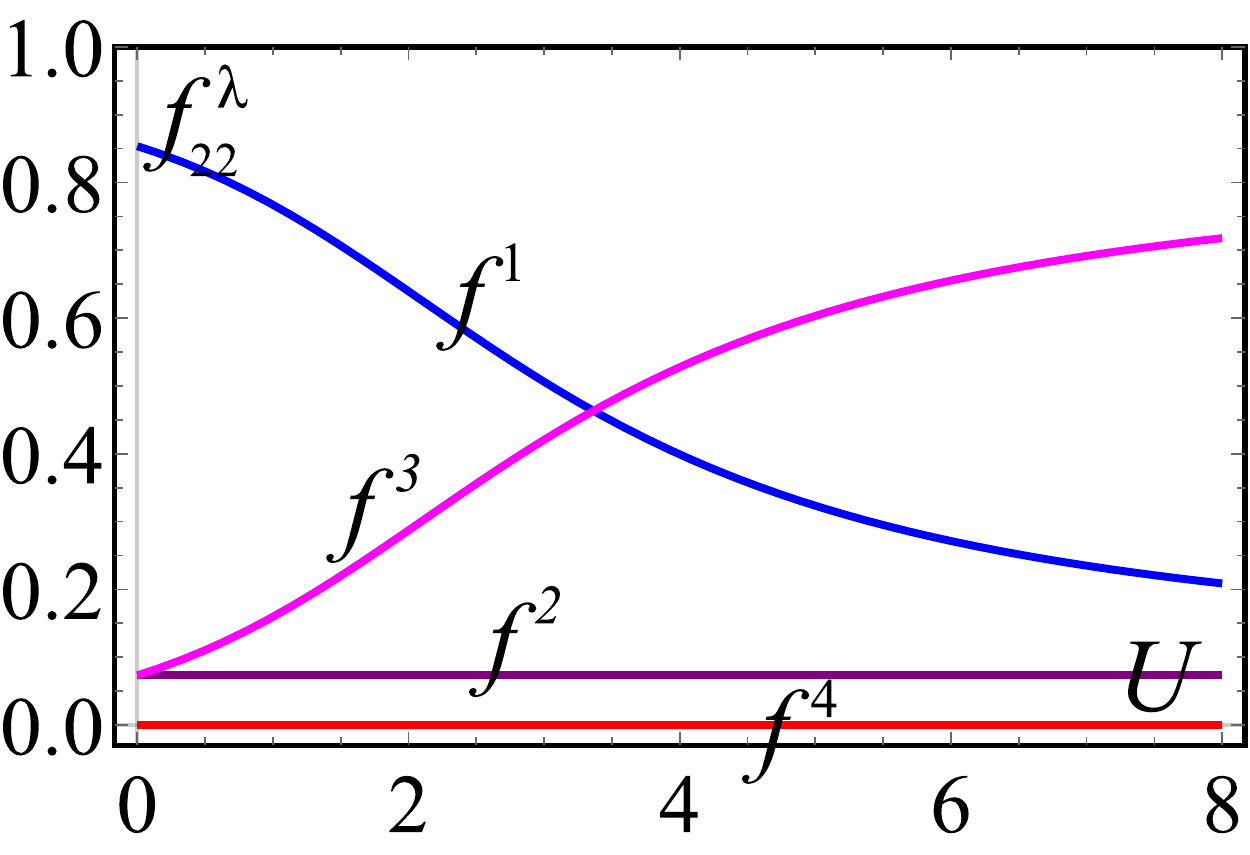}
	~
	\includegraphics[width=0.19\textwidth]{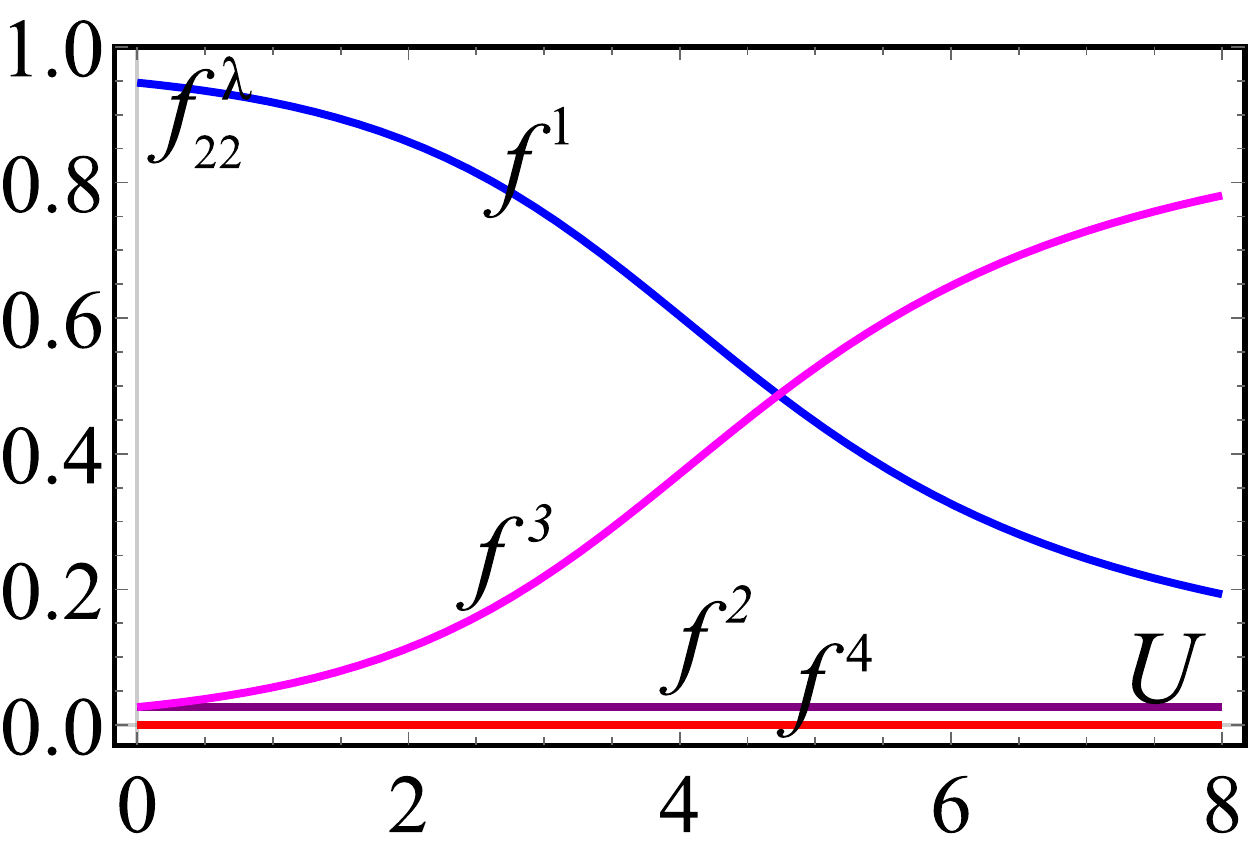}
	\\			$D=0.0$\qquad\qquad\qquad\quad\;\,$D=0.5$\quad\;\,\qquad\qquad\qquad$D=1.0$\qquad\qquad\qquad\quad\;\,$D=2.0$\qquad\quad\;\,\qquad\qquad$D=4.0$\\
	\caption{Weights $f_{22}^{\lambda}$.}
	\label{fig:asymSF_amplitudes_22}
\end{subfigure}
	
\vspace{0.5em}
		\caption{\emph{Position $\omega_{\lambda}$ and weights $f_{ii}^{\lambda}$ (for $i=1$, above, and $i=2$, below) of the poles of the spin--down Green's function, Eq. \eqref{eq:GFN1down}, as a function of $U$, in units of $t$, for different values of $D$, as indicated.}}
		\label{fig:reowp}
	\end{minipage}
\end{figure*}

\subsection{The exact Green's function}

The time ordered Green's function at zero temperature is defined as:
\begin{equation}
G_{ij,\sigma}(t,t')=-i\bra{\rm GS}\hat{T}\hat{c}_{i\sigma}(t)\hat{c}_{j\sigma}^{\dag}(t')\ket{\rm GS}.
\end{equation}

The Green's function is trivial for the spin--up case: the single electron in the ground state can be removed, or another spin--up electron can be added to the system and it will go to the antibonding orbital, where it will not interact with the first electron. As a result, the spin--up Green's function $G$ has two poles and is always equal to its non--interacting counterpart: $G_{ij,\uparrow}(\omega)=G_{ij,\uparrow}^0(\omega)$. The latter reads:
\begin{equation}
G^0_{ij,\sigma}(\omega)=
\frac{
	f_{ij}^{-}}
{\omega-e_--i\eta{\rm sign}\sigma}
+
\frac{
	f_{ij}^{+}}
{\omega-e_++i\eta},
\label{eq:asym_up}
\end{equation}
with weights defined as
\begin{align*}
f_{ij}^{-}&=\left[\delta_{i1}\cos\rho+\delta_{i2}\sin\rho\right]
\left[\delta_{j1}\cos\rho+\delta_{j2}\sin\rho\right]
\\
f_{ij}^{+}&=\left[\delta_{i1}\sin\rho-\delta_{i2}\cos\rho\right]
\left[\delta_{j1}\sin\rho-\delta_{j2}\cos\rho\right]
\end{align*}
and the convention that $\rm sign\sigma=+1$ ($-1$) for $\sigma=\uparrow$ ($\downarrow$).
The spin--down Green's function is far more interesting, and it is derived in App. \ref{sec:gasym}. It does not show any removal energy, as no spin--down electron is present in the system, but there are four addition \textit{channels} describing the different processes an incoming spin--down electron can undergo:
\begin{equation}
G_{ij,\downarrow}(\omega)= \sum_{\lambda=1}^{4} \frac{f_{ij}^{\lambda}}{\omega-\omega_{\lambda}+i\eta}.
\label{eq:GFN1down}
\end{equation}
The poles $\omega_{\lambda}$ and the relative weights $f_{ii}^{\lambda}$ of the spin--down Green's function are represented in Fig. \ref{fig:reowp} for different asymmetry values $D$ as a function of the interaction strength $U$. 
We can characterize these excitations by referring to 
the bonding/anti\-bonding orbital where the additional spin--down electron goes. 
The first and the fourth poles $\omega_1$ and $\omega_4$ are the excitation energies corresponding to the addition of a spin--down electron  to the bonding state
and the second and the third poles $\omega_2$ and $\omega_3$ to an antibonding state.
We will come back to the physical interpretation of the poles in Sec. \ref{sec:sym}.

\begin{figure*}[t]
	\centering
	\begin{minipage}{0.90\textwidth}
		\centering
		\begin{subfigure}[t]{0.24\textwidth}
			\centering
			\includegraphics[width=0.99\textwidth]{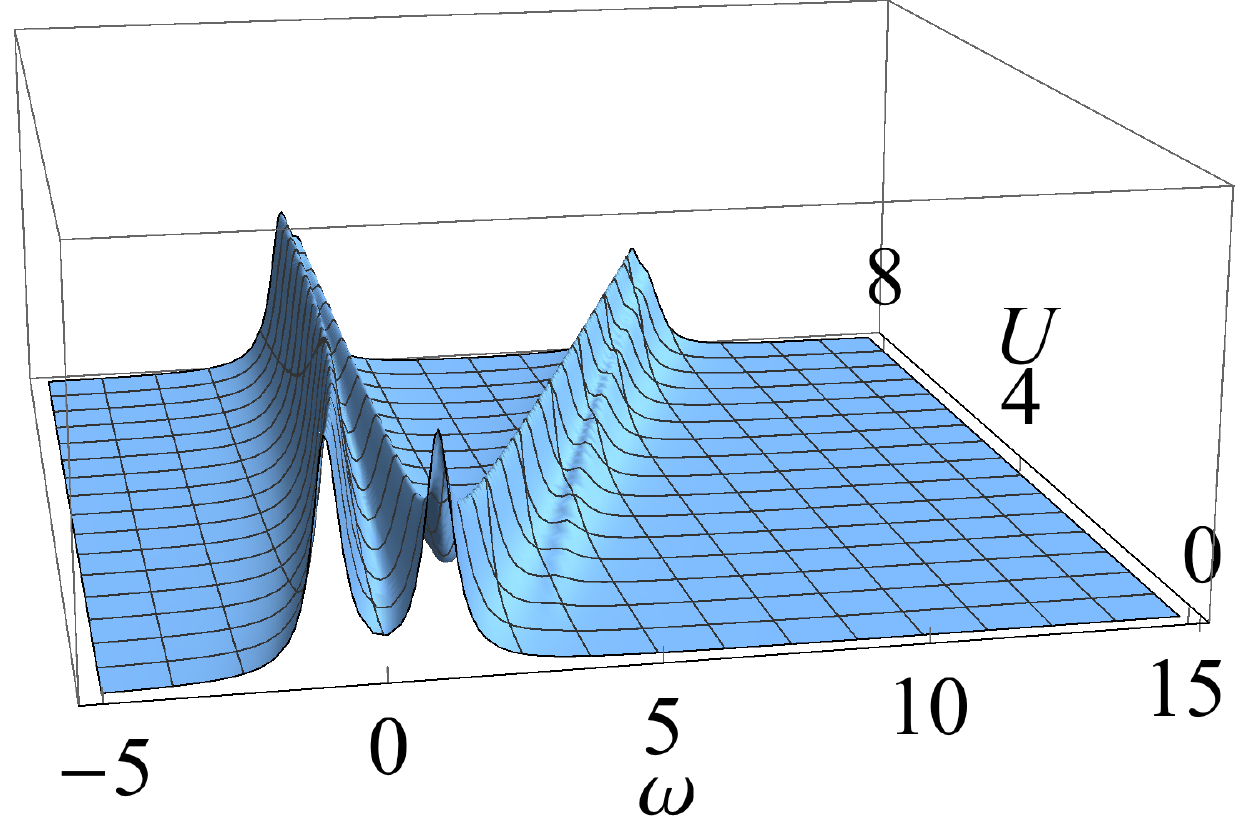}
			\caption{$D=0.0$}
		\end{subfigure}
		~
		\begin{subfigure}[t]{0.24\textwidth}
			\centering
			\includegraphics[width=0.99\textwidth]{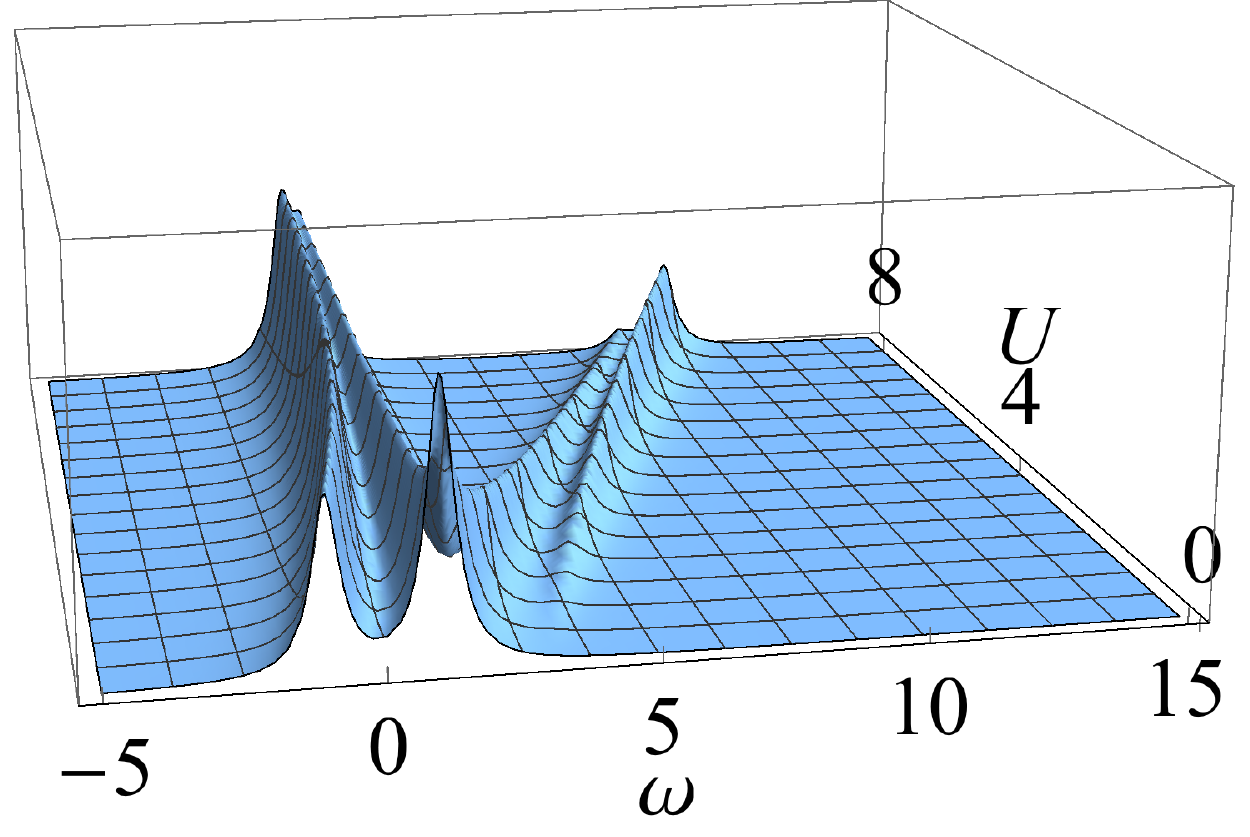}
			\caption{$D=0.5$}
		\end{subfigure}
		~
		\begin{subfigure}[t]{0.24\textwidth}
			\centering
			\includegraphics[width=0.99\textwidth]{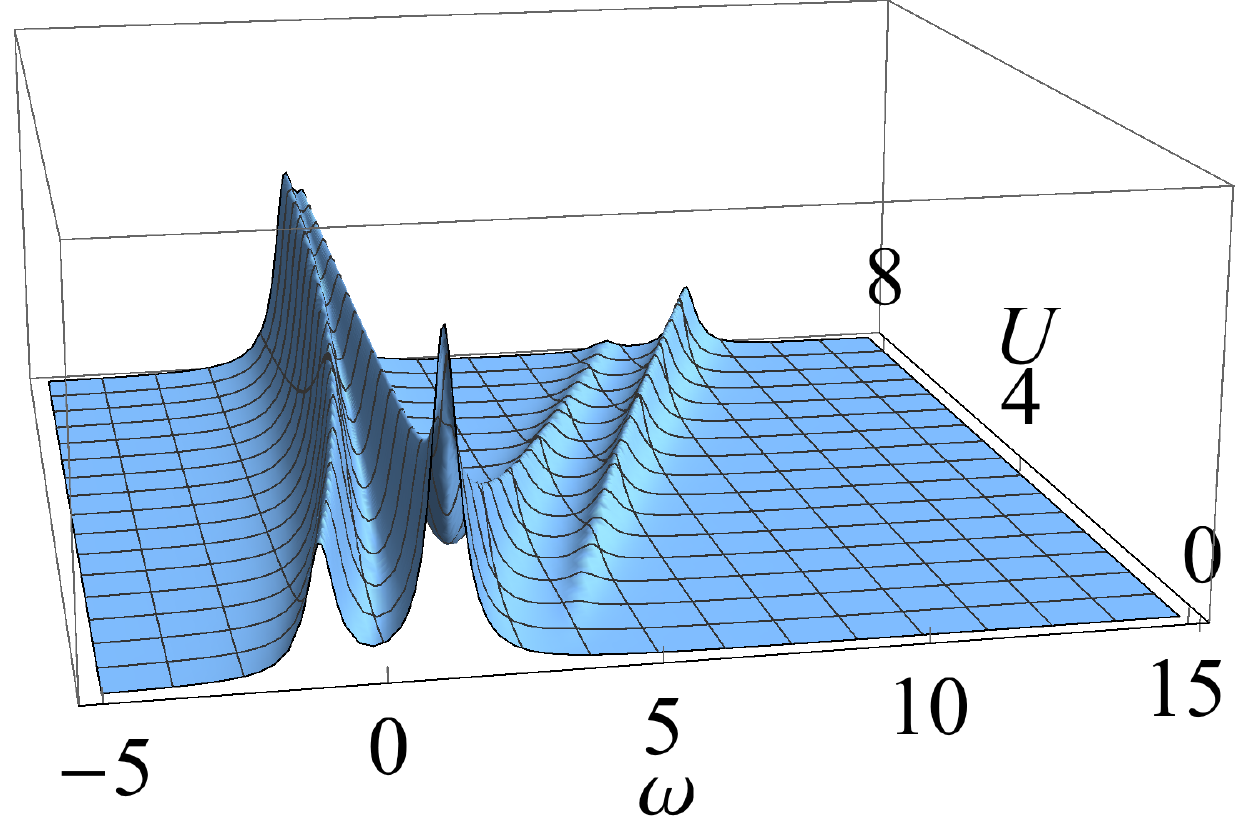}
			\caption{$D=1.0$}
		\end{subfigure}
		~
		\begin{subfigure}[t]{0.24\textwidth}
			\centering
			\includegraphics[width=0.99\textwidth]{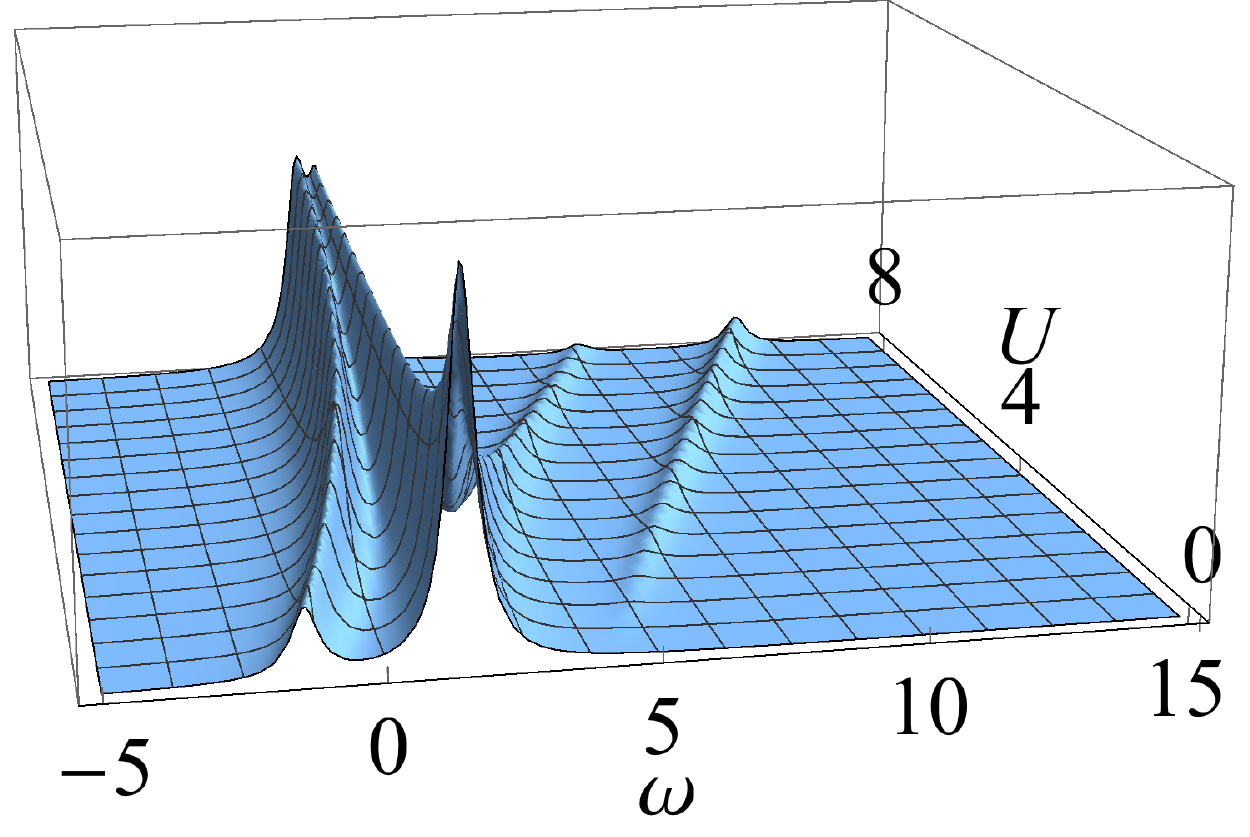}
			\caption{$D=2.0$}
		\end{subfigure}
	\\
	$A_{11,\downarrow}(\omega)$
	\\ \vspace{1em}
	\begin{subfigure}[t]{0.24\textwidth}
		\centering
		\includegraphics[width=0.99\textwidth]{figures/EPJ_D00}
		\caption{$D=0.0$}
	\end{subfigure}
	~
	\begin{subfigure}[t]{0.24\textwidth}
		\centering
		\includegraphics[width=0.99\textwidth]{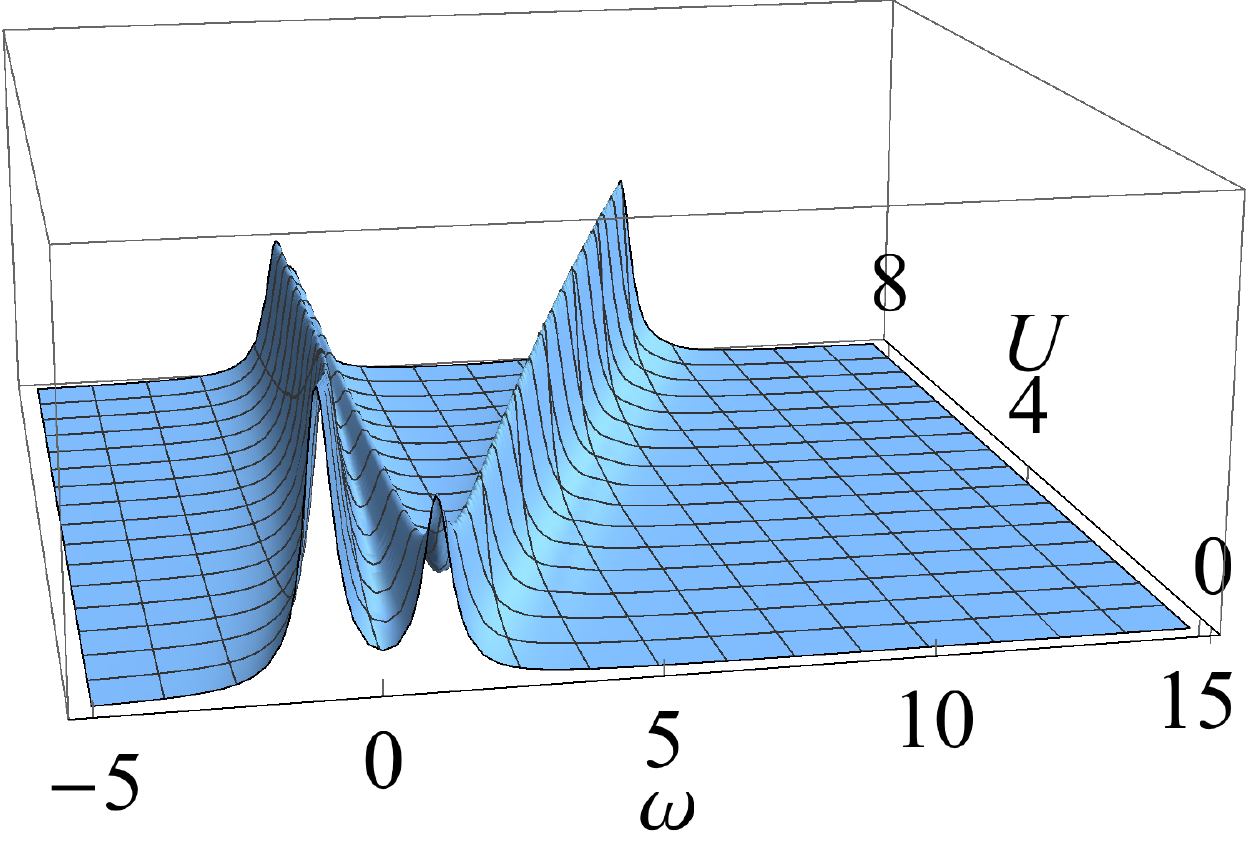}
		\caption{$D=0.5$}
	\end{subfigure}
	~
	\begin{subfigure}[t]{0.24\textwidth}
		\centering
		\includegraphics[width=0.99\textwidth]{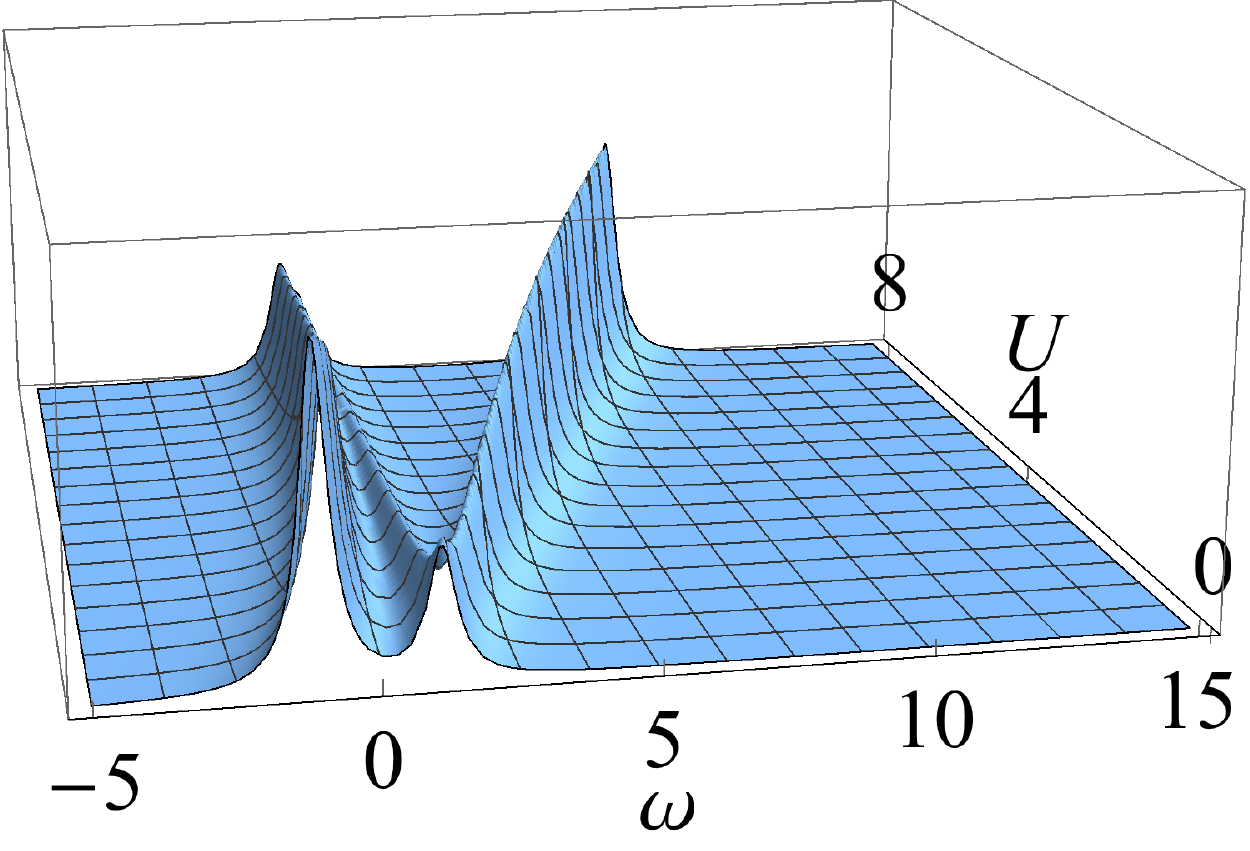}
		\caption{$D=1.0$}
	\end{subfigure}
	~
	\begin{subfigure}[t]{0.24\textwidth}
		\centering
		\includegraphics[width=0.99\textwidth]{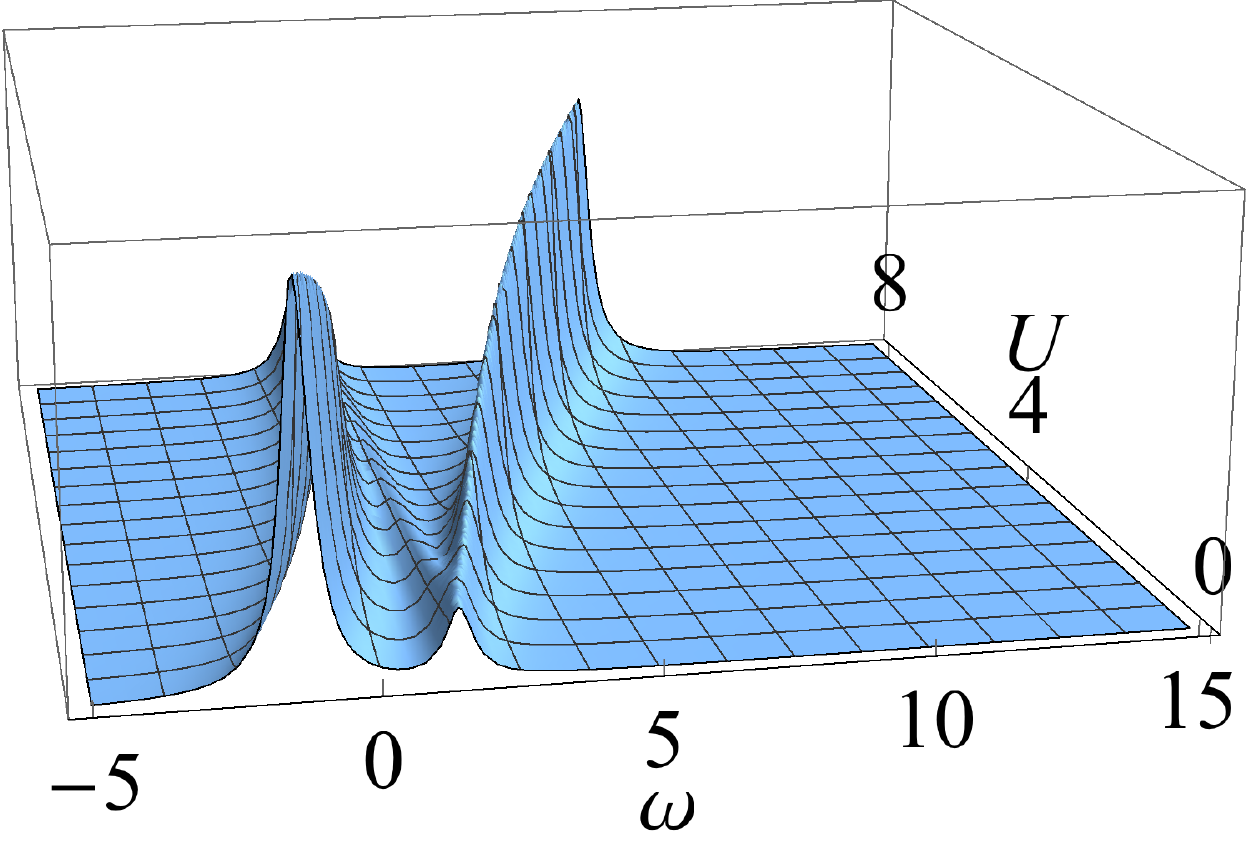}
		\caption{$D=2.0$}
	\end{subfigure}
\\
$A_{22,\downarrow}(\omega)$
\\ \vspace{0.5em}
		\caption{\emph{Diagonal of the spectral function $A_{ii,\downarrow}(\omega)$ for $i=1$ (line above) and $i=2$ (line below) as a function of $\omega$ and $U$, in units of $t$, for different values of the asymmetry parameter $D$, as indicated. The pictures are drawn with a lorentzian broadening $\eta=0.3$, for clarity reasons. The behaviour of the poles as a function of $U$ is the one of Fig. \eqref{fig:asymSF_poles}, while the amplitudes of the peaks follows fig. \ref{fig:asymSF_amplitudes} and fig. \ref{fig:asymSF_amplitudes_22} for $i=1$ and $i=2$ respectively.}}
		\label{fig:asymSF}
	\end{minipage}
\end{figure*}

In the non--interacting limit $U\to0$,  $\omega_1$ accounts for the bonding state $e_-$, while
$\omega_2$ and $\omega_3$ merge to the antibonding pole $e_+$. 
The fourth pole $\omega_4$ remains separate from the others also for $U\to0$. 
However it corresponds to an excitation that is not visible in the non--interacting Green's function.
Indeed the amplitude of the associated peak (see Fig. \ref{fig:asymSF_amplitudes} and Fig. \ref{fig:asymSF_amplitudes_22}) goes to zero  for $U\to0$, in such a way that 
the non--interacting Green's function \eqref{eq:asym_up} has thus just the two expected peaks at $\varepsilon_{\pm}$.

The behaviour of the diagonal elements of the spectral function
\begin{equation}
A_{ij,\sigma}(\omega)=-\tfrac{1}{\pi}{\rm sign}\left(\omega-\mu\right){\rm Im}G_{ij,\sigma}(\omega)
\end{equation}
is shown in Fig. \ref{fig:asymSF} for the spin--down case. 
From the fact that in the non--interacting limit all the weight of the bonding peaks is in $\omega_1$ and nothing in $\omega_4$, the pole $\omega_1$ can be considered as the quasiparticle, while the one at $\omega_4$ its satellite.
Moreover, in the non--interacting limit the poles $\omega_2$ and $\omega_3$ are degenerate  and have the same amplitudes.

Increasing $U$, at $D=0$, the peak in $\omega_1$ loses weight as the one in $\omega_4$ rises up. 
For large $U$ the four peaks  of the symmetric $D=0$ dimer merge into two pairs separated by a distance of the order of $U$: they become the two ``Hubbard bands''. This is the atomic limit, where the two possible excitation energies (which have the same probability) correspond to adding one electron to an isolated empty atom or to an isolated atom with one electron.
As soon as $D\neq0$, the left-right symmetry is broken and for increasing $D$ the bonding state tends to localize in the site 2, while the antibonding one in the site 1. For larger and larger $D$, it is more and more likely that the site 2 be occupied, and site 1 empty.

For non-zero values of $D$, instead, at large $U$ ($U\gg D$), for site 1, the poles  $\omega_1$ and $\omega_2$ become the dominant excitations, while $\omega_3$ and $\omega_4$ are no more degenerate and lose progressively weight with increasing $D$. By contrast, for site 2, $\omega_3$ becomes the most probable excitation.
When $U<D$, for the site 1 the most prominent peaks are $\omega_2$ and $\omega_3$, which are related to the electron addition to the antibonding orbital. In this parameter range, for the site 2 the most probable excitation is instead $\omega_1$, corresponding to the electron addition to the bonding state. Finally, for very large $D$ with respect to $U$, this picture reduces to the non-interacting situation.

Therefore, we see that by changing the ratio between the asymmetry $D$ and the interaction strength $U$ it is possible to explore the different regimes, ranging from a situation where the static correlation (i.e. left-right degeneracy) is essential to the one where the inhomogeneity is the dominant factor.

\subsection{Approximations to the self energy}
\label{sec:self}

The poles of the Green's function $G$ can alternatively be obtained from the Dyson equation:
\beq
G^{-1}_{ij,\sigma}(\omega) = G^{0 \;-1}_{ij,\sigma}(\omega) - \Sigma_{ij,\sigma}(\omega).
\eeq
The exact self energy $\Sigma_{ij,\sigma}(\omega)$ yields the poles $\omega_{\lambda}$ as the solutions of the pole equation $\omega-\omega_{\lambda}(\omega)=0$,
where $\omega_{\lambda}(\omega)$ are the eigenvalues of $h^0_{ij}+\Sigma_{ij,\sigma}(\omega)$
with
\beq
h^0_{ij}=\left(\begin{matrix}
	\tfrac{D}{2} & -1 \\ -1 & -\tfrac{D}{2}
\end{matrix}\right).
\eeq
From the the fact that the off--diagonal elements of the self-energy are equal, the pole equation reads:
\begin{multline}
\omega-\Biggl\{\frac{\Sigma_{11,\sigma}(\omega)+\Sigma_{22,\sigma}(\omega)}{2} +\Biggr.\\
\pm
\Biggl. \sqrt{\Biggl[1-\Sigma_{12,\sigma}(\omega)\Bigr]^2+\left[\frac{D}{2}+\frac{\Sigma_{11,\sigma}(\omega)-\Sigma_{22,\sigma}(\omega)}{2}\right]^2}\Biggr\}
=0.
\label{eq:fgSigma}
\end{multline}
If the exact self energy is not at hand, as it is most often the case, one resorts to \textit{approximate} expressions for it:  $\Sigma_{ij,\sigma}(\omega)\approx\Sigma^a_{ij,\sigma}(\omega)$. 

In the following, we will consider the position of the poles $\omega^a_{\lambda}$, solutions to Eq. \eqref{eq:fgSigma} within a particular approximation\linebreak $\Sigma^a_{ij,\sigma}(\omega)$ for the self energy, and compare them to the exact result. We will focus our analysis to the spin--down case, which is the most interesting one since, as pointed out above, the spin--up Green's function is always non--interacting. 

\begin{figure}[t!]
	\centering
	\begin{minipage}{0.99\columnwidth}
		\centering
		\begin{subfigure}[t]{0.48\textwidth}
			\centering
			\includegraphics[width=0.99\textwidth]{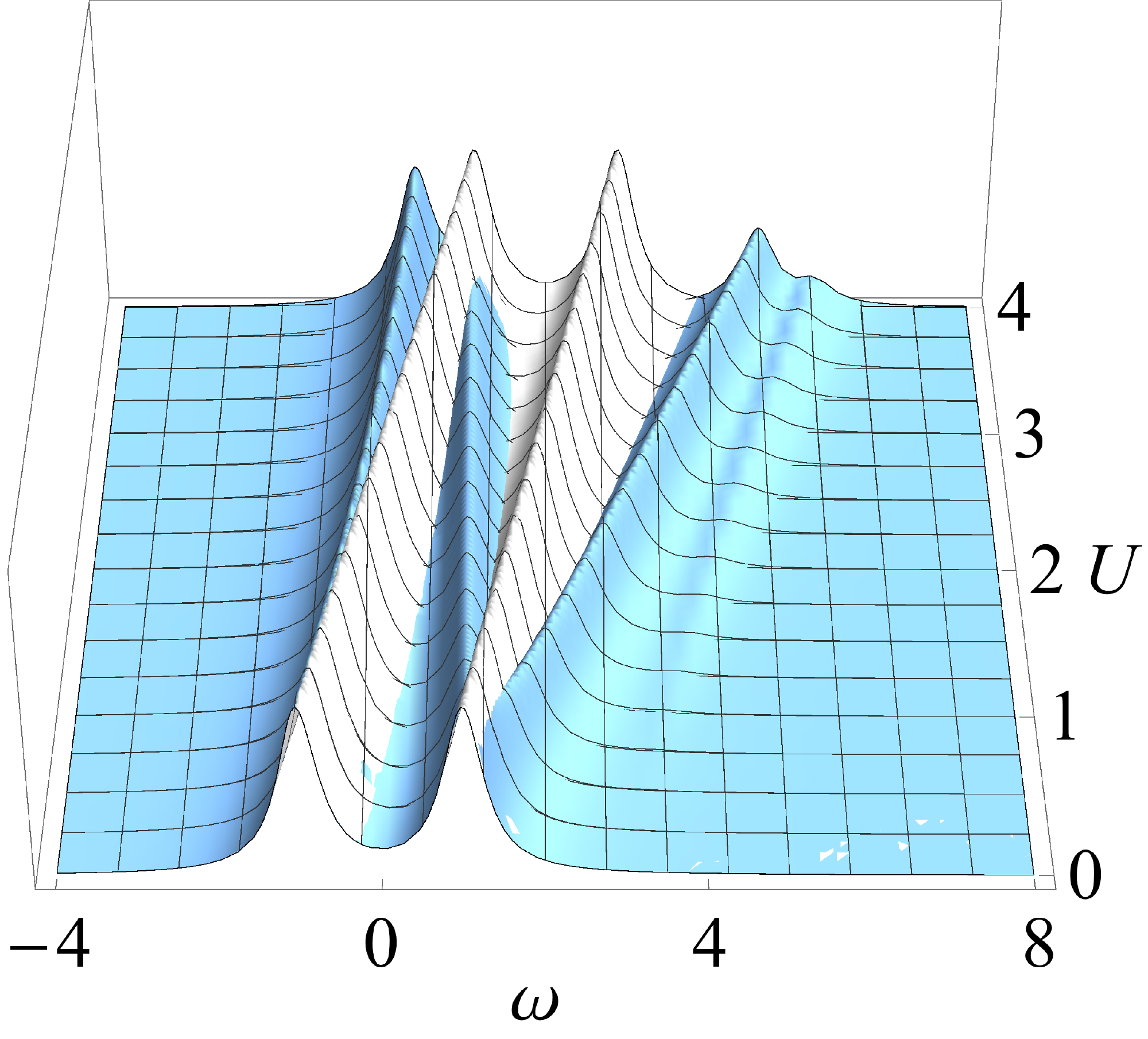}
			\caption{$D=0.0$}
		\end{subfigure}
		~
		\begin{subfigure}[t]{0.48\textwidth}
			\centering
			\includegraphics[width=0.99\textwidth]{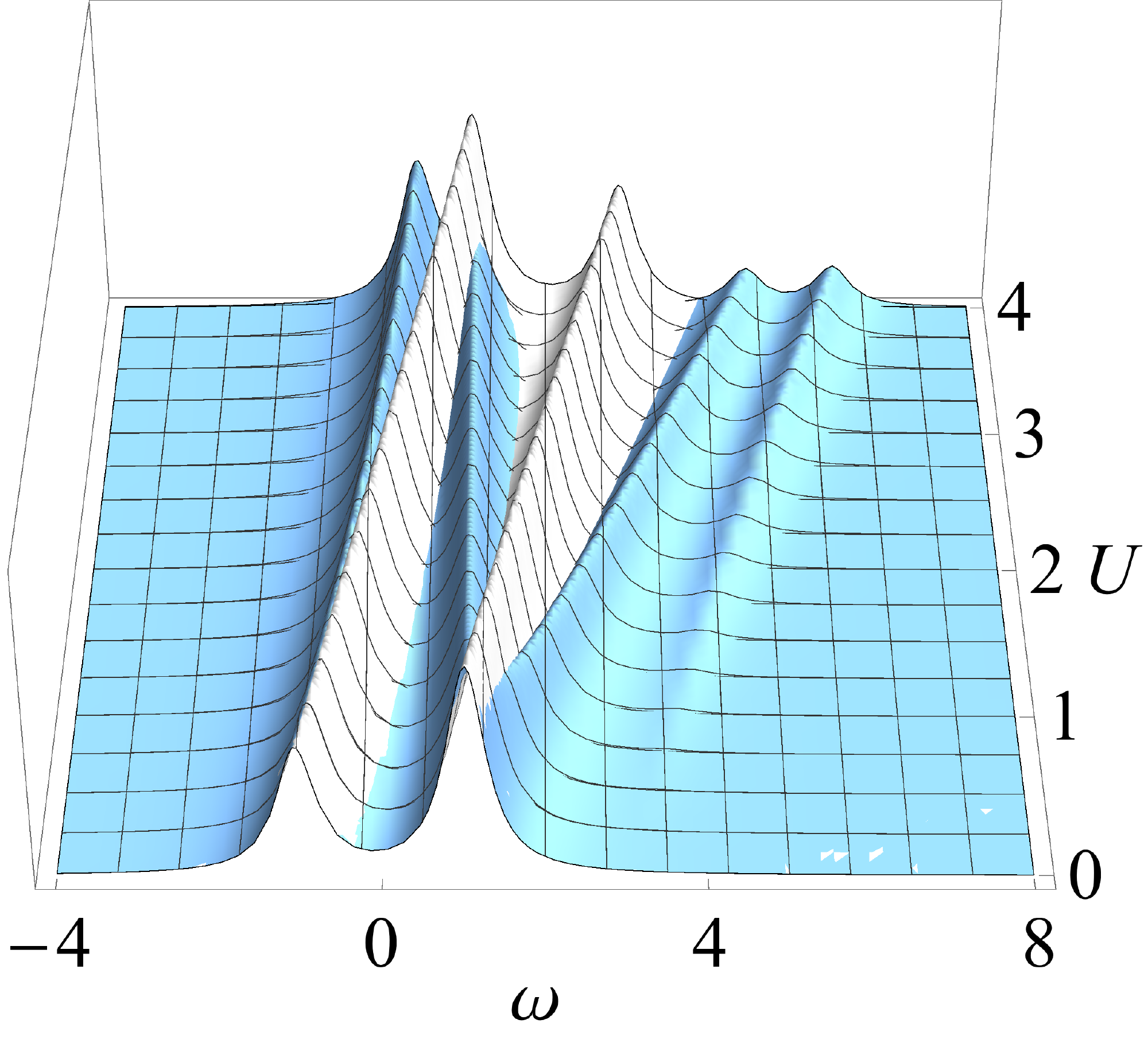}
			\caption{$D=0.5$}
		\end{subfigure}
		
        \vskip 1em
		\begin{subfigure}[t]{0.48\textwidth}
			\centering
			\includegraphics[width=0.99\textwidth]{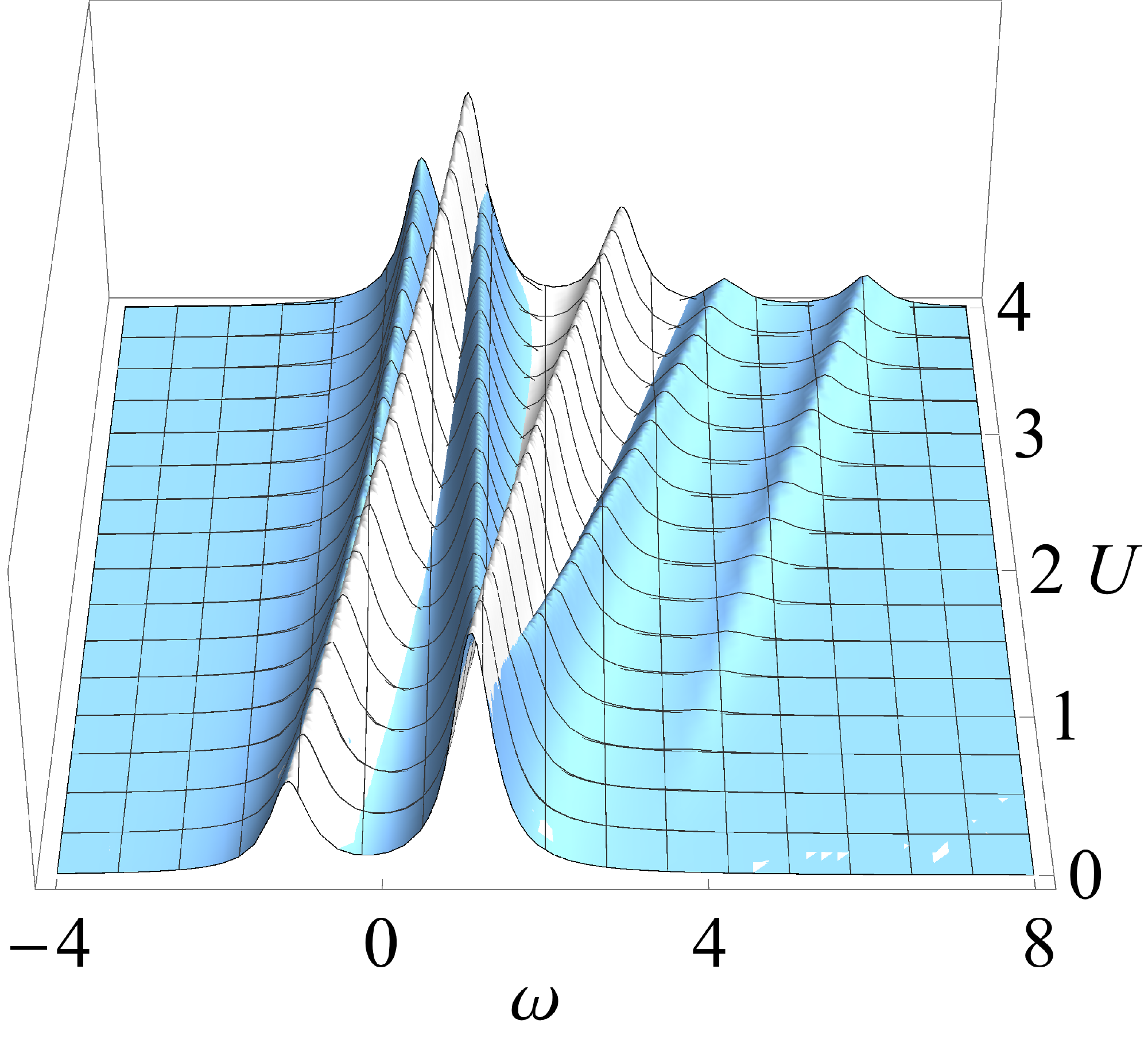}
			\caption{$D=1.0$}
		\end{subfigure}
		~
		\begin{subfigure}[t]{0.48\textwidth}
			\centering
			\includegraphics[width=0.99\textwidth]{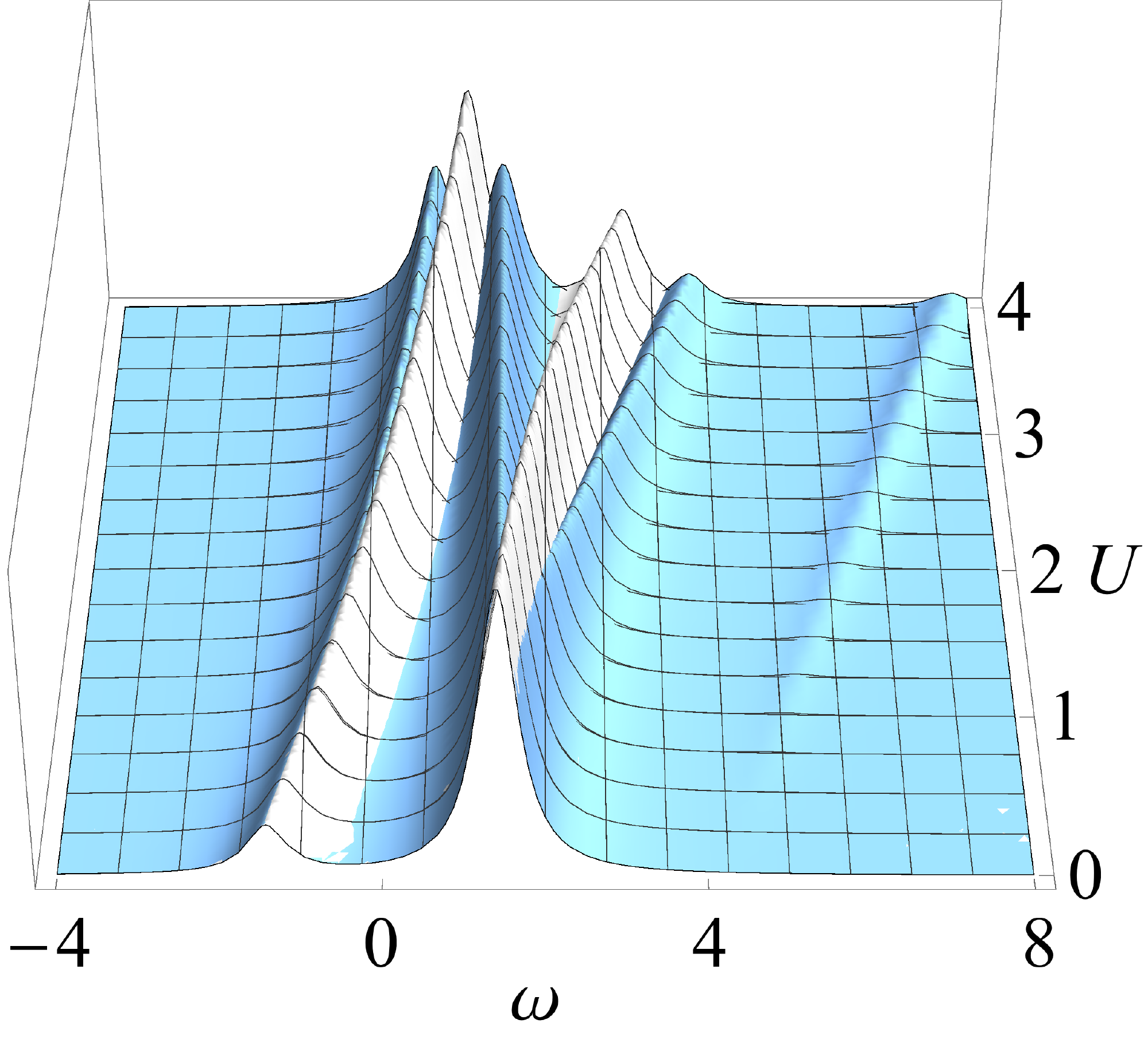}
			\caption{$D=2.0$}
		\end{subfigure}
		\caption{\emph{The Hartree approximation for $A_{11,\downarrow}(\omega)$. The blue surface is the exact spectral function, the same as in the upper row of Fig. \ref{fig:asymSF}, while in white the one evaluated with the Hartree approximation, Eq. \eqref{eq:asym_hartree}.}}
		\label{fig:hartreeSFasym}
	\end{minipage}
\end{figure}

\paragraph{Hartree approximation}
The simplest approximation to the full self-energy is the Hartree approximation.
The Hartree potential is defined as:\footnote{The Hartree potential is here spin-independent, as usual in ab initio calculations \cite{Romaniello2009}. An alternative definition of a spin-dependent Hartree potential is also possible  (see e.g. \cite{Wang2008,vonFriesen2009,vonFriesen2010}). We refer to \cite{marco_phd} for an extended discussion.}
\begin{equation}
v_i^{\rm H} 
=Un_i.
\label{eq:asym_hartree}
\end{equation}

\begin{figure*}[t!]
	\centering
	\begin{minipage}{\textwidth}
		\centering
		\begin{subfigure}[t]{0.24\textwidth}
			\centering
			\includegraphics[width=0.99\textwidth]{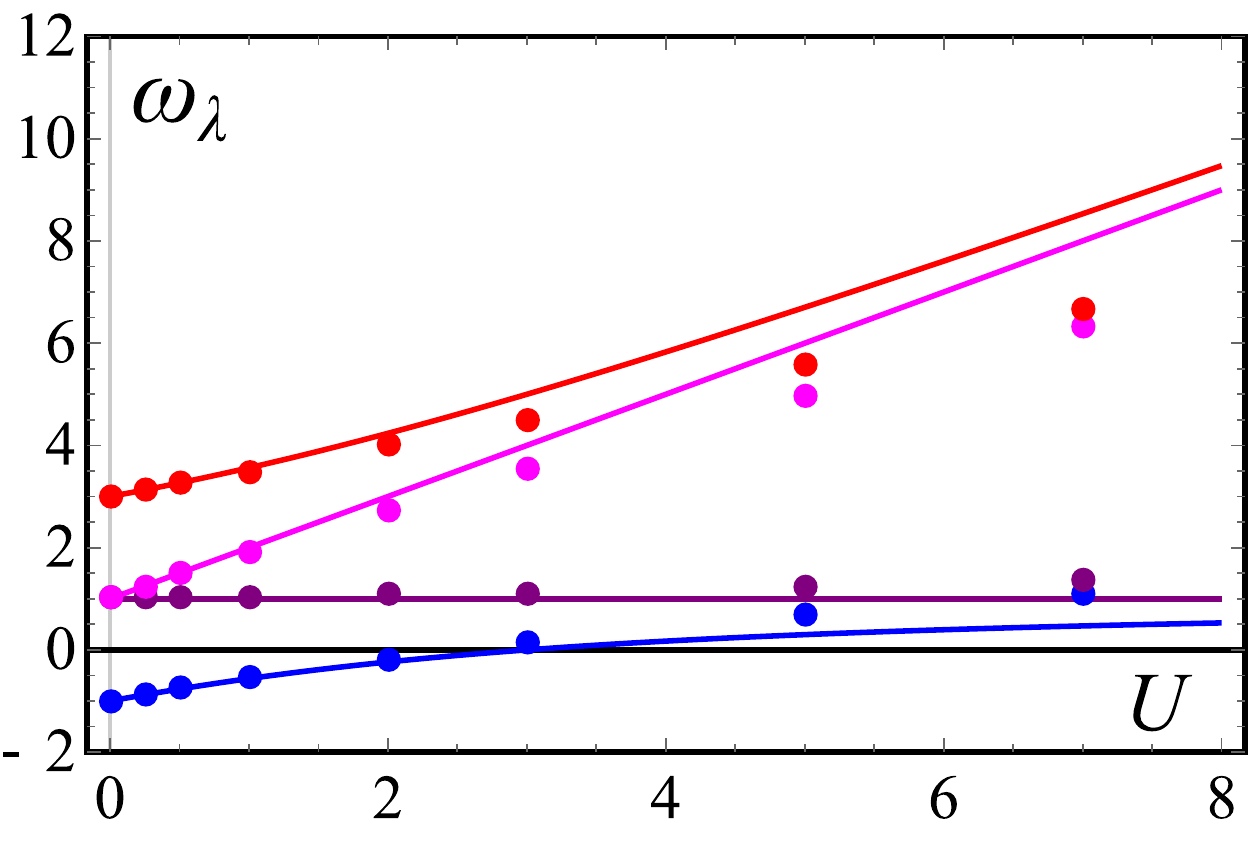}
			\caption{$D=0.0$}
		\end{subfigure}
		~
		\begin{subfigure}[t]{0.24\textwidth}
			\centering
			\includegraphics[width=0.99\textwidth]{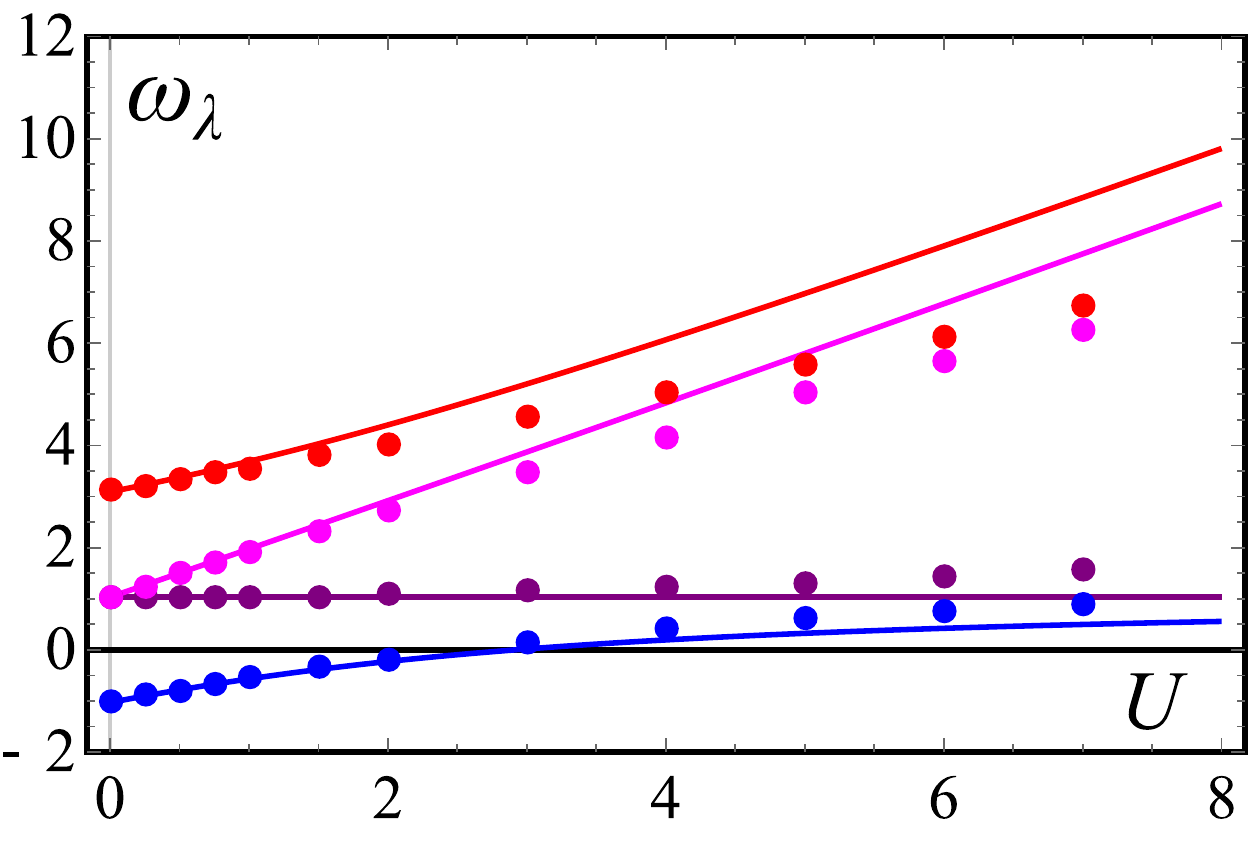}
			\caption{$D=0.5$}
		\end{subfigure}
        ~
		\begin{subfigure}[t]{0.24\textwidth}
			\centering
			\includegraphics[width=0.99\textwidth]{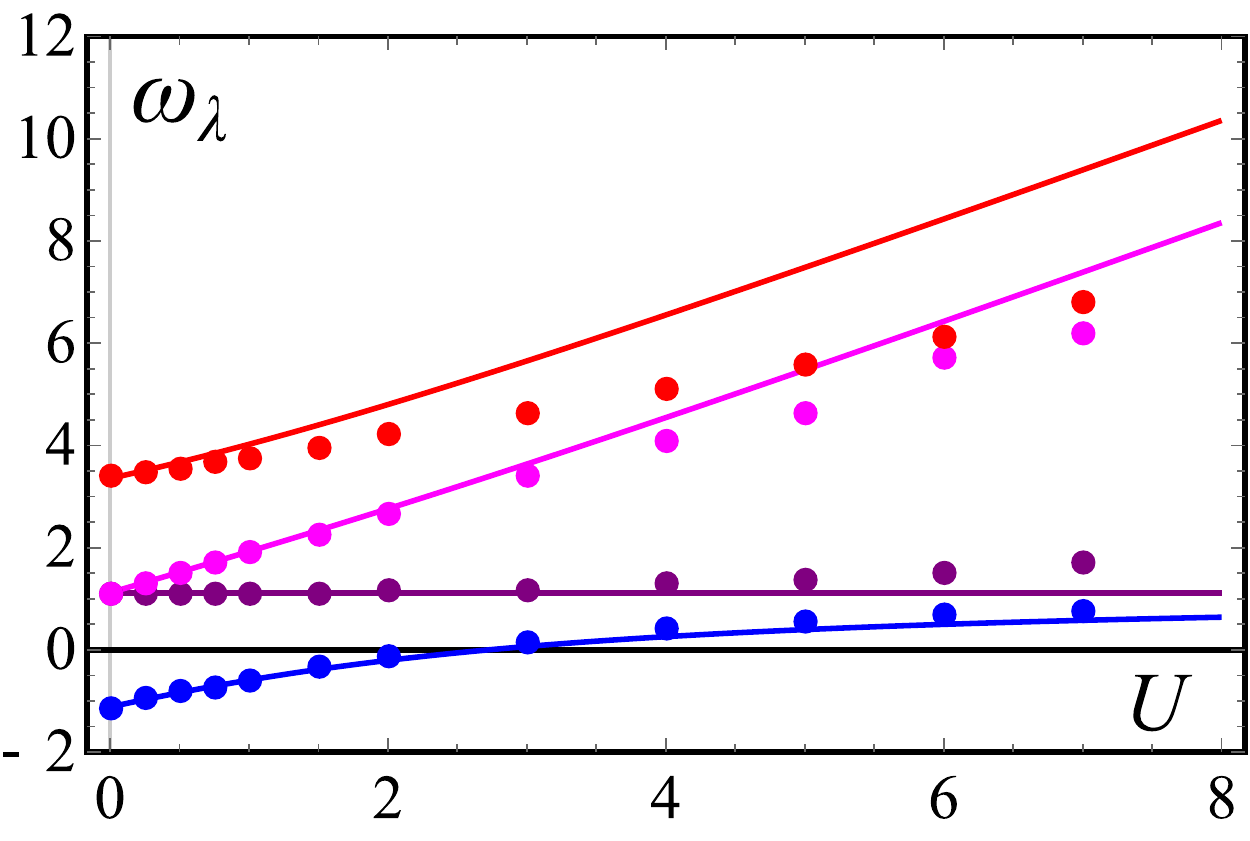}
			\caption{$D=1.0$}
        \end{subfigure}
        ~
		\begin{subfigure}[t]{0.24\textwidth}
			\centering
			\includegraphics[width=0.99\textwidth]{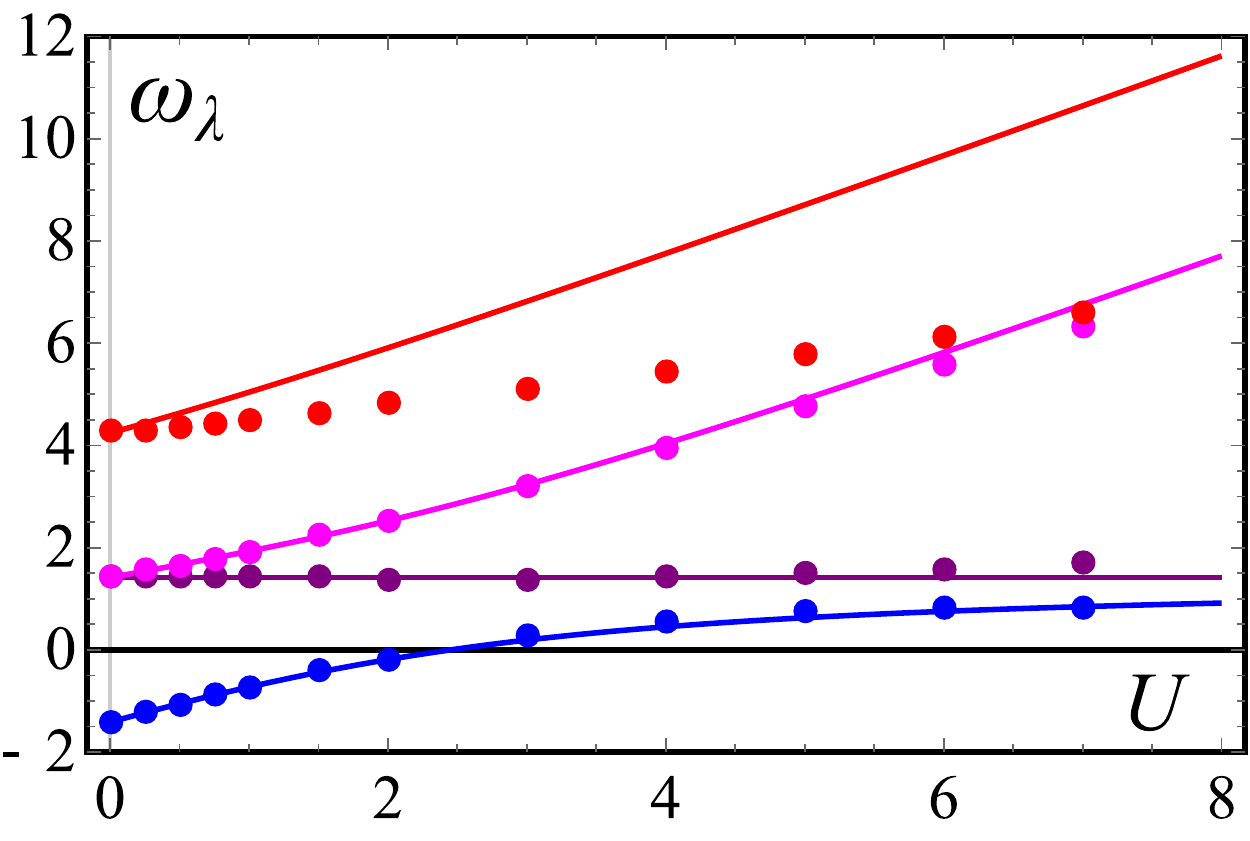}
			\caption{$D=2.0$}
        \end{subfigure}
		\caption{\emph{Position of the poles of the spin--down GW Green's function as a function of $U$. Solid line, exact results, as in Fig. \ref{fig:asymSF_poles}; dots, GW poles. }} 
		\label{fig:GW_pina_SFasym_D}
	\end{minipage}
\end{figure*}
From Eq. \eqref{eq:fgSigma} with $\Sigma^a_{ij}(\omega)=v_i^{\rm H}\delta_{ij}$, the poles are:
\begin{equation}
\omega_{\pm}^{\rm H}=\frac{U}{2}\pm\sqrt{1+h^2},
\label{eq:hartreepoles}
\end{equation}
with 
\beq
h =\frac{D}{2}+U\frac{n_1-n_2}{2}.
\label{def:h}
\eeq
The resulting spectral function is shown in Fig. \ref{fig:hartreeSFasym}. 
The Hartree potential is static, so it can at most shift the position of the two peaks $\varepsilon_{\pm}$ of the non--interacting spectral function, but it is unable to split them.
Nonetheless, it is an improvement with respect to the free--particle approximation $U=0$, in which the position of the poles would be $U$--independent\footnote{Note, however, that a spin--independent Hartree potential (and, more generally, any non--zero spin--up self energy) spoils the spin--up part of the Green's function, which is exact at the non--interacting level.}. Since the Hartree potential $v_i^{\rm H}$ depends on $U$, the two peaks $\omega_{\lambda}^{\rm H}$ of the Hartree spectral function interpolate well between the four peaks of the exact spectral function. In particular, the Hartree approximation is a pretty good approximation for small $U$, independently of $D$. By contrast, for large $U$, where correlation effects become important, the correspondence is worse.

\paragraph{The GW approximation}
In the GW approximation (GWA) the self energy $\Sigma^a_{ij}(\omega)$ is obtained from 
the convolution in frequency space of the Green's function 
and the screened interaction $W$ calculated in the random-phase approximation (RPA).\footnote{As already noted for the Hartree approximation,
	it is possible to consider a GWA self energy constructed starting from a spin--dependent interaction \cite{marco_phd}. 
	In that case one obtains that the exchange self energy is zero and the spin--up Green's function is always exact. Moreover, the GWA would 
	solve exactly the symmetric $D=0$ model.  
	Here instead we employ a spin--independent interaction, which is closer to usual GWA in solids. 
	It does treat spins on the same footing, adding additional poles to the spin--up Green's function \eqref{eq:asym_up}. 
	Moreover, it does not solve exactly the $D=0$ system, and this is precisely the interest to employ this formulation, and not the former, for the following discussion.}
In particular, here we consider the non--self--consistent version of the GWA
where the Green's function is the non--interacting $G^0$
and the RPA polarization $\Pi_0^{RPA}\sim-iG^0G^0$,
which dresses the bare interaction $U$, is also evaluated with non--interacting Green's functions:
\begin{equation}
\Sigma_{ij,\sigma}^{G_0W_0}(\omega)=i\int\frac{d\omega'}{2\pi}e^{i\omega'\eta}G^0_{ij,\sigma}(\omega+\omega'){W_{ij}^{RPA}}(\omega').
\end{equation}
The GW self energy, in addition to the Fock exchange term $\Sigma^X_{ij,\sigma}=-\delta_{\sigma,\uparrow}\delta_{ij}Un_{i}$ (which however acts on the spin-up channel only), contains a non--local, complex and frequency dependent contributions, which is derived in App. \ref{sec:gwasym}. The final result for the spin--down case
reads: 
\begin{multline*}
\Sigma_{ij,\downarrow}^{G_0W_0}(\omega)
=\frac{(-1)^{(i-j)}\frac{U^2}{2l}}{\sqrt{1+\frac{D^2}{4}}}
\left[
\frac{f_{ij}^{+}}{\omega-\left({e}_++l\right)+i\eta}+\right.
 \\
+ 
\left. \frac{f_{ij}^{-}}{\omega-\left({e}_-+l\right)+i\eta}
\right],
\end{multline*}
with 
\beq 
l^2=4\left(1+\frac{D^2}{4}\right)+\frac{2U}{\sqrt{1+\frac{D^2}{4}}}.
\label{def:l}
\eeq
The poles of the Green's function evaluated with the GW self energy are 
obtained as the solutions of Eq. \eqref{eq:poles_asym_GW} in App. \ref{sec:gwasym}.
In Fig. \ref{fig:GW_pina_SFasym_D}, they are represented 
for the spin--down Green's function for four values of $D$, as a function of $U$. 

The behavior of the GWA has been discussed in detail for the symmetric case $D=0$ in Refs. \cite{Romaniello2009,Romaniello2012}.
In that case, the GWA works well for small interaction $U$, while it tends to close the gap between the Hubbard bands for large value of $U$. 
Here we find that, apart from the fourth pole, its performance improves for larger values of $D$, 
showing that the asymmetry counteracts the effect of the interaction. The $D\to\infty$ limit corresponds to the non-interacting $U\to0$ limit, for which the GWA is exact.

\section{The connector strategy}
\label{sec:connector}

We now come back to the real and frequency--dependent spectral potential $v_{{\rm SF}}(\bfr,\omega)$ introduced in Eq. \eqref{eq:SSE_SF}, which is local in real space. In a lattice model it becomes a site-dependent potential $v_{{\rm SF}\,i}(\omega)$.
It defines the Green's function $G_{{\rm SF}\,ij}$ of the auxiliary system:
\begin{equation}
	G^{-1}_{{\rm SF}\,ij}(\omega)=G^{0\,-1}_{ij}(\omega)-v_{{\rm SF}\,i}(\omega),
\label{eq:GF_aux_D}
\end{equation}
which is built to exactly yield the diagonal elements of the spectral function $A_{ii}(\omega)$:
\begin{equation}
-\frac{1}{\pi}{\rm sign}(\omega-\mu)\operatorname{Im}G_{{\rm SF}\,ii}(\omega)\equiv A_{{\rm SF}\,ii}(\omega)\stackrel{!}{=}A_{ii}(\omega)
\label{eq:GSSE_simpl_dimer_1}
\end{equation}
In  a discrete system this condition is equivalent to reproduce the position of the poles, together with the intensities of the corresponding peaks. 
In particular, in the dimer we are interested only in  the spin--down part of the spectral function.
Moreover, since the poles are independent of the particular basis, it is useful to express the previous relation in the basis \eqref{eq:twrhirn}, where the non--interacting Green's function $G^0$ is diagonal, and the spectral potential reads:
\begin{equation}
	v_{{\rm SF}\,\alpha\beta}(\omega)=V_{\rm SF}(\omega)\delta_{\alpha\beta}+\frac{\Delta v_{\rm SF}(\omega)}{\sqrt{D^2+4}}
	\left(
	\begin{matrix}
		-\frac{D}{2}
		& 1
		\\
		1
		& \frac{D}{2}
	\end{matrix}\right),
\label{eq:SP_D_bab}
\end{equation}
with 
\begin{equation*}
\begin{gathered}
V_{\rm SF}(\omega):=\tfrac{1}{2}\bigl[v_{{\rm SF}\,1}(\omega)+v_{{\rm SF}\,2}(\omega)\bigr] \\
\Delta v_{\rm SF}(\omega):=v_{{\rm SF}\,1}(\omega)-v_{{\rm SF}\,2}(\omega). 
\end{gathered}
\end{equation*}
A local potential in the site basis, whose value depends on the particular site, is not local anymore in the $\alpha\equiv\pm$ basis. In this basis, the equation that defines the auxiliary system becomes:
\begin{equation}
G^{-1}_{{\rm SF}\,\alpha\beta}(\omega)=G^{0\;-1}_{\alpha}(\omega)\delta_{\alpha\beta}-v_{{\rm SF}\,\alpha\beta}(\omega),
\label{eq:GF_aux_D_bab}
\end{equation}
with $G^{0\;-1}_{\alpha}(\omega)=\omega-e_{\alpha}$, $e_{\alpha}=e_{\pm}=\pm\sqrt{1+\frac{D^2}{4}}$. The previous equation defines the frequency--dependent effective Hamiltonian in the auxiliary system, namely:
\begin{equation}
	H_{{\rm SF}\,\alpha\beta}(\omega)=e_{\alpha}\delta_{\alpha\beta}+v_{{\rm SF}\,\alpha\beta}(\omega).
\label{eq:Heffauxhub}
\end{equation}
This Hamiltonian, which is not diagonal due to the presence of the local spectral potential, can be diagonalized
and the poles of the Green's function $G_{{\rm SF}\;ij}(\omega)$ can be determined by the conditions (analogous to Eq. \eqref{eq:fgSigma}):
\begin{equation}
	\begin{cases}
		\omega-\Bigl[V_{\rm SF}(\omega)-\sqrt{1+\bigl(\frac{D+\Delta v_{\rm SF}(\omega)}{2}\bigr)^2}\Bigr]=0
		\\
		\omega-\Bigl[V_{\rm SF}(\omega)+\sqrt{1+\bigl(\frac{D+\Delta v_{\rm SF}(\omega)}{2}\bigr)^2}\Bigr]=0
	\end{cases}.
\label{eq:fg}
\end{equation} 
If $v_{{\rm SF}\,i}(\omega)$ is the exact spectral potential, these two equations must possess the four\footnote{Other solutions are allowed if the derivative of the potential diverges, see the discussion of section \ref{sec:symmvSF}
.} solutions $\omega_{\lambda}$, i.e. the poles of the Green's function of the real system. 
We could therefore find the exact spectral potential  by solving those equations for $v_{{\rm SF}\,i}(\omega)$. Those conditions (together with those deriving from the the requirement to match the intensities of the peaks) would be equivalent to solve the generalized Sham-Schl\"uter equation \eqref{eq:SSE_SF} for the dimer.

However, in  real applications one does not dispose of the exact solution of the Hamiltonian.
Therefore, in the following we will consider a different strategy.
We will aim to build directly the spectral potential without making use of the self-energy (or a corresponding spectral function) and we will use it in Eq. \eqref{eq:fg} to calculate the poles $\omega^{{\rm SF}}_{\lambda}$ of the asymmetric dimer. To benchmark our approach, we will compare the resulting $\omega^{{\rm SF}}_{\lambda}$ to the ones obtained from the exact solution of the Hamiltonian or from the different approximations to the self-energy that have been discussed in Sec. \ref{sec:self}.

\subsection{The model system}

In order to directly build  approximations to the spectral potential, we take inspiration from Kohn-Sham DFT where in the LDA the xc potential $V_{{\rm xc}}(\bfr)$ is imported from a model system, namely the homogeneous electron gas.
In our case the natural candidate to play the role of model system is the symmetric Hubbard dimer. 
In the same way as for the homogeneous electron gas,  
in the symmetric dimer inhomogeneities (or asymmetries) are absent, and an exact solution is easier to obtain.

Once the potential is at hand in the model system, one has to import it in the auxiliary system via a suitable ``\textit{connector}''. 
The connector is a very general prescription that states \textit{what} to import and \textit{how} to do that. 
For the LDA, at each point in space $\bfr$ the LDA connector is the local density $n(\bfr)$ that identifies the uniform density defining the corresponding homogeneous electron gas, from which $V_{{\rm xc}}$ for that point is imported.
Here for the spectral potential in the dimer we adopt as a connector a \textit{pole--by--pole} correspondence. 
We can imagine that, switching on $D$ from a $D=0$ initial situation, the \textit{nature} of the poles be unchanged, and the potential needed to reproduce a certain pole $\omega_{\lambda}$ can be mapped continuosly from the potential at $\omega^{(D=0)}_{\lambda}\equiv\omega_{\lambda}^s$ (where $s$ stands for the symmetric dimer), even if $\omega_{\lambda}\neq\omega^{s}_{\lambda}$. Therefore, it is not the energy $\omega_{\lambda}$ that matters, but the \textit{state} $\lambda$. We use the latter as a connector, namely we set: 
\begin{equation}
v_{{\rm SF}\,i}\bigl(\omega_{\lambda}\bigr)=v_{\rm SF}^{s}\bigl(\omega^{s}_{\lambda}\bigr),
\label{eq:conn_hub}
\end{equation}
We note that the right hand side does not depend on the site $i$, as the model system that we have chosen is homogeneous. Therefore, $v_{{\rm SF}}$ in the asymmetric dimer does not depend on the site either. 
We note also that the same argument, namely a continuous behaviour of the position of the poles of the auxiliary system as a function of $D$, pushes us to consider, also for $D\neq0$, $\omega_1$ and $\omega_4$ as bonding poles, i.e. zeros of the first of Eq. \eqref{eq:fg}, while $\omega_2$ and $\omega_3$ as antibonding poles, i.e. zeros of the second of Eq. \eqref{eq:fg}.

The first task hence becomes calculating the spectral potential exactly (and at different levels of approximation) for the symmetric dimer. 
This will be the subject of the next section.

\section{The spectral potential for the symmetric dimer}
\label{sec:sym}

The Hamiltonian for the symmetric Hubbard dimer\footnote{The Green's function and the GWA for the Hubbard dimer with one electron have already been discussed elsewhere \cite{Romaniello2009,Romaniello2012,marco_phd}. Here we gather the main results for consistency, noting that in the present case the on-site energy is $e_1=e_2=0$ and the energy is measured in units of $t$ (i.e. in practice we set $t=1$).} (with one spin-up electron) is obtained from Eq. \eqref{eq:weihojbw} by setting $D=0$:
\begin{equation}
\hat{H}_s= - \sum_{\sigma}\left(\hat{c}_{1\sigma}^{\dag}\hat{c}_{2\sigma}+\hat{c}_{2\sigma}^{\dag}\hat{c}_{1\sigma}\right) +
U\sum_{i}\hat{n}_{i\uparrow}\hat{n}_{i\downarrow}.
\label{eq:sym_dimer}
\end{equation}
The Hamiltonian has eigenvalues $\varepsilon^s_{\pm}= \pm 1$, corresponding to the bonding-antibonding eigenstates:
\begin{equation}
\ket{\pm,\sigma}=\frac{1}{\sqrt{2}}
\biggl[
\ket{1,\sigma}\mp \ket{2,\sigma}
\biggr].
\label{eq:hub_sym_vect}
\end{equation} 
The spin--up electron occupies the bonding state $\ket{\rm GS}\equiv\ket{-,\uparrow}$; therefore, the chemical potential is $\mu^s=\varepsilon^s_-=-1$ and the antibonding excited state is well separated with energy $\varepsilon^s_+= +1$. 

\subsection{The solution of the model}\label{sec:solmode}

\paragraph{The Green's function} We consider the bonding--antibonding basis, where 
the Green's function $G_{\alpha\beta}^s(t,t')=
\delta_{\alpha\beta}G^s_{\alpha}(t-t')$ is diagonal because the two sites $i=1$ and $i=2$ have the same on--site energy. 
Here, again, since the spin--up Green's function is always non-interacting, we are interested in the spin--down Green's function only:
\begin{equation}
\begin{split}
G^s_{-,\downarrow}(\omega)&=
\frac{\frac{1}{2}+\frac{2}{c}}{\omega-\omega_1^s+i\eta}
+
\frac{\frac{1}{2}-\frac{2}{c}}{\omega-\omega_4^s+i\eta} 
\\
G^s_{+,\downarrow}(\omega)&=
\frac{\frac{1}{2}}{\omega-\omega_2^s+i\eta}
+\frac{\frac{1}{2}}{\omega-\omega_3^s+i\eta}
\end{split}
\label{eq:Gdownba}
\end{equation}
with 
$c = \sqrt{16+U^2}$.
In the site basis the spin--down Green's function in the Lehmann representation:
\begin{equation}
G^s_{ij,\downarrow}(\omega)=\frac{(-1)^{i-j}}{2}\sum_{\lambda=1}^4\frac{f^s_{\lambda}}{\omega-\omega^s_{\lambda}+i\eta}
\end{equation}
has four poles $\omega_{\lambda}^s$, which are
summarized in table \ref{eq:poles} together with their amplitudes\footnote{In the notation of Eq. \eqref{eq:GFN1down}, $f_{ij}^{s\,\lambda}\equiv\frac{1}{2}(-1)^{i-j}f_{\lambda}^s$} $f_{\lambda}^s$.

\begin{table}[t]
	\begin{minipage}{\columnwidth}
		\centering
		\begin{tabular}{c c c}
			\toprule
			 & \textit{peak position} & \textit{peak amplitude}  \\
			\toprule
			1st pole & $\omega^s_1= 1+ \frac{U-c}{2}$ & $f^s_1=\frac{1}{2}+\frac{2}{c}$ \\
			\midrule
			2nd pole & $\omega^s_2=1 $ & $f^s_2=\frac{1}{2} $ \\
			\midrule
			3rd pole & $\omega^s_3=1+U$ & $f^s_3= \frac{1}{2} $ \\
			\midrule
			4th pole & $\omega^s_4=1+ \frac{U+c}{2}$ & $f^s_4=\frac{1}{2}-\frac{2}{c}$ \\
			\bottomrule
		\end{tabular}
		\caption{\emph{Peak positions $\omega_{\lambda}^s$ and amplitudes $f_{\lambda}^s$. Note that we have defined $c = \sqrt{16+U^2}$.}}
		\label{eq:poles}
	\end{minipage}
\end{table} 

The nature of the poles remains the same as in the asymmetric case, however their physical interpretation  (making reference 
to the eigenvalues $\varepsilon_{\lambda}^{s\,(N=2)}$ of the Hamiltonian with $N=2$ electrons -- see App. \ref{sec:gasym}) is more intuitive than in the asymmetric case:
\begin{itemize}
	\item The pole $\omega^s_1$ represents the addition of a spin--down electron to the already--occupied bonding orbital; 
	since they are on the same orbital, electrons interact with an effective interaction\footnote{$U$ is the interaction term for two electrons on the same site, $\tilde{U}$ is the interaction for two electrons in the same orbital: sites and orbitals are just two different \textit{basis}, and if the electrons are not interacting in a basis, they are not in the other basis either: for this reason, $\tilde{U}$ always goes to zero in the limit of zero bare interaction $U$.} 
$\tilde{U}_{--}=2 +\frac{1}{2}\bigl(U-c\bigr)$. Hence $\varepsilon^{s\,(N=2)}_{\lambda=1}=-2+\tilde{U}_{--}=\frac{1}{2}\bigl(U-c\bigr)$ and $\omega^s_1=1+\frac{1}{2}\bigl(U-c\bigr)$.
	
	\item The pole $\omega^s_2$ describes two electrons sitting on two different sites and occupying two different orbitals (therefore, no interaction), namely a bonding 
	and an antibonding 
	orbital, giving a total energy $\varepsilon^{s\,(N=2)}_{\lambda=2}=-1+1=0$, and $\omega^s_2=+1$.
	
	\item The pole $\omega^s_3$ is associated with two electrons occupying the same site 
	but in two different orbitals; the total energy is therefore $\varepsilon^{s\,(N=2)}_{\lambda=3}=-1+1+U=U$, and the position of the pole is $\omega_3^s=1+U$.
	
	\item Finally, the pole $\omega^s_4$ represents two electrons occupying the same antibonding orbital 
	with an effective interaction $\tilde{U}_{++}=-2+\frac{1}{2}\bigl(U+c\bigr)$: a spin--down electron enters the system in the bonding orbital, where a spin--up electron was already sitting; the former excites 
	the latter, and both end up in an excited state, the antibonding state, where they interact via $\tilde{U}_{++}$. This process results in a total energy of the two--electron state equals to $\varepsilon^{s\,(N=2)}_{\lambda=4}=2(-1)+4+\tilde{U}_{++}=\frac{1}{2}\bigl(U+c\bigr)$; the pole is $\omega^s_4=1+\frac{1}{2}\bigl(U+c\bigr)$.
\end{itemize}

The effective interaction that we have just introduced is hence
$\tilde{U}_{\alpha\beta}=0$  if electrons occupy different sites with
$\left<\hat{n}_{i,\uparrow}\hat{n}_{i,\downarrow}\right>=0$, whereas
if the electrons have a non--zero probability to be on the same site
$\left<\hat{n}_{i,\uparrow}\hat{n}_{i,\downarrow}\right>\neq0$, the effective interaction reads:
\begin{equation}
\tilde{U}_{\alpha\beta}=
\left(\begin{matrix}
\tilde{U}_{--} & \tilde{U}_{-+}
\\
\tilde{U}_{+-} & \tilde{U}_{++}
\end{matrix}\right)=
\left(\begin{matrix}
2+\frac{U-c}{2} & U
\\
U & -2+\frac{U+c}{2}
\end{matrix}\right).
\label{eq:U_eff}
\end{equation}

\paragraph{The spectral function}
Since both sites are equal, the Green's function is symmetric under exchange of the site indices.  
The diagonal elements of the spectral function are the same and equal to:
\begin{equation}
\begin{aligned}
A^s_{ii,\downarrow}(\omega)=\sum_{\lambda}\frac{f^s_{\lambda}}{2}\delta\bigl(\omega-\omega_{\lambda}^s\bigr).
\label{eq:SFdiagdimer}
\end{aligned}
\end{equation}

\paragraph{The self energy}
As in the asymmetric case, the spin--up self-energy is zero as an additional spin--up electron cannot interact, while
the spin--down self energy can be obtained more easily in the bonding-antibonding basis where the Green's functions are diagonal.
From the inverted Dyson equation $\Sigma^s_{\alpha}(\omega)={G^s_0}^{-1}_{\alpha}(\omega)-{G^s}^{-1}_{\alpha}(\omega)$, with ${G^s_0}^{-1}_{\alpha}(\omega)=\omega-\varepsilon^s_{\alpha}$,
one has:
\begin{equation}
\begin{gathered}
\Sigma^s_{-,\downarrow}(\omega)=\frac{U}{2}+\frac{\frac{U^2}{4}}{\omega-\bigl(3+\tfrac{U}{2}\bigr)+i\eta}
\\
\Sigma^s_{+,\downarrow}(\omega)=\frac{U}{2}+\frac{\frac{U^2}{4}}{\omega-\bigl(1+\tfrac{U}{2}\bigr)+i\eta}
\end{gathered}.
\label{eq:sigmabadown}
\end{equation}
From the relation $\omega-\varepsilon^s_{\alpha}-\operatorname{Re}\Sigma^s_{\alpha}(\omega)=0$ one obtains $\omega^s_1$ and $\omega^s_4$ when considering the bonding state $\alpha=-$, and $\omega^s_2$ and $\omega^s_3$ when considering the antibonding $\alpha=+$. The weights $f_{\lambda}^s$ of the Green's function in eq. \eqref{eq:Gdownba} are nothing but the renormalization factors $Z_{\lambda}^{s}:=\left(1-\partial\operatorname{Re}\Sigma^s_{\alpha}(\omega)/\partial\omega\right)^{-1}_{\omega=\omega^s_{\lambda}}$. They are:
\begin{equation}
Z^s_1=\frac{1}{2}+\frac{2}{c} \qquad
Z^s_2=Z^s_3=\frac{1}{2} \qquad
Z^s_4=\frac{1}{2}-\frac{2}{c}.
\label{eq:Zfactors}
\end{equation}Moving to the site basis, the self energy reads:
\begin{equation}
\begin{aligned}
\Sigma^s_{ij,\downarrow}(\omega)=& \frac{U}{2}\delta_{ij}+\frac{U^2}{8}\left[\frac{(-1)^{i-j}}{\omega-\bigl(1+\frac{U}{2}\bigr)+i\eta} \right. \\ 
 & + \left. \frac{1}{\omega-\bigl(3+\frac{U}{2}\bigr)+i\eta}\right].
\label{eq:selfdimer}
\end{aligned}
\end{equation}
As expected, $\Sigma^s_{ij,\downarrow}$ goes to zero in the limit of $U\to0$ 
except for $\Sigma^s_{-,\downarrow}(\omega_4)\stackrel{U\to0}{\longrightarrow}4$, which is the energy needed to excite the system to the pole $\omega_4$, a process which is suppressed for $U=0$ but is nonetheless present for small interaction $U$.
Note that this is a truly \textit{non--local} self energy in the site basis, with a non-zero imaginary part.

In the next section, we will exactly get the diagonal of the spectral function $A^s_{ii,\downarrow}(\omega)$ \eqref{eq:SFdiagdimer} by replacing the non--local and complex--valued self energy \eqref{eq:selfdimer} with a \textit{real} and \textit{local} (in the site basis) potential.

\subsection{The exact spectral potential \label{sec:symmvSF}}

The auxiliary system is requested to provide the same local spectral function  as Eq. \eqref{eq:SFdiagdimer}. 
Since the spin--symmetry is broken by the choice of a spin--up ground state, we will furthermore consider a spin--dependent spectral potential; for reproducing the spin--up spectral function, a zero spectral potential will trivially do the job, as $\Sigma^s_{ij,\uparrow}(\omega)=0$.
We will henceforth focus on the spin--down sector, dropping the $\downarrow$ notation. The auxiliary system is defined by the following inverted Dyson equation:
\begin{equation}
	G_{{\rm SF}\; ij}^{s\,-1}(\omega)={G_{{\rm 0}\;ij}^{s\,-1}}(\omega)-v_{{\rm SF}\;i}^s(\omega)\delta_{ij},
	\label{eq:aux_GF_dimer}
\end{equation}
where we have introduced the \textit{local} spectral potential $v^s_{{\rm SF}\,i}(\omega)$; since the two sites are equivalent, the potential takes the same value $v^s_{\rm SF}(\omega)$ on both sites, and the equation can be written as $G_{{\rm SF}\; ij}^{s\,-1}(\omega)={G_{0\;ij}^{s\,-1}}\bigl(\omega-v^s_{\rm SF}(\omega)\bigr)$: 
\begin{equation}
\begin{aligned}
G^s_{{\rm SF}\;ij}(\omega)=&\frac{\frac{1}{2}}{\omega-\left(-1+v^s_{\rm SF}(\omega)\right)+i\eta}+\\ 
+ &\frac{(-1)^{(i-j)}\frac{1}{2}}{\omega-\left(1+v^s_{\rm SF}(\omega)\right)+i\eta}.
\end{aligned}
\label{eq:Gaux}
\end{equation}

Instead of working in the site basis, we can move to the bonding--antibonding basis $\ket{\pm}$ where, by virtue of the symmetry of the problem, everything is diagonal. Moreover, in the bonding--antibonding basis the value of the potential is the same\footnote{Indeed, considering a local potential $v_i$, we have:
\begin{equation*}
\begin{aligned}
\left(\begin{matrix}
v_{--} & v_{-+}
\\
v_{+-} & v_{++}
\end{matrix}\right)=& 
\frac{1}{2}
\left(\begin{matrix}
1 & 1
\\
1 & -1
\end{matrix}\right)
\left(\begin{matrix}
v_1 & 0
\\
0 & v_2
\end{matrix}\right)
\left(\begin{matrix}
1 & 1
\\
1 & -1
\end{matrix}\right) \\
= & 
\frac{1}{2}
\left(\begin{matrix}
v_1+v_2 & v_1-v_2
\\
v_1-v_2 & v_1+v_2
\end{matrix}\right)=
\left(\begin{matrix}
v & 0
\\
0 & v
\end{matrix}\right)
\end{aligned}
\end{equation*}
where in the last equality we implemented the site--symmetry property $v_1=v_2:=v$; therefore, the mixed terms are zero and both the bonding $v_{--}$ and antibonding $v_{++}$ potentials are equal to $v$, too.}, as $v^s_{\rm SF}(\omega)$ can be considered as a frequency--dependent energy shift, no matter the basis. Therefore, the bonding--antibonding character is settled by the non--interacting Green's function\linebreak $G^{s\,-1}_{0\;\pm}(\omega)=\omega-\varepsilon^s_{\pm}$ only, and the inverted Green's function in the bonding--antibonding basis simply reads:
\begin{equation}
G_{{\rm SF}\; \pm}^{s\,-1}(\omega)=\omega-\varepsilon^s_{\pm}-v_{\rm SF}^s(\omega).
\label{eq:defGauxba}
\end{equation}
By definition the Green's function in the site basis, Eq. \eqref{eq:Gaux}, must have the same local spectral function as the one defined in terms of the full Green's function \eqref{eq:SFdiagdimer}:
\begin{equation}
-\frac{1}{\pi}{\rm sign}(\omega-\mu)\operatorname{Im}G^s_{{\rm SF}\,ii}(\omega)\equiv A^s_{{\rm SF}\,ii}(\omega)\stackrel{!}{=}A^s_{ii}(\omega).
\label{eq:GSSE_simpl_dimer}
\end{equation}
Since we are in a discrete system, this equation means that both the positions and the amplitudes of the peaks must be reproduced by the auxiliary system.

\begin{figure}[t!]
	\centering
	\begin{minipage}{0.99\columnwidth}
		\centering
		\begin{subfigure}[t]{0.66\textwidth}
			\centering
			\includegraphics[width=0.99\textwidth]{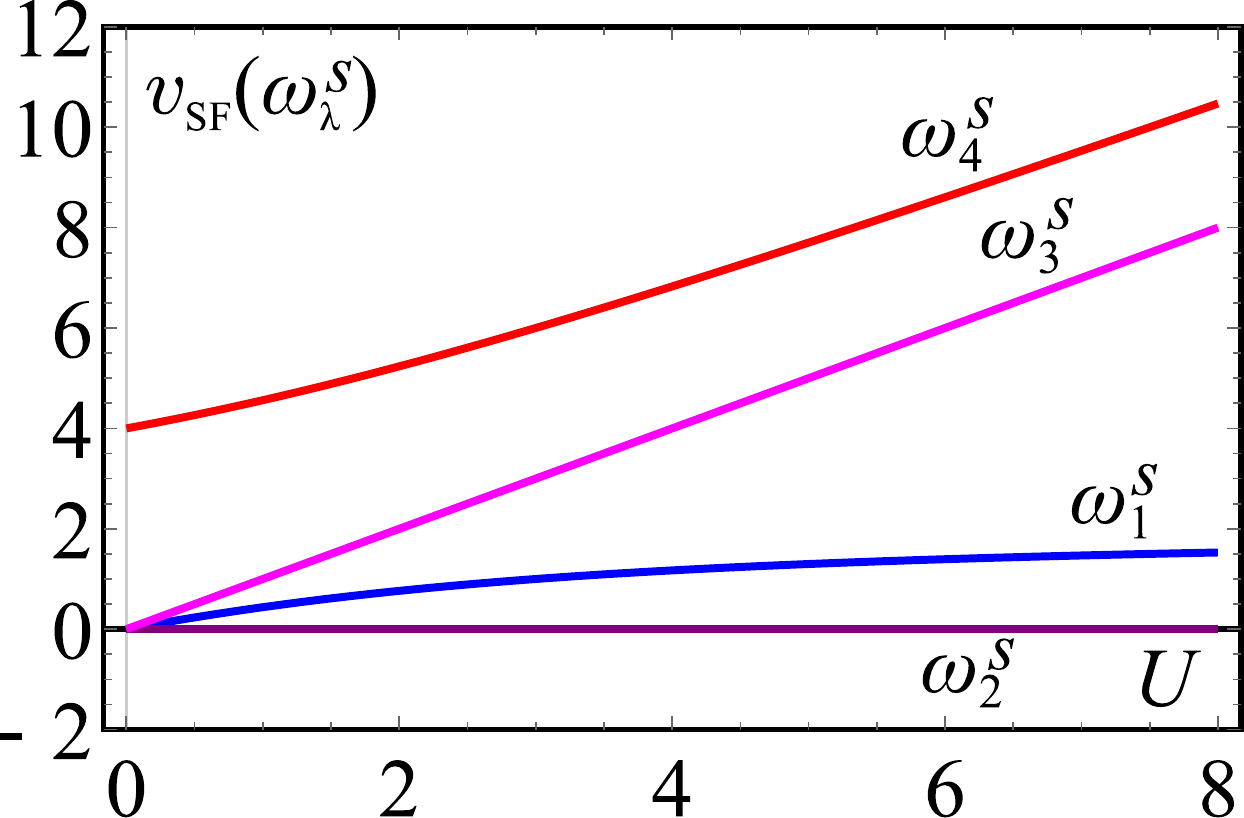}
			\caption{\textit{Spectral potential at the poles}}
			\label{fig:vpoles_a}
		\end{subfigure}
		~\qquad
		\begin{subfigure}[t]{0.66\textwidth}
			\centering
			\includegraphics[width=0.99\textwidth]{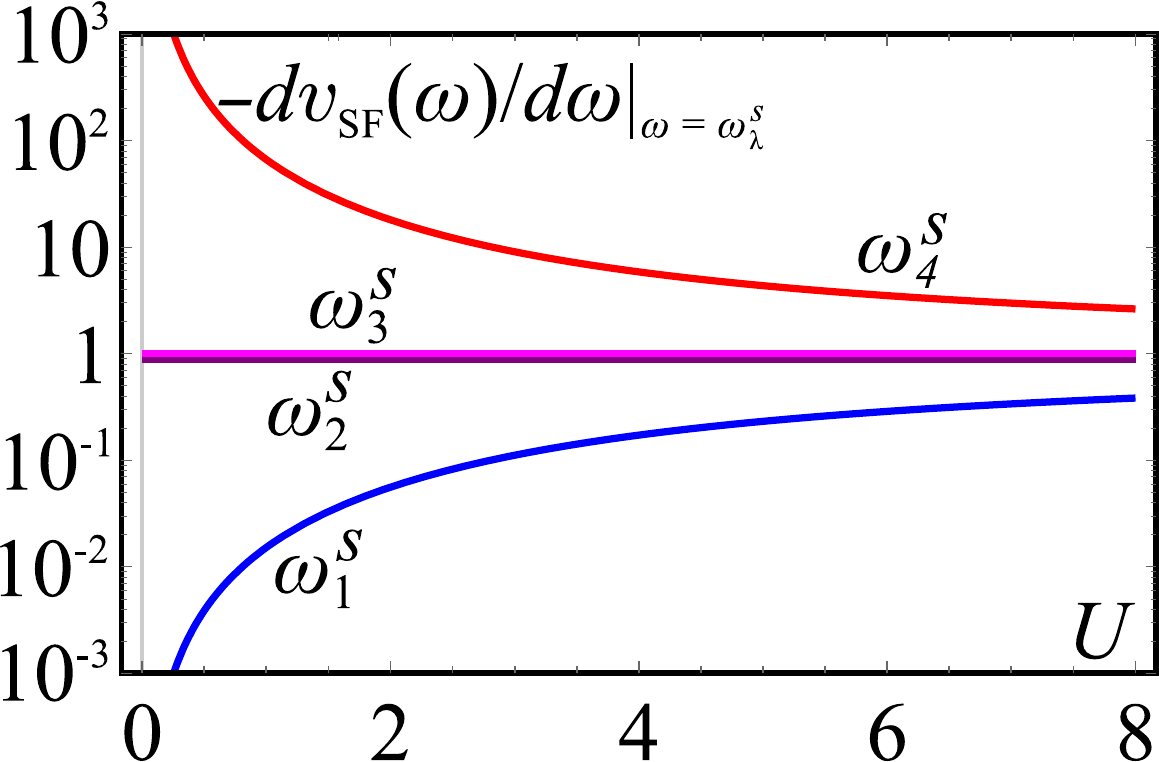}
			\caption{\textit{(minus) Its derivative at the poles}}
			\label{fig:vpoles_b}
		\end{subfigure}
		\caption{\emph{Spectral potential $v_{\rm SF}(\omega)$ and (minus) its first derivative (in logarithmic scale) $-dv_{\rm SF}(\omega)/d\omega$ at the poles $\omega_{\lambda}^s$.}}
		\label{fig:vpoles}
	\end{minipage}
\end{figure}

\paragraph{Position of the peaks}
Since their position does not depend on the basis, and we are in a discrete system, the poles of $G^s_{\rm SF\;\pm}(\omega)$ and $G^s_{\pm}(\omega)$ must be the same, namely:
\begin{equation}
\left.\omega-\varepsilon^s_{\pm}-v^s_{\rm SF}(\omega)\right|_{\omega=\omega_{\lambda}^s}=0
\label{eq:QPdimer1}
\end{equation}
with $\omega_{\lambda}^s$ the four poles of table \ref{eq:poles}.  
For small interaction $U \ll1$, the effect of the potential will be to \textit{slightly} move the poles from their $U=0$ position;  
we assume that, since its effects are small, the spectral potential be small too in this regime.  
It is therefore natural to assume that the \textit{nature} of the poles be unchanged, namely that (anti)bonding poles of the real system be reproduced by (anti)bonding poles of the auxiliary system. 
Therefore, the previous relation could be split into the following two:
\begin{equation}
\begin{aligned}
\left.\omega-\varepsilon_-^s-v^s_{\rm SF}(\omega)\right|_{\omega=\omega^s_1,\omega^s_4}&=0
\\
\left.\omega-\varepsilon_+^s-v^s_{\rm SF}(\omega)\right|_{\omega=\omega^s_2,\omega^s_3}&=0
\label{eq:QPdimer}
\end{aligned}
\end{equation}
(with $\varepsilon_\pm^s=\pm1$) from which the value of $v^s_{\rm SF}(\omega)$ at the poles is:
\begin{equation}
\begin{split}
&v^s_{\rm SF}(\omega^s_1)=2+\tfrac{U-c}{2} 
\\
&v^s_{\rm SF}(\omega^s_2)=0
\\
&v^s_{\rm SF}(\omega^s_3)=U
\\
&v^s_{\rm SF}(\omega^s_4)=2+\tfrac{U+c}{2} 
\end{split}
\label{eq:pot_poles}
\end{equation}
which are shown in Fig. \ref{fig:vpoles_a}. Note that two equations analogous to Eq. \eqref{eq:QPdimer} hold with $\Sigma^s_{\pm}(\omega)$ in place of $v^s_{\rm SF}(\omega)$:
\begin{equation}
\begin{aligned}
\left.\omega-\varepsilon_-^s-\Sigma^s_-(\omega)\right|_{\omega=\omega^s_1,\omega^s_4}&=0
\\
\left.\omega-\varepsilon_+^s-\Sigma^s_+(\omega)\right|_{\omega=\omega^s_2,\omega^s_3}&=0
\label{eq:QPdimer_real}
\end{aligned}\quad.
\end{equation}
Indeed, for a discrete system (not in the thermodynamic limit), the self energy is real at the poles \cite{Farid1999}, 
and in particular the spectral potential is nothing but the self energy at the poles:
\begin{equation}
v^s_{\rm SF}(\omega_{\lambda}^s)=
\begin{cases}
\Sigma^s_-(\omega_{\lambda}^s) &\text{if }\omega_{\lambda}^s=\omega^s_1,\omega^s_4
\\
\Sigma^s_+(\omega_{\lambda}^s) &\text{if }\omega_{\lambda}^s=\omega^s_2,\omega^s_3
\end{cases}
\label{eq:v_sFpot}
\end{equation}
On the contrary, $v^s_{\rm SF}(\omega)\neq{\rm Re}\Sigma^s_{ii}(\omega)$, as one could have naively guessed, because in the site basis the self-energy is non-local. 

From these relations or directly from Eq. \eqref{eq:pot_poles}, the spectral potential can be interpreted as the additional energy which the auxiliary system needs to mimic the behaviour of the full solution. In particular, $v^s_{\rm SF}(\omega_{\lambda}^s)$ is related to the effective interactions that we introduced in Eq. \eqref{eq:U_eff}: 
\begin{equation}
\begin{aligned}
	v^s_{\rm SF}(\omega^s_1)&=\tilde{U}_{--}^{\left<\hat{n}_{i,\uparrow}\hat{n}_{i,\downarrow}\right>\neq0}
	\\
	v^s_{\rm SF}(\omega^s_2)&=\tilde{U}_{\alpha\beta}^{\left<\hat{n}_{i,\uparrow}\hat{n}_{i,\downarrow}\right>=0}
	\\
	v^s_{\rm SF}(\omega^s_3)&=\tilde{U}_{+-}^{\left<\hat{n}_{i,\uparrow}\hat{n}_{i,\downarrow}\right>\neq0}
	\\
	v^s_{\rm SF}(\omega^s_4)&=4+\tilde{U}_{++}^{\left<\hat{n}_{i,\uparrow}\hat{n}_{i,\downarrow}\right>\neq0}.
\end{aligned}
\end{equation}
Only $v^s_{\rm SF}(\omega^s_4)$ differs from the corresponding effective interaction $\tilde U_{++}$ by $4$ (in units of $t$): indeed, $4t$ is the energy that must be provided to the two electrons to go from the bonding to the antibonding state, where they are then free to interact with an energy $\tilde U_{++}$; the spectral potential, like the self-energy, provides the system with both the activation energy $4t$ and the interaction $\tilde U_{++}$, so that (in units of $t$): $v^s_{\rm SF}(\omega^s_4)=\Sigma^s_-(\omega^s_4)=4 +\tilde U_{++}$.
\footnote{Taking the limit is a continuous operation from positive values of $U$ to $U=0$; since the process described by the pole $\omega^s_4$ is actually suppressed for $U=0$, one could decide to redefine ‘‘by hand'' $\Sigma^s_{-,\downarrow}(\omega^s_4)|_{U=0}:=0$ and nothing would change. As a result, also $v^s_{\rm SF}(\omega^s_4)$ would be redefined at $U=0$ as $v^s_{\rm SF}(\omega^s_4)|_{U=0}:=0$.}

\begin{figure*}[t]
	\centering
	\begin{minipage}{0.90\textwidth}
		\centering
		\includegraphics[width=0.79\textwidth]{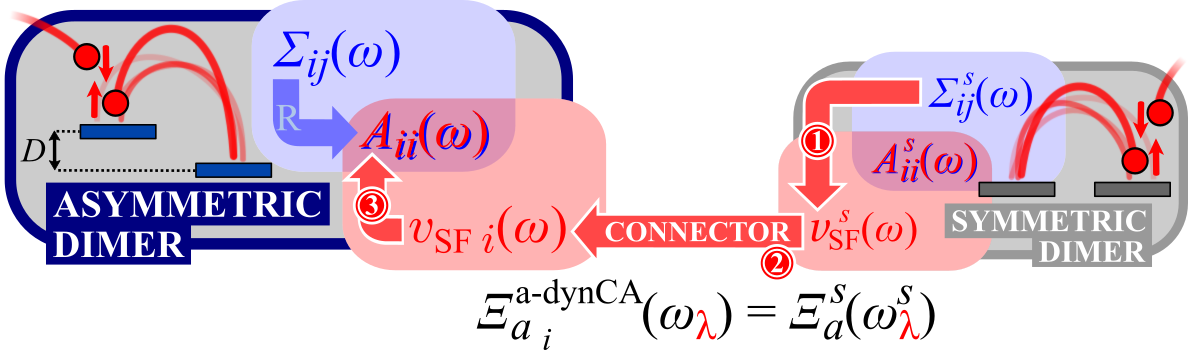}
		\caption{\emph{A schematic picture of the general strategy. }}
		\label{fig:strategy}
	\end{minipage}
\end{figure*}

\paragraph{Amplitude of the peaks}

The spectral potential $v^s_{\rm SF}(\omega)$, besides their positions, 
has to reproduce \textit{also} the \textit{amplitudes} of the peaks of the diagonal spectral function $A^s_{{\rm SF\,}ii}(\omega)$.
For the symmetric dimer, the absolute value of the Lehmann weights $f_{ij}^{s\,\lambda}\equiv\frac{1}{2}(-1)^{(i-j)}f_{\lambda}^s$ is independent of $i$ and $j$, 
and also, modulus $1/2$, independent even of the particular basis.  
Therefore, we can simply match, in the bonding--antibonding basis, the positive weights $f_{\lambda}^s=Z_{\lambda}^s$ with the corresponding ones of the auxiliary system $Z^{s}_{{\rm SF}\,\lambda}$. 
Through the identification $\left(Z^{s}_{\lambda}\right)^{-1}\equiv\left(Z^{s}_{{\rm SF}\,\lambda}\right)^{-1}= \left(1-\frac{d v^s_{\rm SF}(\omega)}{d\omega}\right)_{\omega=\omega_{\lambda}^s}$  
we obtain the values of the derivatives of $v^s_{\rm SF}(\omega)$ evaluated at $\omega_{\lambda}^s$: 
\begin{equation}
\left.\frac{dv^s_{\rm SF}(\omega)}{d\omega}\right|_{\omega=\omega_{\lambda}^s}=
\begin{cases}
\frac{\frac{2}{c}-\frac{1}{2}}{\frac{2}{c}+\frac{1}{2}} & \text{if }\omega_{\lambda}^s=\omega^s_1
\\
-1 & \text{if }\omega_{\lambda}^s=\omega^s_2
\\
-1 & \text{if }\omega_{\lambda}^s=\omega^s_3
\\
\frac{\frac{2}{c}+\frac{1}{2}}{\frac{2}{c}-\frac{1}{2}} & \text{if }\omega_{\lambda}^s=\omega^s_4
\end{cases}.
\label{eq:der_pot_poles}
\end{equation}
Their behaviour as a function of $U$ is shown in Fig. \ref{fig:vpoles_b}.

We note that the requirement that Eq. \eqref{eq:QPdimer1} do not have any other solutions than $\omega=\omega_{\lambda}^s$ can be actually relaxed: indeed, other poles $\omega_{\tilde\lambda}^s$ can show up as additional crossings of the two lines $\omega-\varepsilon^s_{\pm}$ with the function $v^s_{\rm SF}(\omega)$, provided that their weight $Z_{{\rm SF}\,\tilde\lambda}^s$ 
be zero, namely that $\left.\frac{dv^s_{\rm SF}(\omega)}{d\omega}\right|_{\omega=\omega_{\tilde\lambda}}$ 
diverge. Therefore, the potential -- univocally fixed with its derivative by Eq. \eqref{eq:pot_poles} and \eqref{eq:der_pot_poles} wherever the spectral function is non--zero -- can be arbitrarily defined also where $A^s_{ii}(\omega)=0$ provided that, if it crosses the lines $\omega-\varepsilon^s_{\pm}$, its tangent be vertical.

With Eq. \eqref{eq:pot_poles} and Eq. \eqref{eq:der_pot_poles} the problem is solved.

We note that, in particular, the spectral function is reproduced in the \textit{non--interacting limit} $U=0$ (trivial), and also in the \textit{atomic limit} $U \to \infty$. In the latter case, the potential assumes the values $v^s_{\rm SF}(\omega)=0$ in $\omega^s_1$ and $\omega^s_2$, and $v^s_{\rm SF}(\omega)=U$ in $\omega^s_3$ and $\omega^s_4$, yielding the two separated Hubbard bands exactly.
On the contrary, the Kohn-Sham eigenvalues of DFT by definition cannot be interpreted as excitations energies. 
Even within MBPT the GW approximation  fails qualitatively to describe the atomic limit \cite{Romaniello2009}.

\paragraph{More than expected?}
With the choice of Eq. \eqref{eq:pot_poles} and Eq. \eqref{eq:der_pot_poles}, we have obtained the following  results:
\begin{equation}
	\begin{aligned}
		A^s_{\rm SF\,\pm}(\omega)=A^s_{\pm}(\omega)
		\qquad
		A^s_{{\rm SF}\,ij}(\omega)=A^s_{ij}(\omega).
		\label{eq:more}
	\end{aligned}
\end{equation}
Actually, Eq. \eqref{eq:more} contains far more than what we expected: the spectral potential $v^s_{\rm SF}(\omega)$, indeed, has the duty of reproducing only the diagonal of the spectral function in the site basis, namely $A^s_{{\rm SF}\,ii}(\omega)=A^s_{ii}(\omega)$. 
The reason why in this way we actually get also the off--diagonal elements is due to the fact that the matrix of change of basis is fixed; therefore, since the spectral function is reproduced in a basis (i.e. the bonding-antibonding one), it will be fully reproduced also in the other, both for diagonal and off--diagonal elements.

One question remains: why is the bonding-antibonding spectral function fully reproduced? The reasons are three: 1) for the symmetry of the problem (i.e. the two sites are equivalent), $v^s_{\rm SF}(\omega)$ is just a frequency--dependent number in any basis, and in particular in the bonding-antibonding basis; 2) in a discrete system the number of poles is finite and their position is independent of the basis; their assignment to the correct bonding or antibonding character is done on the basis of continuity with the $U=0$ case; 3) in the bonding-antibonding basis the spectral function is diagonal and, in particular, positive, as the target  $A^s_{ii}(\omega)$ is diagonal and positive too; the difference in their weights (the $1/2$ factor) is completely accounted for by the change of basis matrix.

As a consequence, in a discrete translationally invariant system in which $v^s_{\rm SF}(\omega)$ is not site-dependent, the whole spectral function in any diagonal basis (if any) is fully reproduced; therefore, not only the diagonal but even the off-diagonal elements of the site-basis spectral function are exactly reproduced. So is the density matrix, too.

\section{Approximations in practice}
\label{sec:approx}

Now that the spectral potential is available in the symmetric dimer, the question is how to use it in practice in the asymmetric dimer for all possible $D$ values.
In the following we will discuss our strategy, which is schematically represented in Fig. \ref{fig:strategy}.

\paragraph{Corrections to import}

\begin{figure*}[t!]
	\centering
	\begin{minipage}{0.99\textwidth}
		\centering
		\begin{subfigure}[t]{0.32\textwidth}
			\centering
			\includegraphics[width=0.99\textwidth]{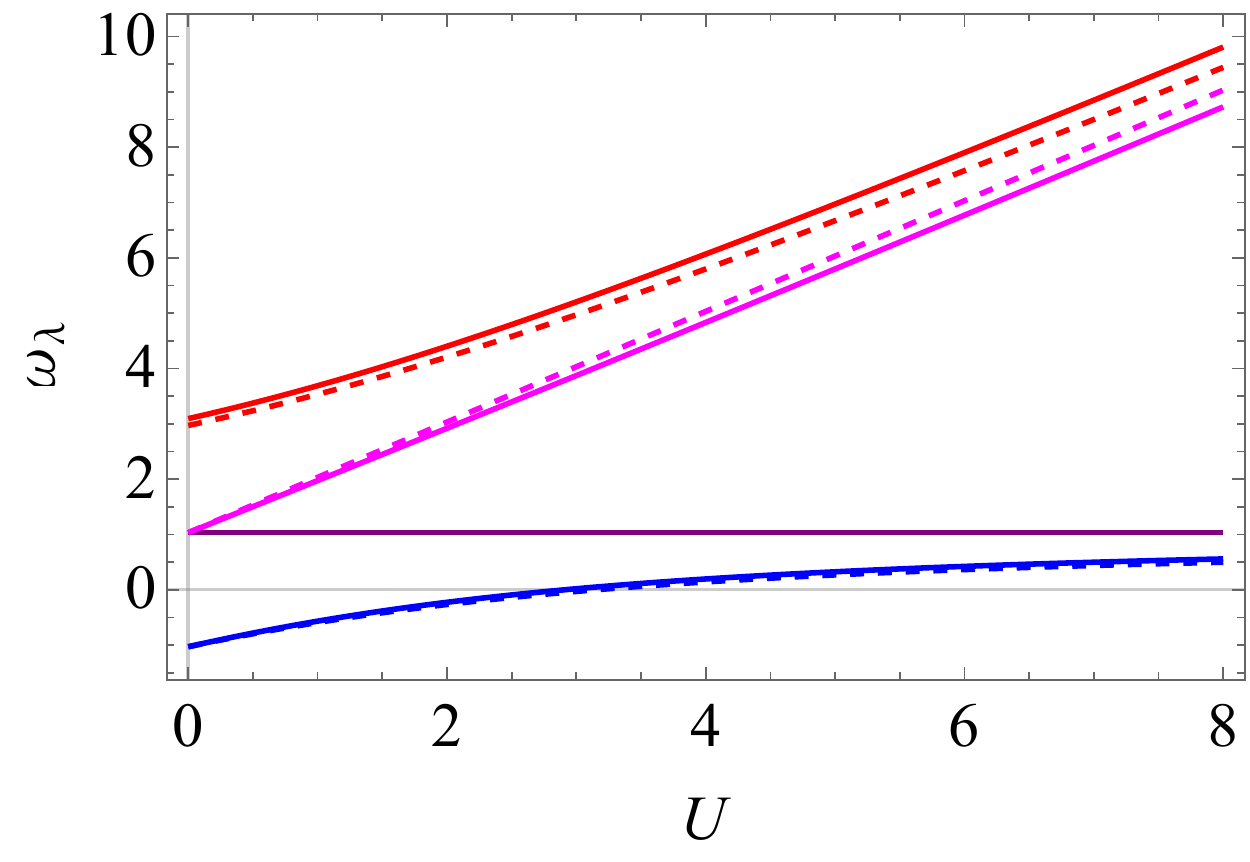}
			\caption{$D=0.5$}
		\end{subfigure}
		~
		\begin{subfigure}[t]{0.32\textwidth}
			\centering
			\includegraphics[width=0.99\textwidth]{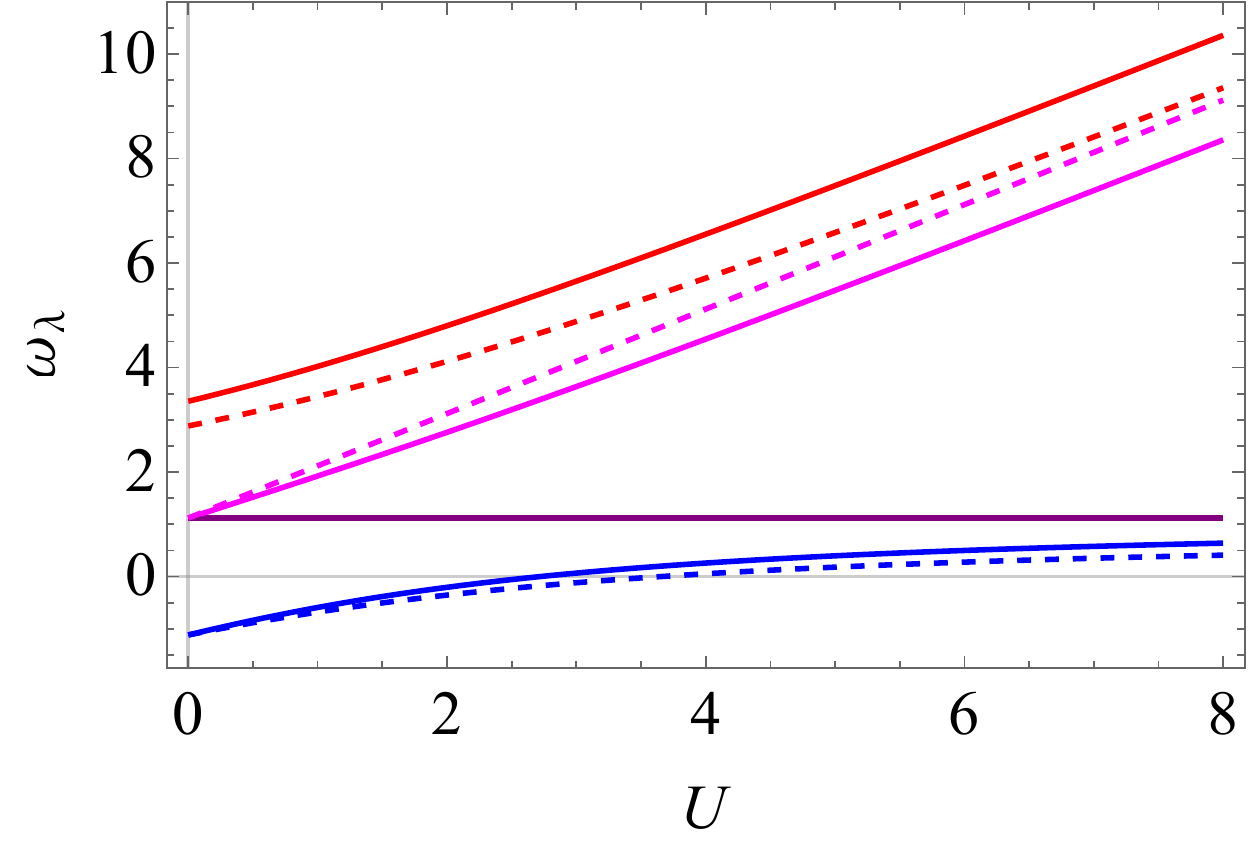}
			\caption{$D=1$}
		\end{subfigure}
		~
		\begin{subfigure}[t]{0.32\textwidth}
			\centering
			\includegraphics[width=0.99\textwidth]{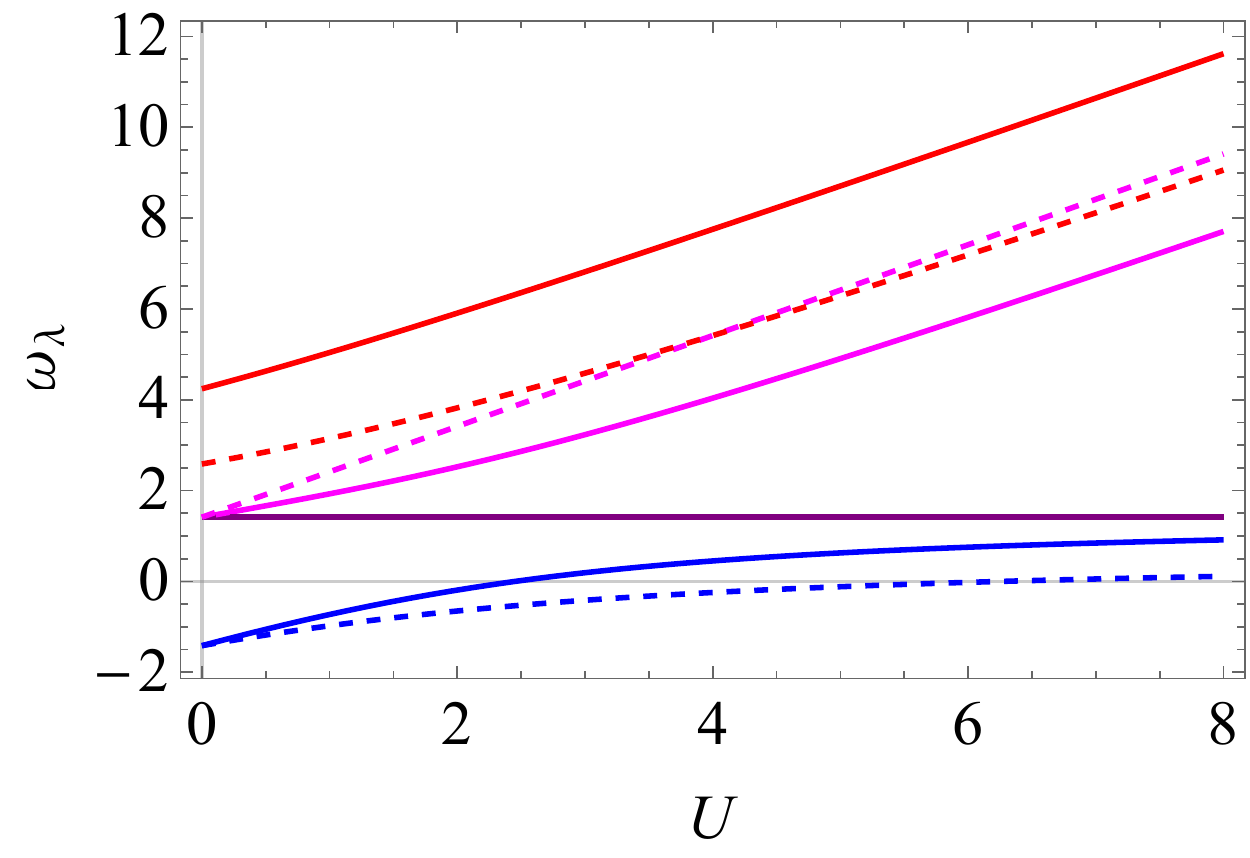}
			\caption{$D=2.0$}
		\end{subfigure}
		\caption{\emph{Position of the poles of the spin--down Green's function as a function of $U$ starting from the free-particle approximation. Solid line, exact results; dashed lines, the poles in Eq. \eqref{eq:dynCA_0_poles}.}}
		\label{fig:dynLDA_0}
	\end{minipage}
\end{figure*}

We will adopt an approximation $a$
yielding the  poles $\omega_{\lambda}^a$ in the asymmetric dimer. These poles can be obtained either by using the approximate self-energy $\Sigma_{a}$ or the corresponding spectral potential $v_{{\rm SF}\,a}$. 
We will then import some corrections to the spectral potential from the model symmetric dimer. 
Finally, with this corrected spectral potential we will calculate the new poles of the asymmetric dimer solving Eq. \eqref{eq:fg}.

Concretely, for each pole $\omega_{\lambda}$ the exact spectral potential $v_{{\rm SF}\,i}$  in the asymmetric dimer can be split into an approximated part $v_{{\rm SF}\,a\,i} $ plus a correction $\Xi_{a\,i}$:
\begin{equation}
v_{{\rm SF}\,i}(\omega_{\lambda})=v_{{\rm SF}\,a\,i}(\omega^{a}_{\lambda})+\Xi_{a\,i}(\omega_{\lambda}).
\label{eq:wrqweribf}
\end{equation}
We note that the correspondence is set by the state $\lambda$ and not by its energy $\omega_{\lambda}$.  Indeed, in general the position of the poles is not the same: $\omega_{\lambda}\neq\omega^{a}_{\lambda}$. 

Analogously, in the symmetric dimer we can  define:
\begin{equation}
v^s_{\rm SF}(\omega_{\lambda}^s)=v^s_{{\rm SF}\,a}(\omega^{a\,s}_{\lambda})+\Xi^{s}_{a}(\omega^s_{\lambda}),
\label{eq:wrqibf}
\end{equation}
where the approximation $a$ is the same as in the asymmetric dimer.
The correction $\Xi_{a\,i}$ in the asymmetric dimer, i.e. in Eq.  \eqref{eq:wrqweribf}, is then imported from  the model symmetric dimer (see Eq. \eqref{eq:conn_hub})
\begin{equation}
\boxed{
\Xi_{a\,i}^{ a-{\rm dynCA}}(\omega_{\lambda})=\Xi^{s}_{a}(\omega^s_{\lambda})
}
\label{eq:conn_states}
\end{equation}
where, as already said, the connection is made through the state label $\lambda$, which is indeed the only quantity which is shared by both sides. 
Moreover, we note that the imported correction  is the same for both sites $i$ as the model is symmetric.
We call this the dynamical connector approximation (dynCA).

We will benchmark the results  obtained in this way with the exact results for the asymmetric dimer from Sec. \ref{sec:asym}.
The performance will of course depend  on the starting approximation chosen in the asymmetric dimer. 
We expect the connector approach to work better in situations in which most of the inhomogeneity is treated exactly within the asymmetric dimer, and all higher orders interaction corrections are provided by the model system.

\begin{figure*}[t!]
	\centering
	\begin{minipage}{0.99\textwidth}
		\centering
		\begin{subfigure}[t]{0.32\textwidth}
			\centering
			\includegraphics[width=0.99\textwidth]{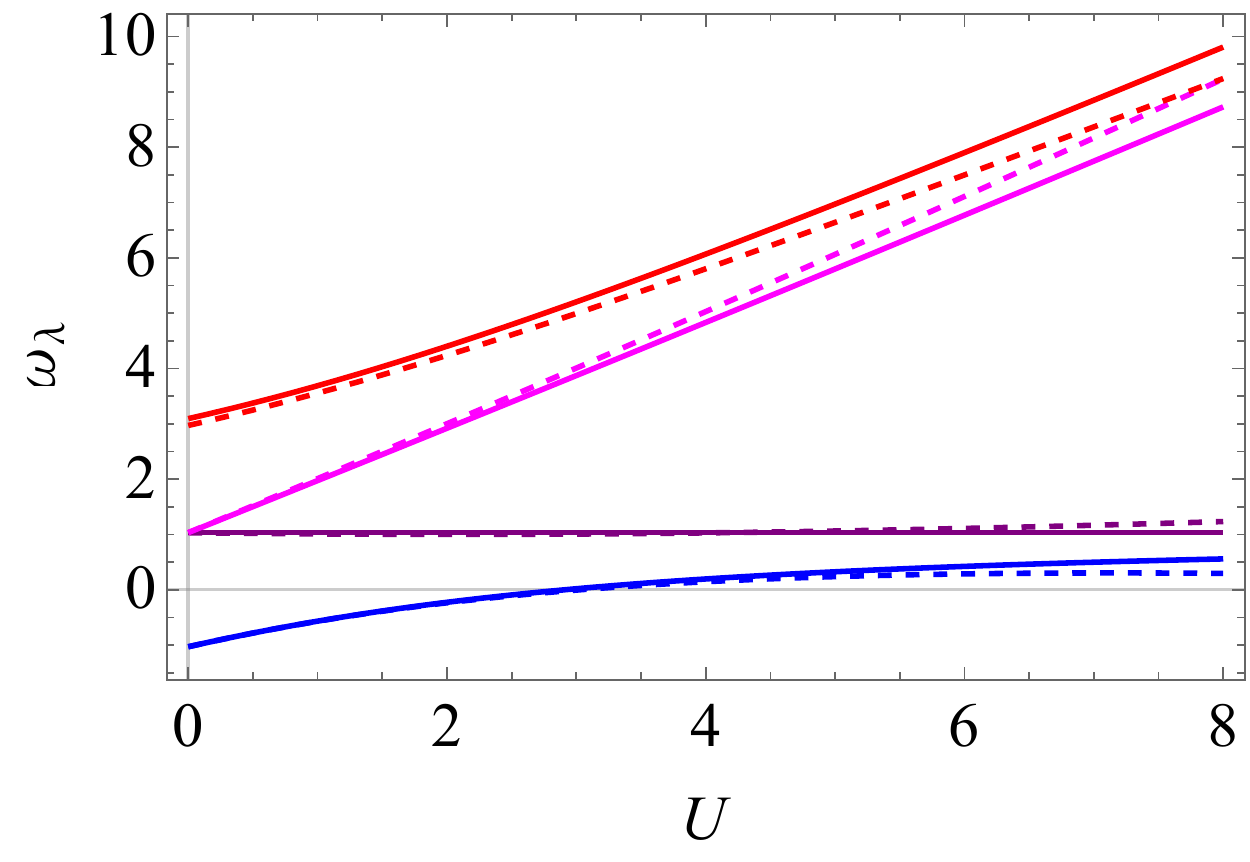}
			\caption{$D=0.5$}
		\end{subfigure}
		~
		\begin{subfigure}[t]{0.32\textwidth}
			\centering
			\includegraphics[width=0.99\textwidth]{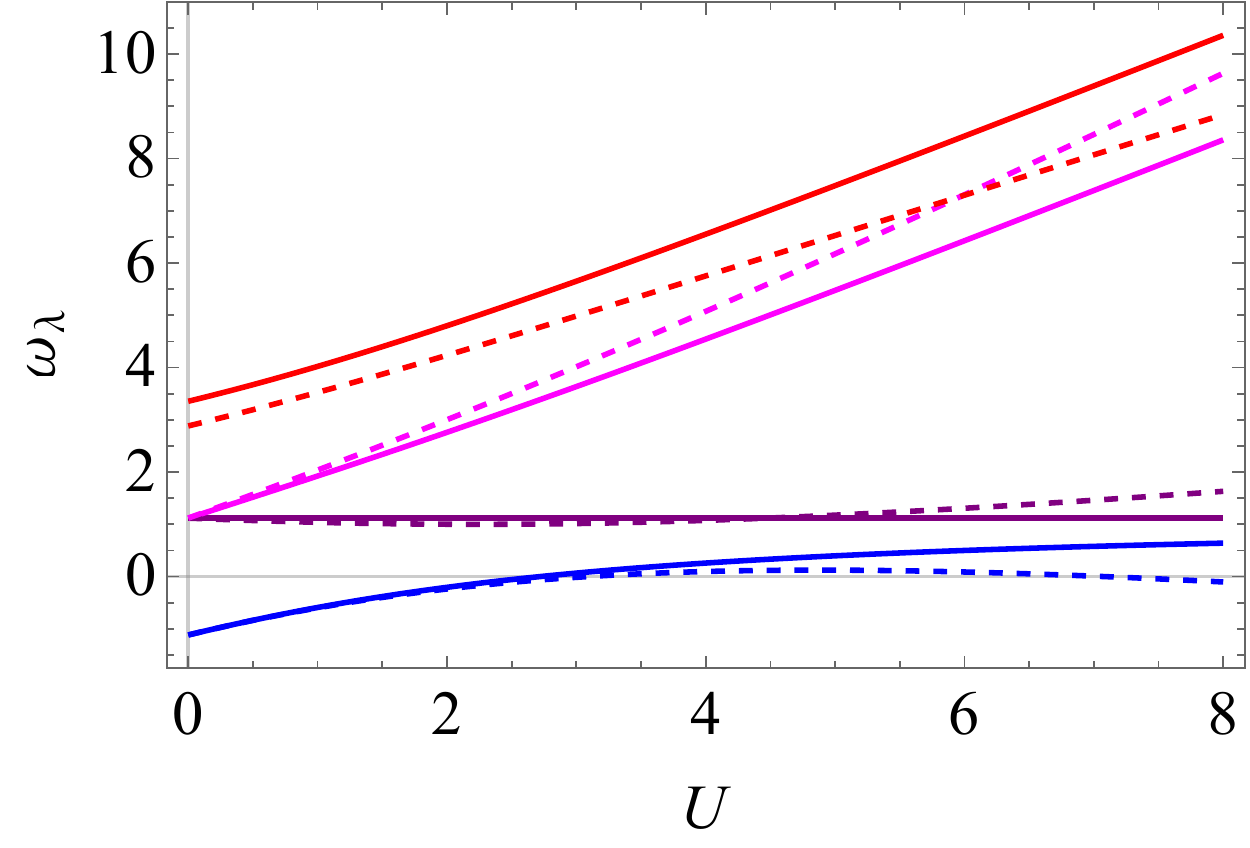}
			\caption{$D=1$}
		\end{subfigure}
		~
		\begin{subfigure}[t]{0.32\textwidth}
			\centering
			\includegraphics[width=0.99\textwidth]{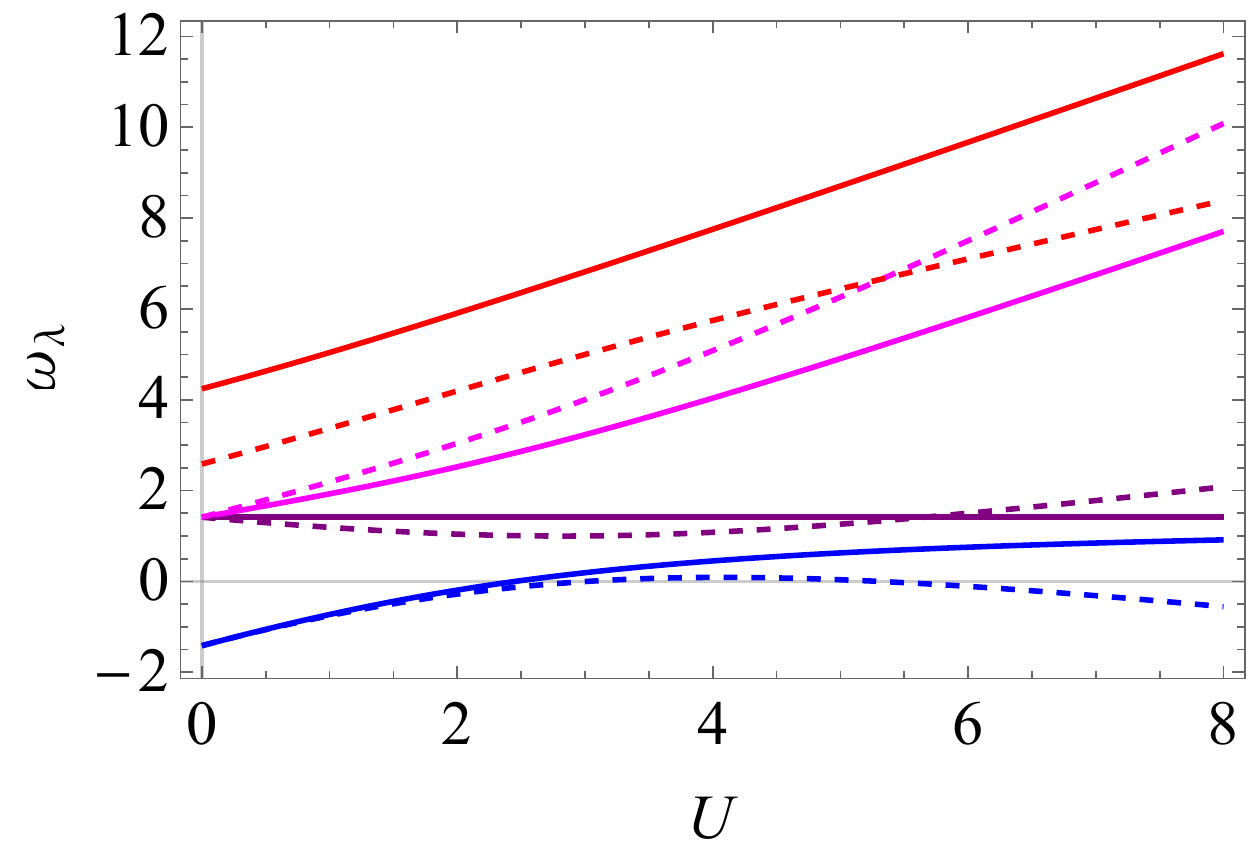}
			\caption{$D=2.0$}
		\end{subfigure}
		\caption{\emph{Position of the poles of the spin--down Green's function as a function of $U$ starting from the Hartree approximation. Solid line, exact results; dashed lines, the poles in Eq. \eqref{eq:dynCA_H_poles}.}}
		\label{fig:dynLDA_H}
	\end{minipage}
\end{figure*}

\paragraph{Starting from a free-particle approximation}
The simplest case is starting from a very crude approximation. We do not take into account any explicit spectral potential in the asymmetric dimer, i.e. we assume $v_{{\rm SF}\,a\,i}(\omega) = 0$.
As a consequence also $v^s_{{\rm SF}\,a}(\omega)=0$ in the symmetric model,
and the correction that we import is the \textit{whole} spectral potential of the model system: $\Xi^{s}_{0}(\omega^s_{\lambda})=v^s_{\rm SF}(\omega_{\lambda}^s)$. The prescription reads:
\begin{equation}
v^{0-{\rm dynCA}}_{{\rm SF}\;i}(\omega_{\lambda})=v^s_{\rm SF}(\omega_{\lambda}^s)
\label{eq:dynCA_0}
\end{equation}
This is a \textit{global} potential, independent of the site $i$. Still, inhomogeneity is accounted for by the external potential term that modifies the free--particle Green's function. By plugging the expression \eqref{eq:dynCA_0} in the pole equation, Eq. \eqref{eq:fg}, and by using the same equation also in the model system, we get the four poles in this approximation:
\begin{equation}
	\omega^{0-{\rm dynCA}}_{\lambda}=\omega_{\lambda}^s\pm\left(\sqrt{1+\tfrac{D^2}{4}}-1\right)
\label{eq:dynCA_0_poles}
\end{equation}
where the upper (lower) sign is for $\omega_2$ and $\omega_3$ ($\omega_1$ and $\omega_4$). This approximation leads to a complete disentanglement of interaction, accounted for by $\omega_{\lambda}^s$, and inhomogeneity, which results from the second term. 
Apart from the pole $\omega_4$, the results are exact in the $U\to0$ limit. Also for small nonvanishing $U$, the performances of this relatively simple approach are pretty good, see Fig. \ref{fig:dynLDA_0}. However, this expression is extremely simple and does not well reproduce the position of the poles for larger values of $D$ or $U$.

\paragraph{Starting from the Hartree approximation}
A better approximation is to have an explicit local dependence in the spectral potential for the asymmetric dimer, and import from the model system a smaller term. This is possible if the Hartree potential $v^H_i=Un_i$ is treated exactly in the asymmetric dimer, and therefore just the exchange--correlation part of the spectral potential is imported from the symmetric dimer. 

Indeed, from the separation $v_{{\rm SF}\,i}(\omega)=v_i^H+ \Xi_{i}^{H}(\omega)$ introduced above  
we take the correction $\Xi_{i}^{H}(\omega)$ from the model system. There, the Hartree potential is simply $U/2$, hence $\Xi_{s}^{H}(\omega)=v_{{\rm SF}}^{s}(\omega)-U/2$. Using the state $\lambda$ as a connector, the equation analogous of Eq. \eqref{eq:wrqweribf} and \eqref{eq:conn_states} reads:
\begin{equation}
v^{\rm H-dynCA}_{{\rm SF}\;i}(\omega_{\lambda})=Un_i+v^{xc\;s}_{\rm SF}(\omega_{\lambda}^s)
\label{eq:dynCA_H}
\end{equation}
This is another connector approximation, which again reduces to the exact result for $D\to0$, and it explicitly treats some inhomogeneities in the interaction through the Hartree potential of the real system.  

\begin{figure*}[t!]
	\centering
	\begin{minipage}{0.99\textwidth}
		\centering
		\begin{subfigure}[t]{0.32\textwidth}
			\centering
			\includegraphics[width=0.99\textwidth]{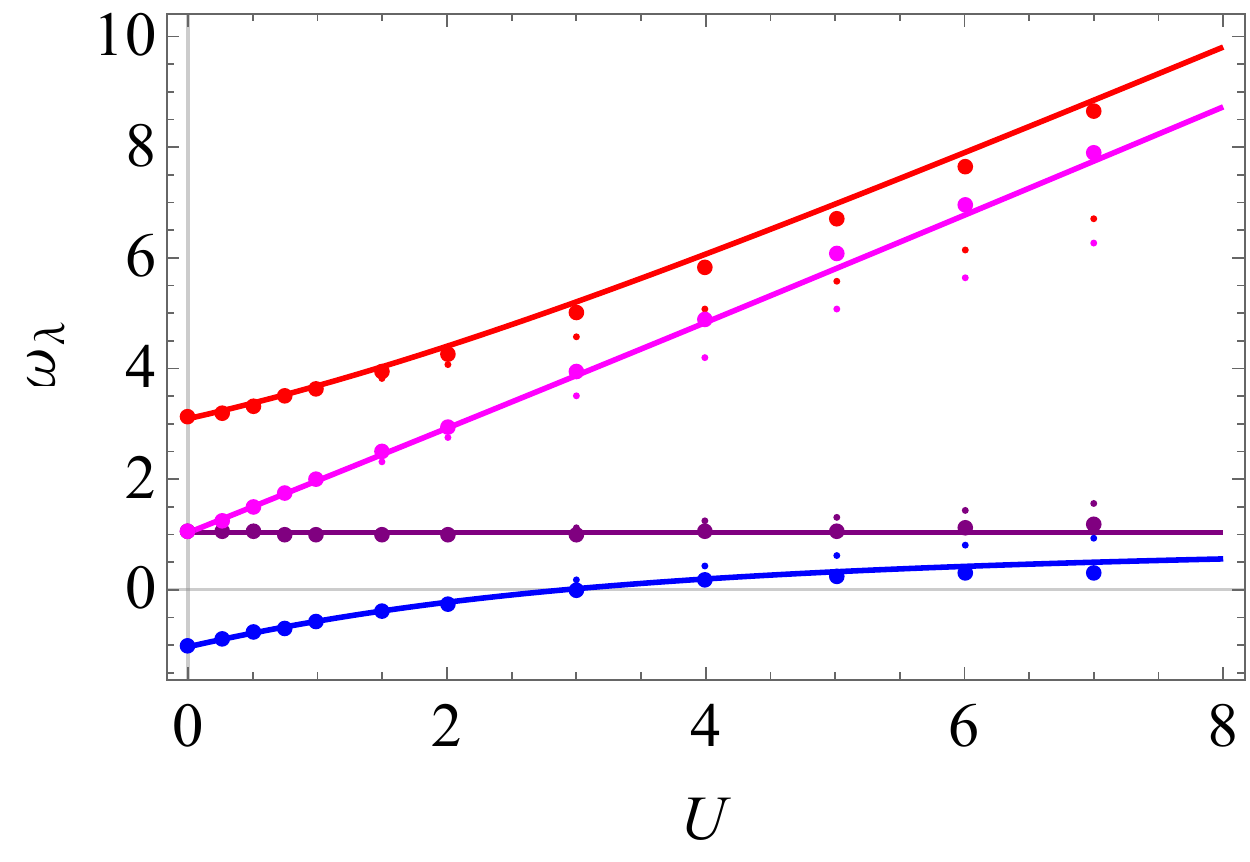}
			\caption{$D=0.5$}
		\end{subfigure}
		~
		\begin{subfigure}[t]{0.32\textwidth}
			\centering
			\includegraphics[width=0.99\textwidth]{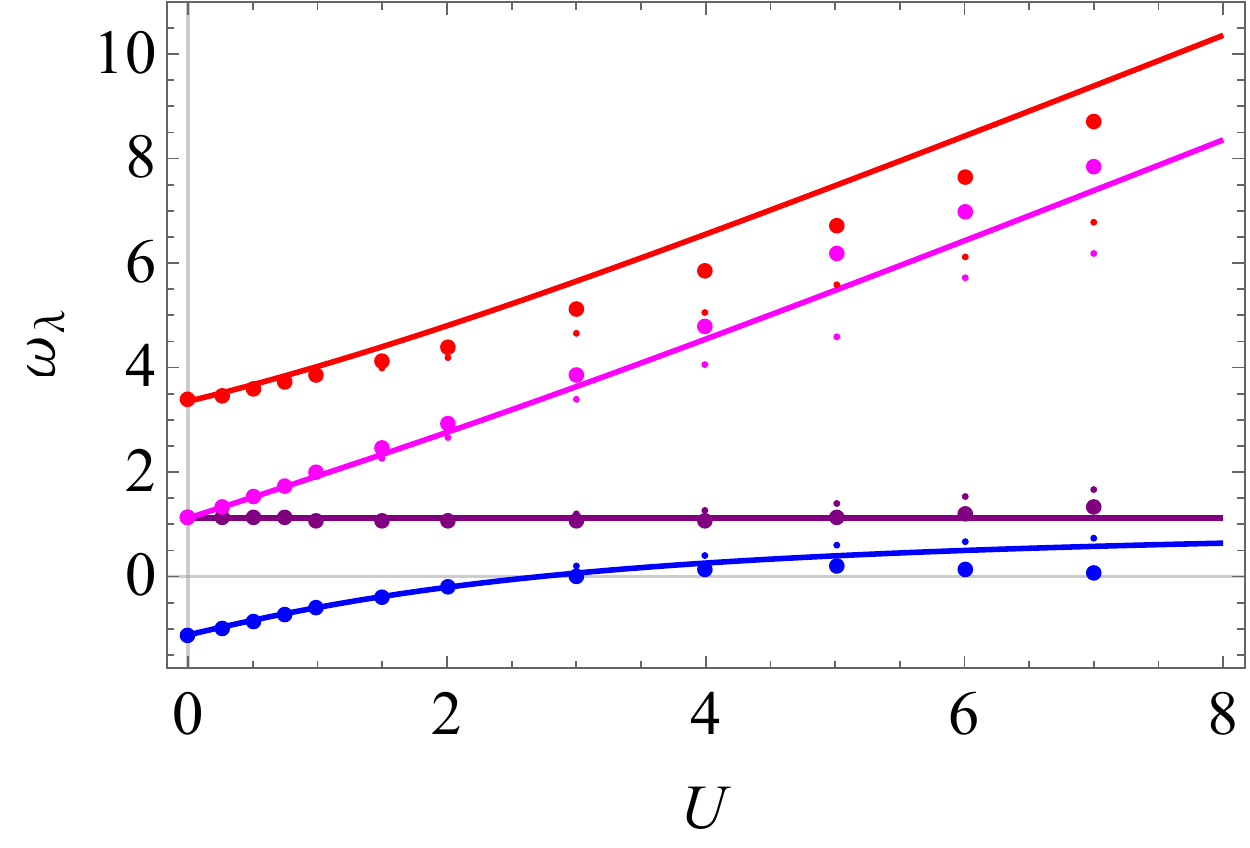}
			\caption{$D=1.0$}
		\end{subfigure}
		~
		\begin{subfigure}[t]{0.32\textwidth}
			\centering
			\includegraphics[width=0.99\textwidth]{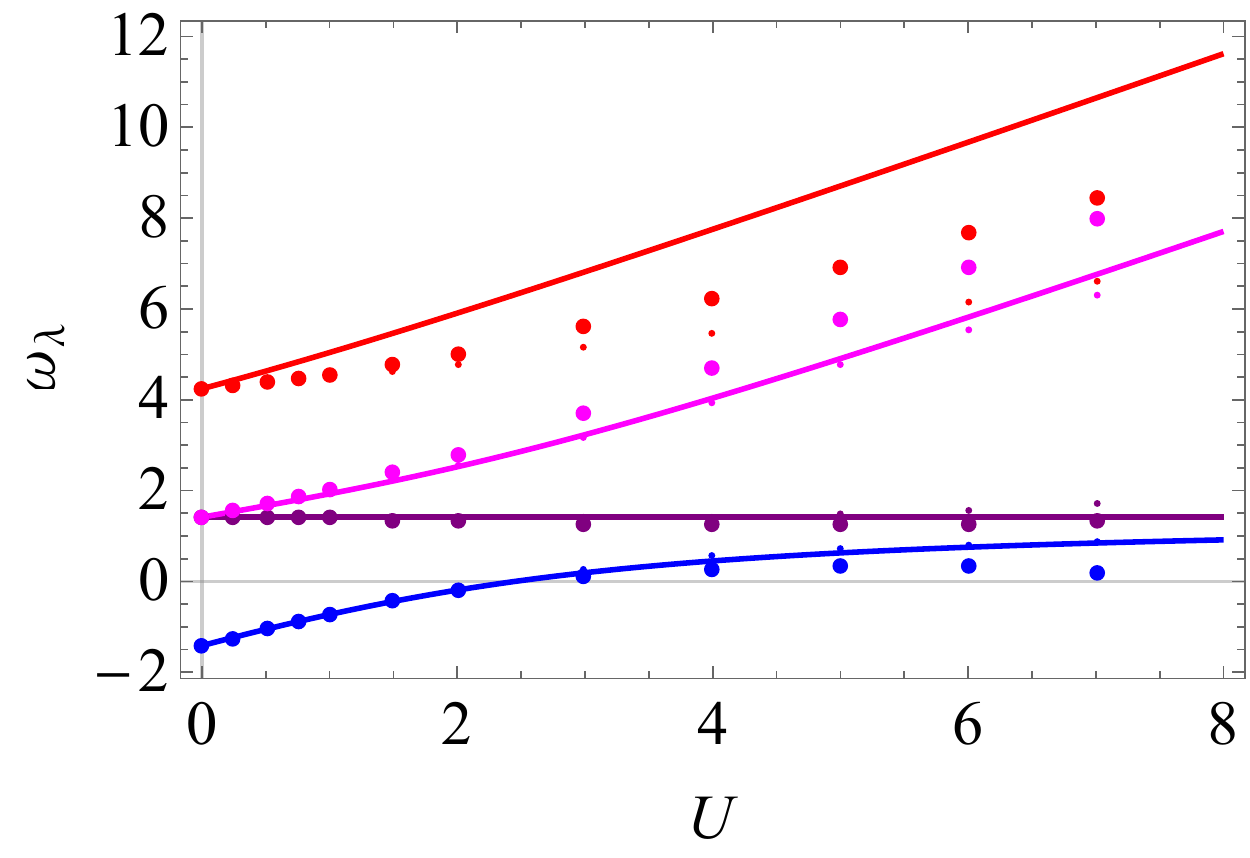}
			\caption{$D=2.0$}
		\end{subfigure}
		\caption{\emph{Position of the poles of the spin--down Green's function as a function of $U$ starting from the GW approximation. Solid line, exact results. Small dots, the $\rm G_0W_0$ poles of Eq. \eqref{eq:poles_asym_GW} and Fig. \ref{fig:GW_pina_SFasym_D}. Large dots, the poles $\omega_{\lambda}^{\rm G_0W_0-dLCA}$ in Eq. \eqref{eq:dynCA_GW_poles}, within the $\rm G_0W_0$--dynCA approach.}}
		\label{fig:dynLDA_GW}
	\end{minipage}
\end{figure*}

Plugging this potential into Eq. \eqref{eq:fg} to find the position of the poles, we get:
\begin{equation}
\omega^{\rm H-dynCA}_{\lambda}=\omega_{\lambda}^s\pm\left(\sqrt{1+h^2}-1\right),
\label{eq:dynCA_H_poles}
\end{equation}
where $h= D/ 2 + [U(n_1-n_2)]/2$, see Eq. \eqref{def:h}. Eq. \eqref{eq:dynCA_H_poles} is very similar to  Eq. \eqref{eq:dynCA_0_poles}, but now the interaction $U$ enters also the square root, creating an interplay between inhomogeneity and interaction in the position of the poles. Still, as in the previous case, the pole $\omega_4$ is not well described by this approximation for nonzero values of $D$.

We finally note that this approximation is an improvement with respect to:
\begin{enumerate}
	\item The Hartree approximation itself, Eq. \eqref{eq:hartreepoles}. Indeed, with a frequency--dependent potential, the two poles \eqref{eq:hartreepoles} can be split. Furthermore, they are in good agreement with the expected result, at least for small $D$, because the correction is imported from the model symmetric dimer.
	\item The free electron starting point, for small values of $U$. For large $U$, starting from the Hartree approximation worsens the result.
\end{enumerate}

\paragraph{Starting from the GW approximation}

We already evaluated the position of the poles in the asymmetric dimer in the GW approximation, see Fig. \ref{fig:GW_pina_SFasym_D} and  Eq. \eqref{eq:poles_asym_GW} in App. \ref{sec:gwasym}. These poles stem from the non--local and complex self energy $\Sigma^{G_0W_0}_{ij}(\omega)$ or, equivalently, from the local and real spectral potential $v^{G_0W_0}_{{\rm SF}\;i}(\omega)$. 

The spectral potential in the symmetric dimer that corresponds to the GW approximation, evaluated at the poles, is given by the relation (see Eq. \eqref{eq:QPdimer}):
\beq 
v^{G_0W_0\,s}_{\rm SF}({\omega}^{G_0W_0\,s}_{\lambda})={\omega}^{G_0W_0\,s}_{\lambda}\mp1.
\label{v_GW_sym}
\eeq
It is reported in table \ref{table:vGW}.
In Eq. \eqref{v_GW_sym} the GW poles ${\omega}^{G_0W_0\,s}_{\lambda}$ of the symmetric dimer are 
are given by Eq. \eqref{eq:wrgiert}:
\begin{equation}
{\omega}^{G_0W_0\,s}_{\lambda}=\frac{l+\frac{U}{2}}{2}\pm\frac{1}{2}\sqrt{\left(l\pm2-\frac{U}{2}\right)^2+\frac{2U^2}{l}}.
\end{equation}
with $l$ defined in Eq. \eqref{def:l}.
They are gathered in table \ref{table:GWpoles}. 

\begin{table}[ht]
	\centering
	\begin{tabular}{ c  c  c }
		\toprule
		$\lambda$ & $\omega^s_{\lambda}$ & ${\omega}^{G_0W_0\,s}_{\lambda}$
		\\
		\midrule
		$1$ & $1+\frac{U-c}{2}$ & $\frac{l+\frac{U}{2}}{2}-\frac{1}{2}\sqrt{\left(l+2-\frac{U}{2}\right)^2+\frac{2U^2}{l}}$ 
		\\ 
		$2$ & $1$ & $\frac{l+\frac{U}{2}}{2}-\frac{1}{2}\sqrt{\left(l-2-\frac{U}{2}\right)^2+\frac{2U^2}{l}}$ 
		\\
		$3$ & $1+U$ & $\frac{l+\frac{U}{2}}{2}+\frac{1}{2}\sqrt{\left(l-2-\frac{U}{2}\right)^2+\frac{2U^2}{l}}$ 
		\\ 
		$4$ & $1+\frac{U+c}{2}$ & $\frac{l+\frac{U}{2}}{2}+\frac{1}{2}\sqrt{\left(l+2-\frac{U}{2}\right)^2+\frac{2U^2}{l}}$
		\\
		\bottomrule
	\end{tabular}
	\caption{\emph{Exact and GW poles of the spin--down Green's function in the model symmetric dimer. Note that $c=\sqrt{U^2+16}$ and $l$ is defined in Eq. \eqref{def:l}.}}
	\label{table:GWpoles}
\end{table}

\begin{table*}[ht]
	\centering
	\begin{tabular}{ c  c  c  c }
		\toprule
		$\lambda$ & $v^s_{\rm SF}(\omega^s_{\lambda})$ & $v^{G_0W_0\,s}_{\rm SF}({\omega}^{G_0W_0\,s}_{\lambda})$ & $\Xi^s_{G_0W_0}({\omega}^{s}_{\lambda})$ \\
		\midrule
		$1$ & $2+\frac{U-c}{2}$ & $\frac{l+\frac{U}{2}}{2}-\frac{1}{2}\sqrt{\left(l+2-\frac{U}{2}\right)^2+\frac{2U^2}{l}}+1$ & $1+\frac{\frac{U}{2}-c-l}{2}+\frac{1}{2}\sqrt{\left(l+2-\frac{U}{2}\right)^2+\frac{2U^2}{l}}$ 
		\\ 
		$2$ & $0$ & $\frac{l+\frac{U}{2}}{2}-\frac{1}{2}\sqrt{\left(l-2-\frac{U}{2}\right)^2+\frac{2U^2}{l}}-1$ & $1-\frac{\frac{U}{2}+l}{2}+\frac{1}{2}\sqrt{\left(l-2-\frac{U}{2}\right)^2+\frac{2U^2}{l}}$ \\ 
		$3$ & $U$ & $\frac{l+\frac{U}{2}}{2}+\frac{1}{2}\sqrt{\left(l-2-\frac{U}{2}\right)^2+\frac{2U^2}{l}}-1$ & $1+\frac{\frac{3}{2}U-l}{2}-\frac{1}{2}\sqrt{\left(l-2-\frac{U}{2}\right)^2+\frac{2U^2}{l}}$ \\
		$4$ & $2+\frac{U+c}{2}$ & $\frac{l+\frac{U}{2}}{2}+\frac{1}{2}\sqrt{\left(l+2-\frac{U}{2}\right)^2+\frac{2U^2}{l}}+1$ & $1+\frac{\frac{U}{2}+c-l}{2}-\frac{1}{2}\sqrt{\left(l+2-\frac{U}{2}\right)^2+\frac{2U^2}{l}}$
		\\
		\bottomrule
	\end{tabular}
	\caption{\emph{Exact and GW potentials that give the poles of table \ref{table:GWpoles}, in the model system. Also their difference $\Xi^s_{G_0W_0}({\omega}^{s}_{\lambda})=v^s_{\rm SF}(\omega^s_{\lambda})-v^{G_0W_0\,s}_{\rm SF}({\omega}^{G_0W_0\,s}_{\lambda})$ is shown.}}
	\label{table:vGW}
\end{table*}

The differences between the exact spectral potential and the one approximated at the level of the GWA:
\begin{equation}
 \Xi^s_{G_0W_0}({\omega}^{s}_{\lambda})=v^s_{\rm SF}(\omega^s_{\lambda})-v^{G_0W_0\,s}_{\rm SF}({\omega}^{G_0W_0\,s}_{\lambda})
 \label{corr_GW_s}
\end{equation}
are the quantities that will be imported from the symmetric dimer. 
Their expressions are contained in table \ref{table:vGW}.

The correction terms $\Xi^s_{G_0W_0}({\omega}^{s}_{\lambda})$ \eqref{corr_GW_s} are imported from the model symmetric dimer and  added  on top of the GW spectral potential $v^{G_0W_0}_{{\rm SF}\;i}\left({\omega}_{\lambda}\right)$ for the asymmetric dimer. 
In this way one obtains:
\begin{equation}
v^{\rm G_0W_0-dynCA}_{{\rm SF}\;i}\left(\omega_{\lambda}\right)=
v^{G_0W_0}_{{\rm SF}\;i}\left({\omega}_{\lambda}\right) +\Xi^s_{G_0W_0}({\omega}^{s}_{\lambda})
\label{eq:dynCA_GW}
\end{equation}
Here the GW spectral potential $v^{G_0W_0}_{{\rm SF}\;i}\left({\omega}_{\lambda}\right)$  is the spectral potential that  yields the GW poles in the asymmetric dimer, see Eq. \eqref{eq:poles_asym_GW}. 
In principle, it is found as the solution of the generalized Sham--Schl\"uter equation when the self energy is $\Sigma_{GW}$. However, in practice we do not need its explicit form to calculate the poles corresponing to the approximate $v^{\rm G_0W_0-dynCA}_{{\rm SF}\;i}\left(\omega_{\lambda}\right)$.
Indeed, plugging Eq. \eqref{eq:dynCA_GW} into Eq. \eqref{eq:fg} and using the fact that $\omega_{\lambda}^{G_0W_0}$ are the solutions of Eq. \eqref{eq:fg} when the spectral potential is $v^{G_0W_0}_{{\rm SF}\;i}\left({\omega}_{\lambda}\right)$, we obtain the following simple expression for the poles:
\begin{equation}
\omega^{\rm G_0W_0-dynCA}_{\lambda}=\omega_{\lambda}^{G_0W_0}+\Xi^s_{G_0W_0}({\omega}^{s}_{\lambda}),
\label{eq:dynCA_GW_poles}
\end{equation}
which is still exact in the limit of $D\to0$. These poles are represented in Fig. \ref{fig:dynLDA_GW}.

For small $D$, the GWA in the asymmetric dimer decreases the gap between the Hubbard bands and yields poles that are blue--shifted in the lower band ($\omega_1$ and $\omega_2$) and red--shifted for the upper band ($\omega_3$ and $\omega_4$) with respect to the exact result, see Fig. \ref{fig:GW_pina_SFasym_D}. 
Adding the correction \linebreak $\Xi^s_{G_0W_0}({\omega}^{s}_{\lambda})$ imported from the model system restores the expected position of the poles, and the agreement between our theory and the exact result is very good. 

On the contrary, for larger values of $D$, the GWA  was already  good for $\omega_1$, $\omega_2$ and $\omega_3$, while it was not so good for $\omega_4$. With the present approach, the agreement is slightly worsened for the first three poles while it is again improved for the fourth pole. 

\begin{figure}[t!]
	\centering
	\begin{minipage}{0.99\columnwidth}
		\centering
		\begin{subfigure}[t]{0.45\textwidth}
			\centering
			\includegraphics[width=0.99\textwidth]{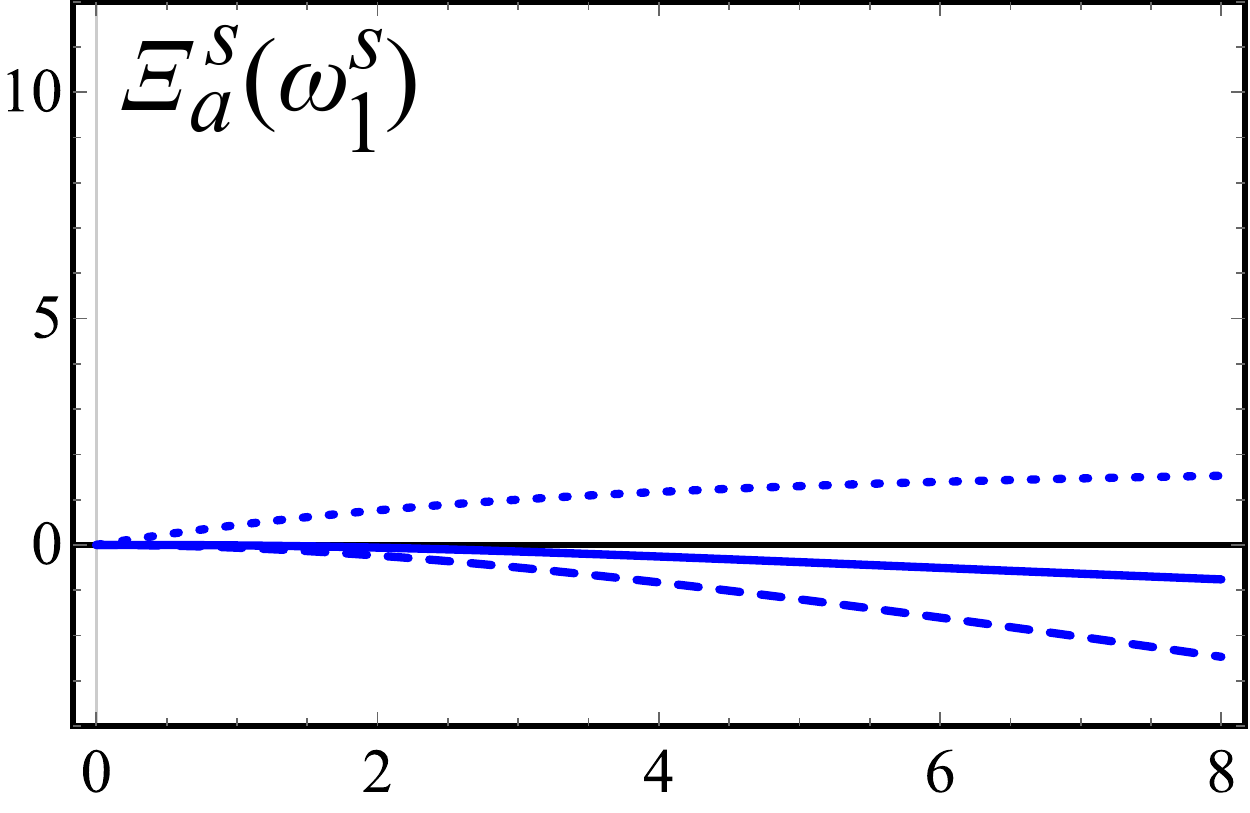}
			\caption{$\lambda=1$}
		\end{subfigure}
		~\quad
		\begin{subfigure}[t]{0.45\textwidth}
			\centering
			\includegraphics[width=0.99\textwidth]{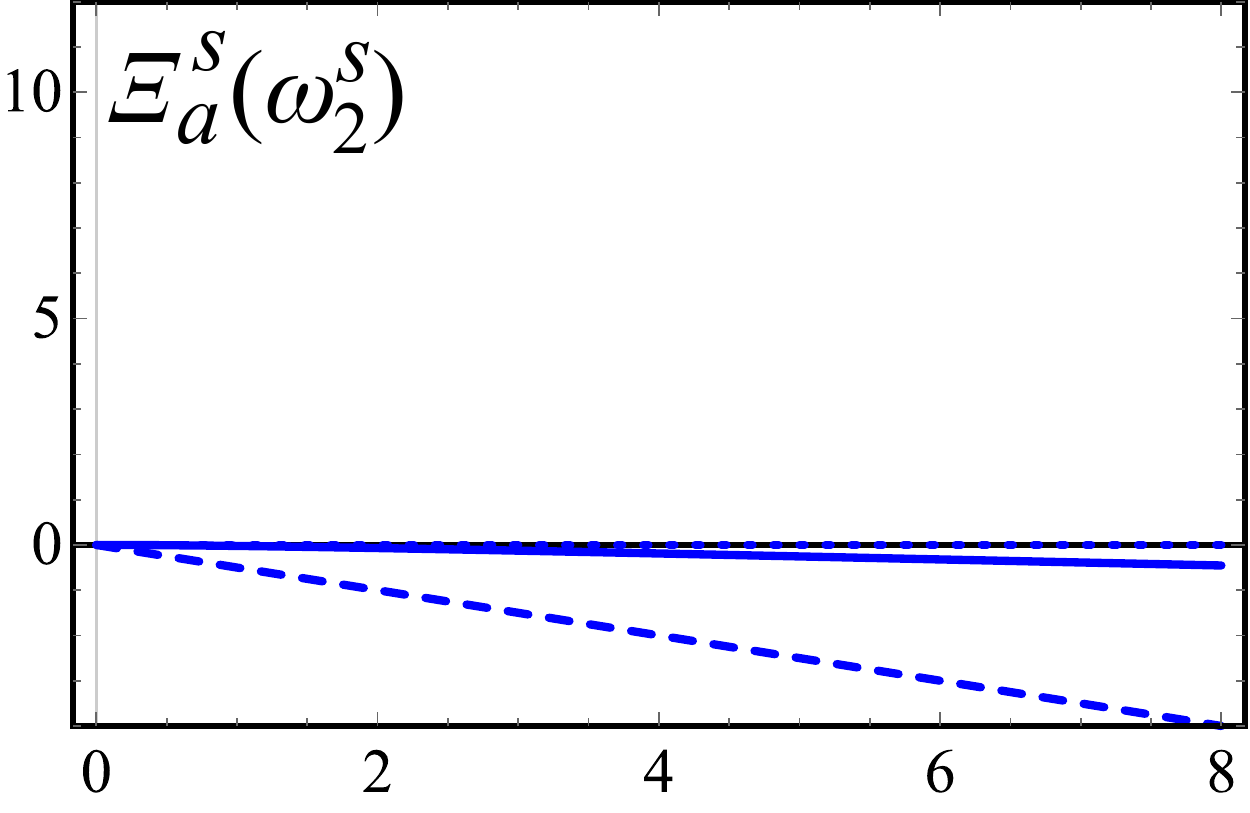}
			\caption{$\lambda=2$}
		\end{subfigure}
		\\ \vspace{1em}
		\begin{subfigure}[t]{0.45\textwidth}
			\centering
			\includegraphics[width=0.99\textwidth]{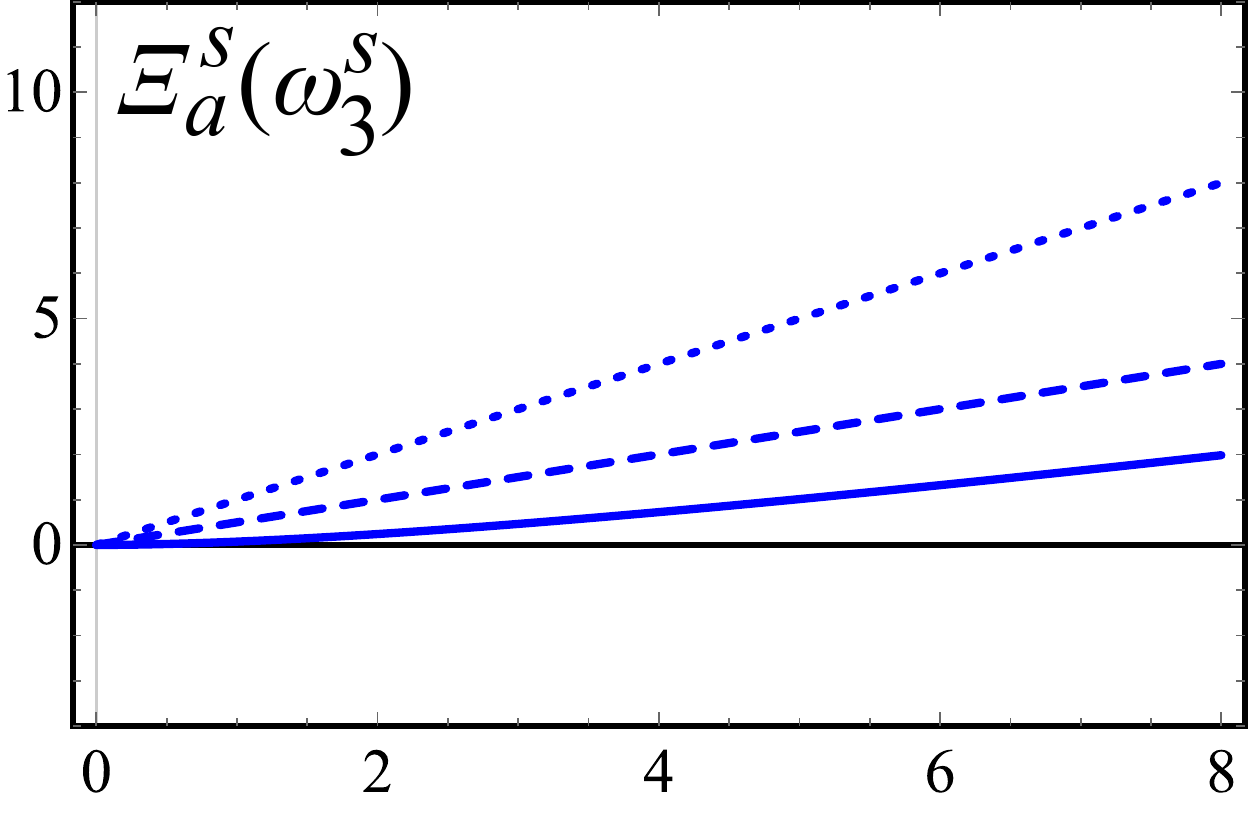}
			\caption{$\lambda=3$}
		\end{subfigure}
		~\quad
		\begin{subfigure}[t]{0.45\textwidth}
			\centering
			\includegraphics[width=0.99\textwidth]{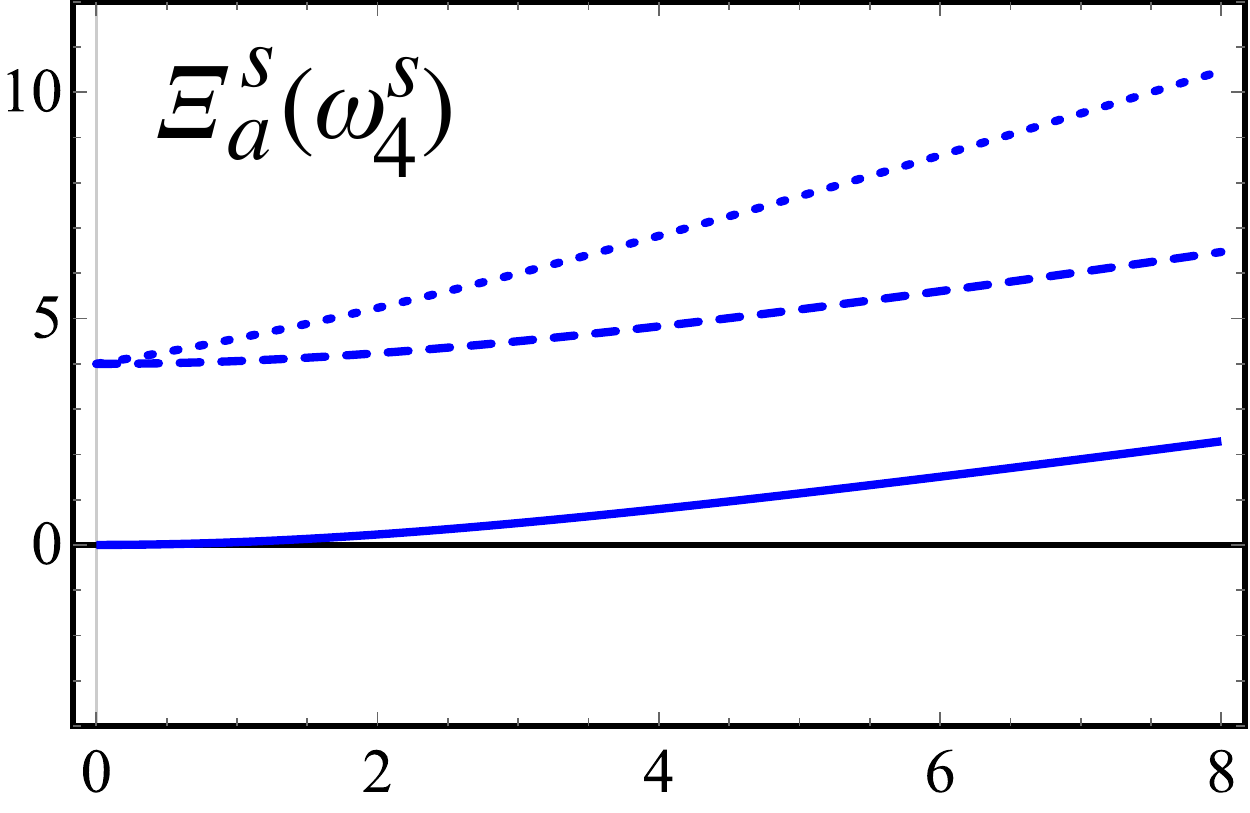}
			\caption{$\lambda=4$}
		\end{subfigure}
		\caption{\emph{Corrections $\Xi^s_{a}(\omega_{\lambda}^s)$ for the potential of the model symmetric dimer as a function of $U$. Dotted lines are the corrections $\Xi^s_{0}({\omega}^{s}_{\lambda})\equiv v_{\rm SF}^{s}(\omega^{s}_{\lambda})$. Dashed lines represent the corrections to the Hartree potential, $\Xi_{H}^{s}(\omega^{s}_{\lambda})\equiv v_{\rm SF}^{xc\;s}(\omega^{s}_{\lambda})$, while the continuous lines are the corrections $\Xi^s_{G_0W_0}({\omega}^{s}_{\lambda})$ to the GW potential of table \ref{table:vGW} }}
		\label{fig:corr_GW}
	\end{minipage}
\end{figure}

\paragraph{Discussion}

In Fig.  \ref{fig:corr_GW} we plot, as a function of $U$, the corrections $\Xi^{s}_{a}({\omega}^{s}_{\lambda})$ that are imported from the model symmetric dimer in the three approximation that we have considered. As one may expect, the corrections are smaller if the level of starting approximation is higher.
For example, it is clear that the GWA starting point is a great improvement over the simpler Hartree approximation. Indeed, the required corrections $\Xi^s_{G_0W_0}({\omega}^{s}_{\lambda})$ are smaller. Moreover, that same correction always tends to zero for $U\to0$ (i.e., the GWA  becomes exact for $U\to0$), even for the satellite $\omega_4$, whose physics is now clearly caught by the RPA polarization within the GWA.

\begin{figure}[t!]
	\centering
	\begin{minipage}{\columnwidth}
		\centering
		\includegraphics[width=0.8\textwidth]{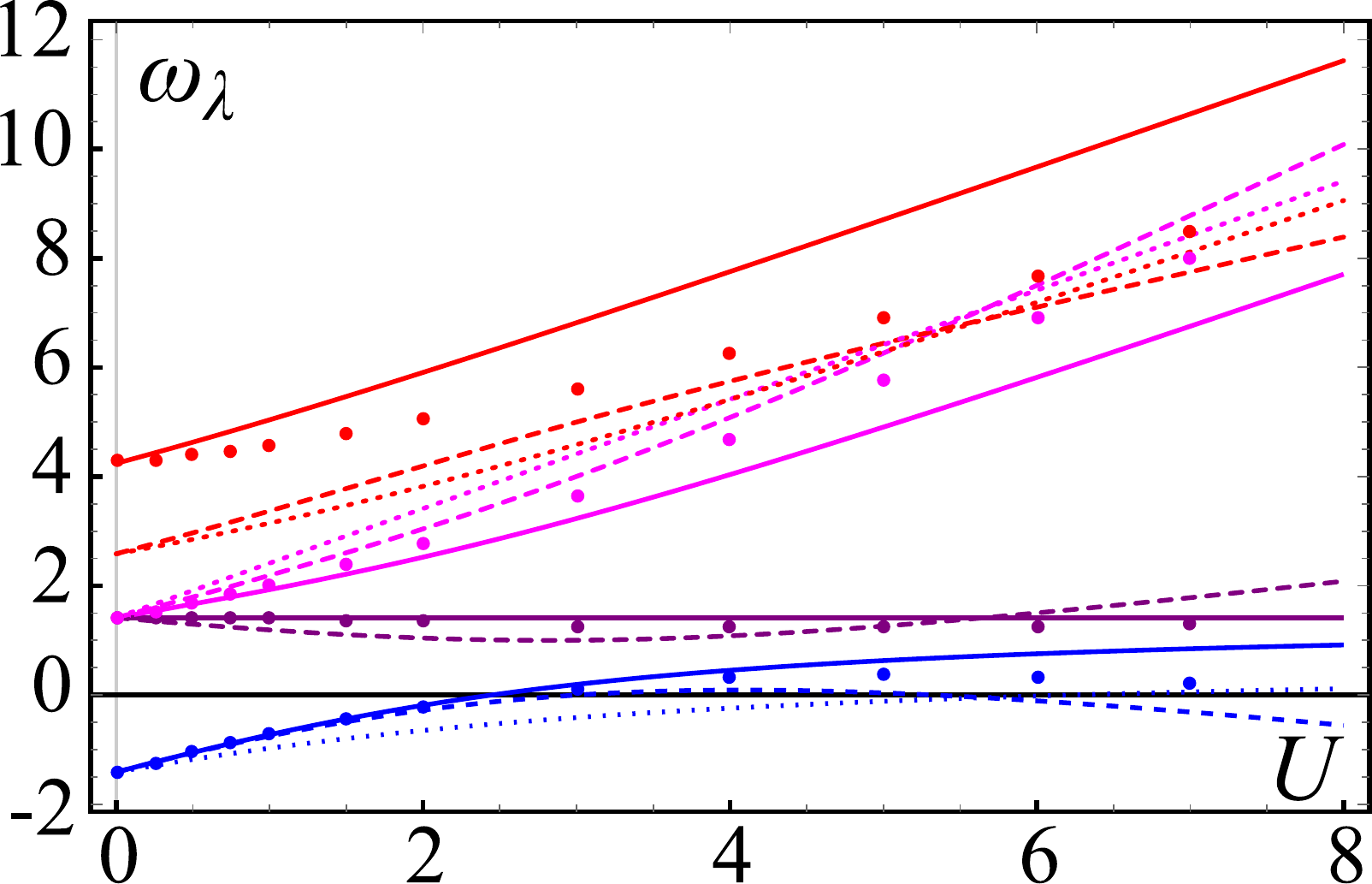}
		\caption{\emph{Position of the poles of the spin--down Green's function as a function of $U$, for $D=2$. Solid line, exact results. Dotted lines  for the poles of Eq. \eqref{eq:dynCA_0_poles}, dynCA on top of free-particle approximation. Dashed lines for the poles of Eq. \eqref{eq:dynCA_H_poles}, dynCA on top of the Hartree approximation. Dots, the poles of Eq. \eqref{eq:dynCA_GW_poles}, dynCA on top of the GWA. This figure summarizes panels (c) of Figs. \ref{fig:dynLDA_0}-\ref{fig:dynLDA_GW}.}}
		\label{fig:dynLDA_Hub}
	\end{minipage}
\end{figure}

As a result, the more pieces of the potential are put into evidence and treated exactly in the auxiliary system, the more accurate is the dynCA.
In Fig. \ref{fig:dynLDA_Hub} we compare the poles of the exact Green's function of the asymmetric dimer for $D=2$ as a function of $U$ (solid lines) with the three approximations that we have considered. 
We find that passing from the poles  calculated from  Eq. \eqref{eq:dynCA_0} (i.e. dynCA on top of the free-particle approximation, dotted lines in Fig. \ref{fig:dynLDA_Hub}) to those calculated from Eq. \eqref{eq:dynCA_H} (i.e. dynCA on top of the Hartree approximation, dashed lines in Fig. \ref{fig:dynLDA_Hub}) the agreement with the exact results improves considerably for not too--large interaction, i.e. $U< 2$.
The same happens making the further step to Eq. \eqref{eq:dynCA_GW} (i.e. dynCA on top of the GW approximation, dots in Fig. \ref{fig:dynLDA_Hub})
for an even larger range of $U$.

Moreover, in general, for small $D$ the connector approximations work well, as the real system is closer to the model system itself, and therefore the dynCA prescription is better suited. For small $D$, the real system is only slightly inhomogeneous, hence an average description as the one proposed here works well. 
On the contrary, for higher values of $D$, when properties are truly site--dependent, a \textit{global} connector $\Xi^s_{a}(\omega_{\lambda}^s)$ shows its limits
because it requires the model system to be similar to the real one. 
This issue could be overcome by introducing a \textit{local} (i.e. site-dependent) connector between the model and the auxiliary system, like the local density of LDA. This is not straightforward, as the model system, the $N=1$ symmetric Hubbard dimer, misses a density that could be tuned: there $n_i=\frac{1}{2}$ is a constant. This is a general difficulty concerning density functional approaches.
A possibility could be to tune the hopping parameter introducing an effective $t_{\rm eff}$ as proposed in \protect\cite{Schindlmayr1995}.

\section{Summary and outlook}
\label{sec:conclusion}

We have discussed a general strategy to directly calculate the observables of interest. It is based on two steps that can be considered to be
a generalization of the successful paradigm of the LDA for the Kohn-Sham formulation of DFT. 
The first step is the definition of an auxiliary system with an effective potential (or kernel) that is designed to yield exactly the target quantity.
The second step is the formulation of a direct approximation to the effective potential, through the introduction of a connector, i.e. a recipe to import the needed information from a model system. 
In this way one can also hope to be able to disentangle material-specific properties from universal effects that are captured by the model system already. 

We have illustrated this general strategy with a toy model, the asymmetric Hubbard dimer with one electron. We have compared the spectral functions obtained from the exact solution of the model with the Hartree and GW approximations to the self-energy and with calculations performed employing different approximations to the spectral potential, i.e. the effective local, real and frequency-dependent potential that is built in such a way to give, in principle, the diagonal of the spectral function exactly.
The approximations to the spectral potential have been obtained introducing a suitable connector to the symmetric Hubbard dimer, which here plays the role of model system. We have discussed the 
performances of the approximations and the limitations inherent to this very simple model.

In real applications, the use of the spectral potential aims to replace computationally expensive MBPT calculations for spectral properties or to add corrections to existing approximations, such as the GWA, in an efficient manner. The key point clearly is the development of accurate connector approximations in real materials, which is the subject of ongoing work \cite{marco_arxiv}.

\section*{Acknowledgements}
We acknowledge Stefano di Sabatino for fruitful discussions.
The research leading to these results has received funding from the European Research Council
under the European Union's Seventh Framework Programme (FP/2007-2013)
/ ERC grant agreement no. 320971 
and from a Marie Curie FP7 Integration Grant.

\appendix
\section{The Sham-Schl\"uter equation for the density matrix}
\label{sec:sse_dm}
In a discrete lattice model the generalized Sham--Schl\"uter equation \eqref{eq:SSE_rdm} for the density matrix $\gamma_{ij}=\sum_{\sigma}\int\frac{d\omega}{2\pi i}e^{i\omega\eta}G_{ij,\sigma}(\omega)$ is:
\begin{multline}
\sum_{\sigma}\sum_{kl}
\int\frac{d\omega}{2\pi i}e^{i\omega\eta}
G^{\rm DM}_{ik,\sigma}(\omega)
\Sigma_{kl,\sigma}(\omega)
G_{lj,\sigma}(\omega)
=\\=
\sum_{\sigma}\sum_{kl}
v^{\rm DM}_{kl}
\int\frac{d\omega}{2\pi i}e^{i\omega\eta}
G^{\rm DM}_{ik,\sigma}(\omega)
G_{lj,\sigma}(\omega)
\label{eq:SS7}
\end{multline}
with the non-local $\gamma$--effective potential $v^{\rm DM}_{ij}$. We take the symmetric Hubbard dimer, Eq. \eqref{eq:sym_dimer}, as a model to which applying the previous equation. As the Green's function, the self energy and the effective potential are diagonal in the bonding--antibonding basis $\{\alpha\}=\{\pm\}$ (see sec. \ref{sec:solmode}), we express the Sham--Schl\"uter equation in that basis:
\begin{multline} 
\sum_{\sigma}\int\frac{d\omega}{2\pi i}e^{i\omega\eta}
G^{\rm DM}_{\alpha,\sigma}(\omega)
\Sigma_{\alpha,\sigma}(\omega)
G_{\alpha,\sigma}(\omega)
=\\=
v^{\rm DM}_{\alpha}\sum_{\sigma}\int\frac{d\omega}{2\pi i}e^{i\omega\eta}
G^{\rm DM}_{\alpha,\sigma}(\omega)
G_{\alpha,\sigma}(\omega)
\label{eq:SSE8}
\end{multline}
Let us check if a solution to Eq. \eqref{eq:SSE8} exists. 
If the ground state is composed of a single electron, the previous equation will always have an undetermined solution $v^{\rm DM}_{\alpha}$. Indeed, with a single spin--up electron, the ground state is a trivial Slater determinant, and the spin-resolved density matrix is  idempotent and independent of the interaction: $\gamma_{\alpha\beta,\sigma}=\delta_{\sigma,\uparrow}\delta_{\alpha\beta}\delta_{\alpha,-}$ or $\gamma_{ij,\sigma}=\frac{1}{2}\delta_{\sigma,\uparrow}$.

Therefore, to check if non--trivial solutions of Eq. \eqref{eq:SSE8} exist, we have to move to the $N=2$ sector (half--filling), where Green's function and self energy, in the bonding--antibonding basis, are spin--independent and read \cite{marco_phd,stef_phd}:
\begin{gather*}
G_{-,\sigma}(\omega)=\frac{\frac{1}{2}-\frac{2}{c}}{\omega-\left(1+\frac{U+c}{2}\right)+i\eta}+
\frac{\frac{1}{2}+\frac{2}{c}}{\omega-\left(1+\frac{U-c}{2}\right)-i\eta}
\\
G_{+,\sigma}(\omega)=\frac{\frac{1}{2}+\frac{2}{c}}{\omega-\left(-1+\frac{U+c}{2}\right)+i\eta}+
\frac{\frac{1}{2}-\frac{2}{c}}{\omega-\left(-1+\frac{U-c}{2}\right)-i\eta}
\\
\Sigma_{-,\sigma}(\omega)=\frac{U}{2}+\frac{\frac{U^2}{4}}{\omega-\left(3+\frac{U}{2}\right)+i\eta}
\\
\Sigma_{+,\sigma}(\omega)=\frac{U}{2}+\frac{\frac{U^2}{4}}{\omega-\left(-3+\frac{U}{2}\right)-i\eta}.
\end{gather*}
with $c=\sqrt{16+U^2}$. In particular, the occupation numbers are fractional: $n_{\pm,\sigma}:=\int\frac{d\omega}{2\pi i}e^{i\omega\eta}G_{\pm,\sigma}(\omega)=\frac{1}{2}\mp\frac{2}{c}$. The auxiliary system Green's function $G^{\rm DM}_{\alpha,\sigma}(\omega)$ can be built from the full Green's function by setting $U=0$ and introducing, in place of the self energy, a state--dependent\footnote{As $v_{\pm}^{\rm DM}$ is static, Eq. \eqref{eq:RDMGF} holds whenever there is no level crossing between the bonding and the antibonding states, with energy $-1+v_{-}^{\rm DM}$ and $1+v_{+}^{\rm DM}$ respectively, namely if $v^{\rm DM}_+>v^{\rm DM}_--2$. In the opposite case, when $v^{\rm DM}_+<v^{\rm DM}_--2$, the occupation of the two states swap and Eq. \eqref{eq:RDMGF} becomes $G^{\rm DM}_{\pm,\sigma}(\omega)=\frac{1}{\omega-\left(\pm 1+v^{\rm DM}_{\pm}\right)\mp i\eta}$.} potential $v^{\rm DM}_{\alpha}$:
\begin{gather}
G^{\rm DM}_{\pm,\sigma}(\omega)=\frac{1}{\omega-\left(\pm 1+v^{\rm DM}_{\pm}\right)\pm i\eta}.
\label{eq:RDMGF}
\end{gather}
Let us plug all these quantities in Eq. \eqref{eq:SSE8}. The sum over spin yields a trivial multiplicative factor. We first consider the bonding state; the right hand side is:
\begin{equation*}
v_{-}^{\rm DM}\int\frac{d\omega}{2\pi i}e^{i\omega\eta}G^{\rm DM}_{-,\sigma}(\omega)G_{-,\sigma}(\omega)=
\frac{\frac{1}{2}\left(1-\frac{4}{c}\right)v_{-}^{\rm DM}}{v_{-}^{\rm DM}-2-\frac{c+U}{2}}.
\end{equation*}
As for the left hand side, it reads:
\begin{multline*}
\int\frac{d\omega}{2\pi i}e^{i\omega\eta}G^{\rm DM}_{-,\sigma}(\omega)\Sigma_{-,\sigma}(\omega)G_{-,\sigma}(\omega)=
\\
=\frac{U}{4}\left(1-\frac{4}{c}\right)\Biggl\{
\frac{1}{v_-^{\rm DM}-2-\frac{c+U}{2}}+\Biggr.
\\
+\frac{\frac{U}{2}}{\left[v_-^{\rm DM}-4-\frac{U}{2}\right]\left[v_-^{\rm DM}-2-\frac{c+U}{2}\right]}+
\\
\quad\Biggl.+\frac{U}{2}\left(\frac{1+\frac{4}{c}}{1-\frac{4}{c}}\right)
\Biggl[\frac{1}{\left[v_-^{\rm DM}-4-\frac{U}{2}\right]\left[v_-^{\rm DM}-2-\frac{U-c}{2}\right]}+\Biggr.\\+\Biggl.
\frac{1}{\left[2+\frac{c}{2}\right]\left[v_-^{\rm DM}-2-\frac{U-c}{2}\right]}\Biggr]
\Biggr\}.
\end{multline*}
After simplifying both sides, the Sham-Schlüter equation \eqref{eq:SSE8} becomes:
\begin{equation}
0=-\frac{1}{2}\left(1-\frac{4}{c}\right)\frac{v_-^{\rm DM}-\left(2+\frac{U+c}{2}\right)}{v_-^{\rm DM}-\left(2+\frac{U+c}{2}\right)}
\label{eq:gwegfrefg}
\end{equation}
and analogously for the one corresponding to the antibonding state:
\begin{equation}
0=\frac{1}{2}\left(1-\frac{4}{c}\right)\frac{v_+^{\rm DM}+\left(2-\frac{U-c}{2}\right)}{v_+^{\rm DM}+\left(2-\frac{U-c}{2}\right)}.
\label{eq:reyhery}
\end{equation}
Since the dependence on the potential cancels, these equations do not have any solution for $v_\pm^{\rm DM}$, apart from the case $U=0$, where $c=4$ and both equations display undetermined solutions. Of course, this is a trivial case as the density matrix is  idempotent. 

As a final observation, it is interesting to compare Eq. \eqref{eq:SSE8} to its DFT counterpart, namely the original Sham--Schl\"uter equation for the density $n_i$ \cite{Sham1983}. The latter involves a \emph{local} potential $v_i^{\rm KS}$, which is a constant for the symmetric dimer, and reads:
	\begin{multline} 
	\int\frac{d\omega}{2\pi i}e^{i\omega\eta}
	\sum_{\alpha,\sigma}G^{\rm KS}_{\alpha,\sigma}(\omega)
	\Sigma_{\alpha,\sigma}(\omega)
	G_{\alpha,\sigma}(\omega)
	=\\=
	v^{\rm KS}\int\frac{d\omega}{2\pi i}e^{i\omega\eta}
	\sum_{\alpha,\sigma}G^{\rm KS}_{\alpha,\sigma}(\omega)
	G_{\alpha,\sigma}(\omega).
	\label{eq:SSEboh}
	\end{multline}
With respect to Eq. \eqref{eq:SSE8}, this equation displays an $\alpha$-independent potential $v^{\rm KS}$ and a sum over $\alpha$ is performed. As a consequence, Eq. \eqref{eq:SSEboh} is equivalent to the sum of Eq. \eqref{eq:SSE8} for $v^{\rm DM}_{-}$ and for $v^{\rm DM}_{+}$, with both taken equals to $v^{\rm KS}$. 

While the two equations \eqref{eq:gwegfrefg} and \eqref{eq:reyhery} for $v^{\rm DM}_{-}$ and $v^{\rm DM}_{+}$ never have a solution for $U\neq0$, their sum
\textit{always} has an undetermined solution. This proves that, while the $\gamma$--effective potential $v^{\rm DM}_{ij}$ exists only for $U=0$, the $n$--effective potential, namely the familiar Kohn-Sham potential $v^{\rm KS}_{i}$, always exists, also for $U\neq0$. This was actually expected from the beginning because, once the ground state of the Kohn--Sham system is fixed to the one of the real one, as it is by Eq. \eqref{eq:RDMGF}, no matter the value of the Kohn--Sham potential the density will stay the same\footnote{If $v^{\rm DM}_+<v^{\rm DM}_--2$, the final results Eq. \eqref{eq:gwegfrefg} and \eqref{eq:reyhery} become:
\begin{gather*}
0=-\frac{1}{2}\left(1+\frac{4}{c}\right)\frac{v_-^{\rm DM}-\left(2+\frac{U-c}{2}\right)}{v_-^{\rm DM}-\left(2+\frac{U-c}{2}\right)}
\\
0=\frac{1}{2}\left(1+\frac{4}{c}\right)\frac{v_+^{\rm DM}+\left(2-\frac{U+c}{2}\right)}{v_+^{\rm DM}+\left(2-\frac{U+c}{2}\right)}
\end{gather*}
which do not have any solutions, not even for $U=0$ (when there would not be any level crossing). As in the other case, the sum of the two equations always has an undetermined solution. However, here the sum is not equivalent to the Sham-Schl\"uter equation for the density, because if $v^{\rm DM}_{-}=v^{\rm DM}_{+}\equiv v^{\rm KS}$, there is no level crossing.}.

\section{The Green's function of the asymmetric Hubbard dimer}
\label{sec:gasym}
In this appendix, we derive the expression \eqref{eq:GFN1down} for the spin--down Green's function of the asymmetric dimer at one fourth filling, via the Lehmann representation. To this purpose, we first present the result of the diagonalization of Hamiltonian \eqref{eq:weihojbw} in the half--filling case $N=2$. The site basis $\ket{i_1\sigma_1,i_2\sigma_2}$, in which the electron 1 (2) occupies the site $i_1$ ($i_2$) with spin $\sigma_1$ ($\sigma_2$), is ordered as 
$\bigl\{\ket{1\uparrow,2\uparrow},\ket{1\uparrow,2\downarrow},\ket{1\uparrow1,\downarrow},\ket{2\uparrow,2\downarrow},\ket{1\downarrow,2\uparrow},\bigr.$ $\bigl.\ket{1\downarrow,2\downarrow}\bigr\}$. As our aim is to derive the spin--down Green's function for the case in which $\ket{\rm GS}\equiv\ket{-,\uparrow}$, we can further restrict the basis to elements with opposite spins, namely $\bigl\{\ket{1\uparrow,2\downarrow},\bigr.$ $\bigl.\ket{1\uparrow,1\downarrow},\ket{2\uparrow,2\downarrow},\ket{1\downarrow,2\uparrow}\bigr\}$. In this basis, the Hamiltonian \eqref{eq:weihojbw} reads:
\begin{equation}
\hat{H}^{(N=2,S_z=0)}\longrightarrow
\left(
\begin{matrix}
0 & -1 & -1 & 0 \\
-1 & U+D & 0 & 1 \\
-1 & 0 & U-D & 1 \\
0 & 1 & 1 & 0
\end{matrix}
\right).
\label{eq:jtuyfnmhf}
\end{equation}
The eigenvalue equation $\det\bigl[\hat{H}-e_{\lambda}\hat{1}\bigr]=0$ 
has the triplet solution $e_{\lambda}=0:=e_2$, with associated eigenvector $\ket{\phi_2}=
\frac{1}{\sqrt{2}}\left(\ket{1\uparrow,2\downarrow}+\ket{1\downarrow,2\uparrow}\right)$, plus the three solutions of the third order equation $e^3_{\lambda}-2Ue^2_{\lambda}-e_{\lambda}\left(D^2-U^2+4\right)+4U=0$, that can conveniently be expressed as \cite{Carrascal2015}
\begin{align*}
e_1&=\frac{2}{3}\left[U-r\cos\left(\theta-\frac{\pi}{3}\right)\right]
\\
e_3&=\frac{2}{3}\left[U-r\cos\left(\theta+\frac{\pi}{3}\right)\right]
\\
e_4&=\frac{2}{3}\Bigl[U-r\cos\left(\theta+\pi\right)\Bigr]
\end{align*}
with:
\begin{equation*}
\begin{gathered}
z^2:=U^2+18-9D^2
\\
r^2:=U^2+12+3D^2
\\
\cos 3\theta:=-\frac{z^2U}{r^3}.
\end{gathered}
\end{equation*}
The corresponding normalized eigenvectors are:
\begin{multline*}
\ket{\phi_{\lambda}}=
\frac{1}{\mathcal{N}_{\lambda}}\Bigl[
\Bigl(\ket{1\uparrow,2\downarrow}-\ket{1\downarrow,2\uparrow}\Bigr)\Bigr.
+\Bigl(\frac{2}{e_{\lambda}+(D-U)}+
\\
\Bigl.-e_{\lambda}\Bigr)\ket{1\uparrow,1\downarrow}-
\frac{2}{e_{\lambda}+(D-U)}
\ket{2\uparrow,2\downarrow}
\Bigr]
\end{multline*}
with $\displaystyle \frac{\mathcal{N}^2_{\lambda}}{4}=\frac{1}{2}+\left(\frac{1}{e_{\lambda}+D-U}-\frac{e_{\lambda}}{2}\right)^2+\left(\frac{1}{e_{\lambda}+D-U}\right)^2$.
The four eigenvectors $\phi_{\lambda}$ of the $N=2$ Hamiltonian enter the Lehmann representation of the spin--down $N=1$ Green's function, which reads:
\begin{equation*}
G_{ij,\downarrow}(\omega)=
\sum_{\lambda=1}^4
\frac{\bra{\phi_{\lambda}}\hat{c}^{\dag}_{j\downarrow}\ket{-,\uparrow}\bra{-,\uparrow}\hat{c}_{i\downarrow}\ket{\phi_{\lambda}}}{\omega-\left(e_{\lambda}-e_-\right)+i\eta}:=
\sum_{\lambda=1}^4G^{(e_{\lambda})}_{ij,\downarrow}(\omega).
\end{equation*}
Calling $\omega_{\lambda}:=e_{\lambda}-e_-$, the contributions relative to the transitions 
from the ground state to the $e_1$, $e_3$ and $e_4$ excited state are:
\begin{align*}
G^{(e_{\lambda})}_{11,\downarrow}(\omega)&=\frac{\bra{\phi_{\lambda}}\hat{c}^{\dag}_{1\downarrow}\ket{-,\uparrow}\bra{-,\uparrow}\hat{c}_{1\downarrow}\ket{\phi_{\lambda}}}{\omega-\omega_{\lambda}+i\eta}=
\\
&=\frac{1}{\left|\mathcal{N}_{\lambda}\right|^2} \frac{\left[\cos\rho\left(\frac{2}{e_{\lambda}+(D-U)}-e_{\lambda}\right)+\sin\rho\right]^2}{\omega-\omega_{\lambda}+i\eta}
\\
G^{(e_{\lambda})}_{12,\downarrow}(\omega)&=G^{(e_{\lambda})}_{21,\downarrow}(\omega)=\frac{\bra{\phi_{\lambda}}\hat{c}^{\dag}_{2\downarrow}\ket{-,\uparrow}\bra{-,\uparrow}\hat{c}_{1\downarrow}\ket{\phi_{\lambda}}}{\omega-\omega_{\lambda}+i\eta}=\\
&=\frac{1}{\left|\mathcal{N}_{\lambda}\right|^2} \frac{\left[\cos\rho-\sin\rho\frac{2}{e_{\lambda}+(D-U)}\right]
	}{\omega-\omega_{\lambda}+i\eta}\cdot\\
&\qquad\qquad\cdot\bigl[\cos\rho\bigl(\tfrac{2}{e_{\lambda}+(D-U)}-e_{\lambda}\bigr)+\sin\rho\bigr]
\\
G^{(e_{\lambda})}_{22,\downarrow}(\omega)&=\frac{\bra{\phi_{\lambda}}\hat{c}^{\dag}_{2\downarrow}\ket{-,\uparrow}\bra{-,\uparrow}\hat{c}_{2\downarrow}\ket{\phi_{\lambda}}}{\omega-\omega_{\lambda}+i\eta}
=\\&=\frac{1}{\left|\mathcal{N}_{\lambda}\right|^2} \frac{\left[\cos\rho-\sin\rho\frac{2}{e_{\lambda}+(D-U)}\right]^2}{\omega-\omega_{\lambda}+i\eta}
\end{align*}
while for the second pole $\lambda=2$:
\begin{align*}
G^{(e_2)}_{11,\downarrow}(\omega)&=\frac{\bra{\phi_2}\hat{c}^{\dag}_{1\downarrow}\ket{-,\uparrow}\bra{-,\uparrow}\hat{c}_{1\downarrow}\ket{\phi_2}}{\omega-\omega_{2}+i\eta}
=\frac{\frac{1}{2}\sin^2\rho}{\omega-\omega_{2}+i\eta}
\\
G^{(e_2)}_{12,\downarrow}(\omega)&=G^{(e_2)}_{21,\downarrow}(\omega)=\frac{\bra{\phi_2}\hat{c}^{\dag}_{2\downarrow}\ket{-,\uparrow}\bra{-,\uparrow}\hat{c}_{1\downarrow}\ket{\phi_2}}{\omega-\omega_{2}+i\eta}=
\\
&=\frac{-\frac{1}{2}\sin\rho\cos\rho}{\omega-\omega_{2}+i\eta}
\\
G^{(e_2)}_{22,\downarrow}(\omega)&=\frac{\bra{\phi_2}\hat{c}^{\dag}_{2\downarrow}\ket{-,\uparrow}\bra{-,\uparrow}\hat{c}_{2\downarrow}\ket{\phi_2}}{\omega-\omega_{2}+i\eta}
=\frac{\frac{1}{2}\cos^2\rho}{\omega-\omega_{2}+i\eta}
\end{align*}
Finally, the spin--down Green's function takes the form of Eq. \eqref{eq:GFN1down}, namely $
G_{ij,\downarrow}(\omega)= \sum_{\lambda=1}^{4} \frac{f_{ij}^{\lambda}}{\omega-\omega_{\lambda}+i\eta}
$, with the amplitudes $f_{ij}^{\lambda}$ given by:
\begin{equation*}
f_{ij}^{\lambda=2}=\frac{1}{2}\Bigl[\delta_{i1}\sin\rho-\delta_{i2}\cos\rho\Bigr]\Bigl[\delta_{j1}\sin\rho-\delta_{j2}\cos\rho\Bigr]
\end{equation*}
\begin{multline*}
f^{\lambda\neq2}_{ij}=\frac{1}{\left|\mathcal{N}_{\lambda}\right|^2}
\\
\times \left[\delta_{i1}\left(\cos\rho\left(\frac{2}{e_{\lambda}+(D-U)}-e_{\lambda}\right)+\sin\rho\right)+\right.
\\
\left.+\delta_{i2}\left(\cos\rho-\sin\rho\frac{2}{e_{\lambda}+(D-U)}\right)\right]
 \\ \times
\left[\delta_{j1}\left(\cos\rho\left(\frac{2}{e_{\lambda}+(D-U)}-e_{\lambda}\right)+\sin\rho\right)+\right.
\\
\left.+\delta_{j2}\left(\cos\rho-\sin\rho\frac{2}{e_{\lambda}+(D-U)}\right)\right].
\end{multline*}

\subsection{Symmetric limit}
In the limit $D\to0$,  
the four eigenenergies of the $N=2$ Hamiltonian \eqref{eq:jtuyfnmhf} become
$e_1=\frac{U-c}{2}$, $e_2=0$, $e_3=U$ and $e_4=\frac{U+c}{2}$, hence
the four poles of the spin--down $N=1$ Green's function are:
\begin{align*}
\omega_1&=1+\tfrac{U-c}{2}\\
\omega_2&=1\\
\omega_3&=1+U\\
\omega_4&=1+\tfrac{U+c}{2}.
\end{align*}
If $D=0$, the two sites are completely equivalent and, in particular, the Green's function assumes the same value on both. As a consequence, the Green's function is diagonal in the bonding--antibonding basis, and reads:
\begin{gather*}
G_{-}(\omega)=
\frac{\frac{1}{2}+\frac{2}{c}}{\omega-\omega_1+i\eta}+
\frac{\frac{1}{2}-\frac{2}{c}}{\omega-\omega_4+i\eta}
\\
G_{+}(\omega)=
\frac{\frac{1}{2}}{\omega-\omega_2+i\eta}+
\frac{\frac{1}{2}}{\omega-\omega_3+i\eta}.
\end{gather*}

\section{The GW approximation in the asymmetric Hubbard dimer}
\label{sec:gwasym}

In  a basis of local orbitals, the interactions appearing in the GW expressions should have four indices \cite{Martin2016}.
However, in the context  of the Hubbard model the overlap between orbitals on different sites is neglected, and the expressions simplify \cite{Romaniello2009}. Taking also into account the fact the Green's function is spin-diagonal, the spin-summed polarizabiliyty is 
\begin{equation*}
\Pi_{ij}^{0}(\omega)= \sum_\sigma \int\frac{d\omega'}{2\pi i}e^{i\omega'\eta}G^0_{ij,\sigma}(\omega+\omega')G^0_{ji,\sigma}(\omega'),
\end{equation*}
where $G^0_{ji,\sigma}(\omega)$ is given by Eq. \eqref{eq:asym_up}, or
\begin{equation*}
\begin{split}
\Pi_{ij}^{0}(\omega)&= \sum_\sigma \frac{(-1)^{(i-j)}\delta_{\sigma,\uparrow}}{D^2+4}\left[
\frac{1}{\omega-\Delta e+i\eta}-\frac{1}{\omega+\Delta e-i\eta}
\right],
\end{split}
\end{equation*}
where $\Delta e\equiv e_+-e_-=\sqrt{4+D^2}$ is the gap between the excited (antibonding) and the ground (bonding) states.

The polarization screens the bare interaction $U^0_{ij}=U\delta_{ij}$ via the Dyson equation $W_{ij}=U^0_{ij}+U^0_{ik}\Pi^0_{kl}W_{lj}$. 
Inverting the Dyson equation we obtain the spin--independent screened interaction $W$:
\begin{equation*}
{W}_{ij}(\omega)=U\delta_{ij}+\frac{(-1)^{(i-j)}\frac{U^2}{2l}}{\sqrt{1+\frac{D^2}{4}}}
\left[
\frac{1}{\omega-l+i\eta}-\frac{1}{\omega+l-i\eta}
\right]
\end{equation*}
with $l^2:=4\Bigl(1+\frac{D^2}{4}\Bigr)+\frac{2U}{\sqrt{1+\frac{D^2}{4}}}$. The first term $U\delta_{ij}$ yields, in the self energy, the exchange term, while the rest is the screening due to the polarization of the spin--up electron (only a single bubble). 
From this result, the self energy is found as the convolution of $W$ with the non--interacting Green's function $G_0$:
\begin{align}
{\Sigma}_{ij,\sigma}^{G_0W_0}(\omega)=i\int\frac{d\omega'}{2\pi}e^{i\omega'\eta}{G}^0_{ij,\sigma}(\omega+\omega'){W}_{ij}(\omega')
\label{eq:frewiuhwj}
\end{align}
Evaluating the integral, we obtain:
\begin{multline}
{\Sigma}_{ij,\sigma}^{G_0W_0}(\omega)={\Sigma}_{ij,\sigma}^{X}
+
\frac{(-1)^{(i-j)}\frac{U^2}{2l}}{\sqrt{1+\frac{D^2}{4}}}\left[\frac{f_{ij}^{+}}{\omega-\left({e}_++l\right)+i\eta}+\right.
\\
\left.
+\frac{f_{ij}^{-}}{\omega-\left({e}_--l\operatorname{sign}\sigma\right)-i\eta\operatorname{sign}\sigma}
\right]
\label{eq:SEGW}
\end{multline}
with the exchange self energy ${\Sigma}_{ij,\sigma}^{X}\equiv -\int\frac{d\omega'}{2\pi i}e^{i\omega'\eta}G^0_{ij,\sigma}(\omega+\omega')U^0_{ij}$, given by $-\delta_{\sigma,\uparrow}n_{ij}U^0_{ij}
=-\delta_{\sigma,\uparrow}\delta_{ij}Un_{i}$, which exactly balances the Hartree potential $v_i^H\equiv Un_i$ when considering the removal of the single electron from the ground state. 
From the self energy, the Green's function can be found by inverting the quantity $G_0^{-1}-v^H-\Sigma^{G_0W_0}\equiv G^{-1}_{G_0W_0}$. Its poles are the solution to the equation $\det\left[G_0^{-1}-v^H-{\Sigma}^{G_0W_0}\right]=0$. They can be expressed analytically for the symmetric dimer, $D=0$, in which the self energy reduces to \cite{Romaniello2009}:
\begin{multline*}
{\Sigma}_{ij,\sigma}^{G_0W_0}(\omega)
\stackrel{D=0}{=}{\Sigma}_{ij,\sigma}^{X}
+
\frac{U^2}{4l}\left[\frac{1}{\omega-\left(1+l\right)+i\eta}+\right.
\\
+\left.\frac{(-1)^{(i-j)}}{\omega+\left(1+l\operatorname{sign}\sigma\right)-i\eta\operatorname{sign}\sigma}
\right],
\end{multline*}
and they are, for the spin--up Green's function:
\begin{equation}
{\omega}^{G_0W_0\,\uparrow}_{1,2,3,4}\stackrel{D=0}{=}\pm\frac{l}{2}\pm\frac{1}{2}\sqrt{\Bigl(l+2\Bigr)^2+\frac{2U^2}{l}},
\label{eq:wrgie}
\end{equation}
while for the spin--down part:
\begin{equation}
{\omega}^{G_0W_0\,\downarrow}_{1,2,3,4}\stackrel{D=0}{=}\frac{l+\frac{U}{2}}{2}\pm\frac{1}{2}\sqrt{\left(l\pm2-\frac{U}{2}\right)^2+\frac{2U^2}{l}},
\label{eq:wrgiert}
\end{equation}
where in both expressions the ``$\pm$'' signs are unrelated in order to form four poles each. In the more general $D\neq0$ case, the spin--down poles are the solutions to the following equation:
\begin{multline}
0=\det\left[\left(
\begin{matrix}
\omega-\frac{U}{2}-h & 1
\\
1 & \omega-\frac{U}{2}+h
\end{matrix}
\right)
+\right.\\-
\frac{\frac{1}{\sqrt{1+\frac{D^2}{4}}}\frac{U^2}{2l}}{\left(\omega-l\right)^2-\left(1+\frac{D^2}{4}\right)}
\left.\left(
\begin{matrix}
\omega-l+\frac{D^2}{4} & 1
\\
1 & \omega-l-\frac{D^2}{4}
\end{matrix}
\right)\right].
\label{eq:poles_asym_GW}
\end{multline}

Note that in practice there is no unique recipe for building the non--self--consistent GW self energy. An alternative definition, for instance, is replacing the non--interacting Green's function $G_0$ with the Hartree Green's function $G^{\rm H}$, which reads:
\begin{equation}
G^{\rm H}_{ij,\sigma}(\omega)=
\frac{
	f_{ij}^{{\rm H}\,-}}
{\omega-\omega^{\rm H}_--i\eta{\rm sign}\sigma}
+
\frac{
	f_{ij}^{{\rm H}\,+}}
{\omega-\omega_+^{\rm H}+i\eta},
\label{eq:Har_asym}
\end{equation}
with weights $f_{ij}^{{\rm H}\,\pm}$ defined as their free counterparts $f_{ij}^{\pm}$ with the substitution $\frac{D}{2}\to h$ \cite{marco_phd}.
The Hartree polarizability $\Pi^{\rm H}\sim-iG^{\rm H}G^{\rm H}$ yields a screened interaction which has the same expression as the one above, but with the replacement $\frac{D}{2}\to h$. Finally, also the self energy has the  expression of Eq. \eqref{eq:SEGW} with $\frac{D}{2}\to h$ and the two free eigenenergies $\varepsilon_{\pm}$ replaced by the Hartree eigenenergies $\omega_{\pm}^{\rm H}$. This expression for the self energy yields poles of the Green's function with similar behaviour as the ones resulting from Eq. \eqref{eq:SEGW}; a difference is that in this case the levels do not cross, as in the exact solution.

\bibliographystyle{epj}
\bibliography{biblio}

\end{document}